\documentclass[longauth]{aa}

\usepackage{amsmath}
\usepackage{natbib}
\usepackage{makecell}
\usepackage{graphicx}
\usepackage{tabularx}
\usepackage{xcolor}
\usepackage{txfonts}
\usepackage{lscape}
\usepackage[colorlinks=true, allcolors=blue]{hyperref}
\usepackage[normalem]{ulem}
\usepackage{academicons}
\graphicspath{{figures/}}
\usepackage{subfiles}

\newcommand{\numsrcs}{601}
\newcommand{\startobs}{6 October 2024}
\newcommand{\lastobs}{3 December 2024}
\newcommand{\mapdepthnine}{4.6}
\newcommand{\mapdepthonefive}{3.8}
\newcommand{\mapdepthtwotwo}{13.2}
\newcommand{\numedfs}{334}
\newcommand{\edfsnumsynch}{182}
\newcommand{\edfsnumdusty}{152}
\newcommand{\edfsnumdsfgcand}{114}

\newcommand{\edfsnumextended}{23}

\newcommand{\edfsnumFalse}{311}
\newcommand{\edfsfracsynch}{54.5}
\newcommand{\edfsfracdusty}{45.5}

\newcommand{\edfsfracdsfgcand}{34.1}
\newcommand{\edfsfracextended}{6.9}

\newcommand{\edfsnumdetonlynine}{63}
\newcommand{\edfsnumdetonlyonefive}{26}
\newcommand{\edfsnumdetonlytwotwo}{79}
\newcommand{\edfsnumdetonlynineandonefive}{85}
\newcommand{\edfsnumdetonlyonefiveandtwotwo}{45}
\newcommand{\edfsnumdetonlynineandtwotwo}{0}
\newcommand{\edfsnumdetonlyallbands}{36}

\newcommand{\edfsfracdetonlynine}{18.9}
\newcommand{\edfsfracdetonlyonefive}{7.8}
\newcommand{\edfsfracdetonlytwotwo}{23.7}
\newcommand{\edfsfracdetonlynineandonefive}{25.4}
\newcommand{\edfsfracdetonlyonefiveandtwotwo}{13.5}
\newcommand{\edfsfracdetonlynineandtwotwo}{0.0}
\newcommand{\edfsfracdetonlyallbands}{10.8}

\newcommand{\fracedfs}{55.6}
\newcommand{\numobs}{197}
\newcommand{\pixelmaskarea}{57}
\newcommand{\percentinterparea}{0.4}
\newcommand{\interparea}{0.2}
\newcommand{\numinterpsrcs}{14}
\newcommand{\numdetonlynine}{120}
\newcommand{\numdetonlyonefive}{48}
\newcommand{\numdetonlytwotwo}{146}
\newcommand{\numdetonlynineandonefive}{146}
\newcommand{\numdetonlyonefiveandtwotwo}{79}
\newcommand{\numdetonlynineandtwotwo}{0}
\newcommand{\numdetonlyallbands}{62}

\newcommand{\fracdetonlynine}{20.0}
\newcommand{\fracdetonlyonefive}{8.0}
\newcommand{\fracdetonlytwotwo}{24.3}
\newcommand{\fracdetonlynineandonefive}{24.3}
\newcommand{\fracdetonlyonefiveandtwotwo}{13.1}
\newcommand{\fracdetonlynineandtwotwo}{0.0}
\newcommand{\fracdetonlyallbands}{10.3}

\newcommand{\numsynch}{324}
\newcommand{\numdusty}{277}
\newcommand{\numnans}{80}

\newcommand{\numdsfgcand}{210}

\newcommand{\numnocps}{243}
\newcommand{\numextended}{38}

\newcommand{\numFalse}{563}
\newcommand{\fracsynch}{53.9}
\newcommand{\fracdusty}{46.1}

\newcommand{\fracdsfgcand}{34.9}
\newcommand{\fracextended}{6.3}
\newcommand{\alphacut}{1.5}

\newcommand{\Sminnine}{1.7}
\newcommand{\Sminonefive}{2.0}
\newcommand{\Smintwotwo}{6.5}

\newcommand{\rassocATtwentyG}{30.0}

\newcommand{\rassocSZ}{43.2}

\newcommand{\rassocWISE}{30.0}
\newcommand{\rassocASKAPlow}{30.0}
\newcommand{\rassocASKAPmid}{30.0}
\newcommand{\rassocASKAPhigh}{30.0}

\newcommand{\nummatchesATtwentyG}{18}

\newcommand{\nummatchesSZ}{104}

\newcommand{\nummatchesWISE}{44}
\newcommand{\nummatchesASKAPlow}{328}
\newcommand{\nummatchesASKAPmid}{327}
\newcommand{\nummatchesASKAPhigh}{326}
\newcommand{\nummatchesASKAP}{343}

\newcommand{\PrandomATtwentyG}{0.01}

\newcommand{\PrandomSZ}{0.1}

\newcommand{\PrandomWISE}{0.7}
\newcommand{\PrandomASKAPlow}{1.9}
\newcommand{\PrandomASKAPmid}{1.8}
\newcommand{\PrandomASKAPhigh}{1.9}

\newcommand{\PrandomSPT}{0.5}
\newcommand{\srcdensitySPT}{10.6}
\newcommand{\srcdensitySPTnine}{5.8}
\newcommand{\srcdensitySPTonefive}{5.9}
\newcommand{\srcdensitySPTtwotwo}{5.0}
\newcommand{\srcdensityATtwentyG}{0.3}

\newcommand{\srcdensitySZ}{2.2}

\newcommand{\srcdensityWISE}{32.6}
\newcommand{\srcdensityASKAPlow}{87.0}
\newcommand{\srcdensityASKAPmid}{83.3}
\newcommand{\srcdensityASKAPhigh}{86.2}

\newcommand{\rmsRAoffset}{3.8}
\newcommand{\rmsDECoffset}{1.8}
\newcommand{\meanRAoffset}{$-$0.8}
\newcommand{\meanDECoffset}{$-$0.8}

\newcommand{\internalrassoc}{43.2}

\newcommand{\numclusterstot}{217}
\newcommand{\numclustersedfs}{121}
\newcommand{\numclusterswithredshifts}{188}
\newcommand{\fracclusterswithredshifts}{87\%}
\newcommand{\numclusterswithhighredshifts}{59}

\newcommand{\medclusterz}{0.70}
\newcommand{\minclusterz}{0.07}
\newcommand{\maxclusterz}{1.6+}
\newcommand{\clusterpurity}{87\%}
\newcommand{\minmass}{\mass$=1.03 \times 10^{14}$\msun}
\newcommand{\medmass}{\mass$=1.91 \times 10^{14}$\msun}
\newcommand{\maxmass}{\mass$=7.68 \times 10^{14}$\msun}
\newcommand{\msun}{\ensuremath{M_\odot/h_{70} }}
\newcommand{\mass}{\mbox{$M_{\mbox{\scriptsize 500c}}$}}

\newcommand{\numclustersaboveone}{49}
\newcommand{\orcid}[1]{%
  \href{https://orcid.org/#1}{%
    \raisebox{-0.15ex}{\includegraphics[height=1.0em]{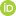}}%
  }%
}

\begin{document}

\title{Millimeter-wave observations of Euclid Deep Field South using \\ the South Pole Telescope}
\subtitle{A data release of temperature maps and catalogs}

\author{
M.~Archipley\textsuperscript{1,2}\corrauth{archipleym@uchicago.edu} \orcid{0000-0002-0517-9842} \and 
A.~Hryciuk\textsuperscript{3,2} \and 
L.~E.~Bleem\textsuperscript{4,2,1}\orcid{0000-0001-7665-5079} \and 
K.~Kornoelje\textsuperscript{1,2,4} \and 
M.~Klein\textsuperscript{5}\orcid{0000-0002-8248-4488} \and 
A.~J.~Anderson\textsuperscript{6,2,1} \and 
B.~Ansarinejad\textsuperscript{7} \and 
M.~Aravena\textsuperscript{8,9}\orcid{0000-0002-6290-3198} \and 
L.~Balkenhol\textsuperscript{10}\orcid{0000-0001-6899-1873} \and 
P.~S.~Barry\textsuperscript{11}\orcid{0000-0001-9103-9354}
% ...
}

\institute{
\textsuperscript{1}{Department of Astronomy and Astrophysics, University of Chicago, 5640 South Ellis Avenue, Chicago, IL, 60637, USA} \\
\textsuperscript{2}{Kavli Institute for Cosmological Physics, University of Chicago, 5640 South Ellis Avenue, Chicago, IL, 60637, USA} \\
\textsuperscript{3}{Department of Physics, University of Chicago, 5640 South Ellis Avenue, Chicago, IL, 60637, USA} \\
\textsuperscript{4}{High-Energy Physics Division, Argonne National Laboratory, 9700 South Cass Avenue, Lemont, IL, 60439, USA} \\
\textsuperscript{5}{University Observatory, Faculty of Physics, Ludwig-Maximilians-Universit\"{a}t, Scheinerstr. 1, 81679 Munich, Germany} \\
\textsuperscript{6}{Fermi National Accelerator Laboratory, MS209, P.O. Box 500, Batavia, IL, 60510, USA} \\
\textsuperscript{7}{School of Physics, University of Melbourne, Parkville, VIC 3010, Australia} \\
\textsuperscript{8}{Instituto de Estudios Astrof\'{\i}cos, Facultad de Ingenier\'{\i}a y Ciencias, Universidad Diego Portales, Av. Ej\'ercito 441, Santiago, Chile} \\
\textsuperscript{9}{Millenium Nucleus for Galaxies} \\
\textsuperscript{10}{Sorbonne Universit\'e, CNRS, UMR 7095, Institut d'Astrophysique de Paris, 98 bis bd Arago, 75014 Paris, France} \\
\textsuperscript{11}{School of Physics and Astronomy, Cardiff University, Cardiff, CF24 3YB, United Kingdom} \\
\textsuperscript{12}{Kavli Institute for Particle Astrophysics and Cosmology, Stanford University, 452 Lomita Mall, Stanford, CA, 94305, USA} \\
\textsuperscript{13}{Department of Physics, Stanford University, 382 Via Pueblo Mall, Stanford, CA, 94305, USA} \\
\textsuperscript{14}{SLAC National Accelerator Laboratory, 2575 Sand Hill Road, Menlo Park, CA, 94025, USA} \\
\textsuperscript{15}{Enrico Fermi Institute, University of Chicago, 5640 South Ellis Avenue, Chicago, IL, 60637, USA} \\
\textsuperscript{16}{Department of Astronomy, University of Florida, Gainesville, FL 32611, USA} \\
\textsuperscript{17}{Department of Physics and Astronomy, University of British Columbia, 6225 Agricultural Rd., Vancouver, V6T 1Z1, Canada} \\
\textsuperscript{18}{National Research Council, Herzberg Astronomy and Astrophysics, 5071 West Saanich Rd., Victoria, V9E 2E7, Canada} \\
\textsuperscript{19}{Department of Physics and Atmospheric Science, Dalhousie University, Halifax, Nova Scotia, Canada} \\
\textsuperscript{20}{University of Chicago, 5640 South Ellis Avenue, Chicago, IL, 60637, USA} \\
\textsuperscript{21}{Department of Physics, University of California, Berkeley, CA, 94720, USA} \\
\textsuperscript{22}{Universit\'e Paris-Saclay, Universit\'e Paris Cit\'e, CEA, CNRS, AIM, 91191, Gif-sur-Yvette, France} \\
\textsuperscript{23}{Department of Astronomy, University of Illinois Urbana-Champaign, 1002 West Green Street, Urbana, IL, 61801, USA} \\
\textsuperscript{24}{High Energy Accelerator Research Organization (KEK), Tsukuba, Ibaraki 305-0801, Japan} \\
\textsuperscript{25}{Wits Centre for Astrophysics, University of the Witwatersrand, 1 Jan Smuts Avenue, 2000, Johannesburg, South Africa} \\
\textsuperscript{26}{Department of Physics, University of Pretoria, Hatfield, Pretoria, 0028, South Africa} \\
\textsuperscript{27}{Department of Physics and McGill Space Institute, McGill University, 3600 Rue University, Montreal, Quebec H3A 2T8, Canada} \\
\textsuperscript{28}{Canadian Institute for Advanced Research, CIFAR Program in Gravity and the Extreme Universe, Toronto, ON, M5G 1Z8, Canada} \\
\textsuperscript{29}{Joseph Henry Laboratories of Physics, Jadwin Hall, Princeton University, Princeton, NJ 08544, USA} \\
\textsuperscript{30}{Department of Astrophysical and Planetary Sciences, University of Colorado, Boulder, CO, 80309, USA} \\
\textsuperscript{31}{Department of Physics, University of Illinois Urbana-Champaign, 1110 West Green Street, Urbana, IL, 61801, USA} \\
\textsuperscript{32}{Department of Physics and Astronomy, University of California, Los Angeles, CA, 90095, USA} \\
\textsuperscript{33}{Department of Physics and Astronomy, Michigan State University, East Lansing, MI 48824, USA} \\
\textsuperscript{34}{Institute of Cosmology and Gravitation, University of Portsmouth, Dennis Sciama Building, Portsmouth PO1 3FX, UK} \\
\textsuperscript{35}{Department of Physics \& Astronomy, University of California, One Shields Avenue, Davis, CA 95616, USA} \\
\textsuperscript{36}{Universit\"{a}t Innsbruck, Institut f\"{u}r Astro- und Teilchenphysik, Technikerstrasse 25, 6020 Innsbruck, Austria} \\
\textsuperscript{37}{Cosmic Dawn Center (DAWN), Technical University of Denmark, DTU Space, Elektrovej 327, 2800 Kgs Lyngby, Denmark} \\
\textsuperscript{38}{Department of Physics and Astronomy, University College London, Gower Street, London, WC1E 6BT, UK} \\
\textsuperscript{39}{Department of Physics and Astronomy, Northwestern University, 633 Clark St, Evanston, IL, 60208, USA} \\
\textsuperscript{40}{CASA, Department of Astrophysical and Planetary Sciences, University of Colorado, Boulder, CO, 80309, USA} \\
\textsuperscript{41}{Department of Physics, University of Colorado, Boulder, CO, 80309, USA} \\
\textsuperscript{42}{Department of Physics, Case Western Reserve University, Cleveland, OH, 44106, USA} \\
\textsuperscript{43}{Center for AstroPhysical Surveys, National Center for Supercomputing Applications, Urbana, IL, 61801, USA} \\
\textsuperscript{44}{Dunlap Institute for Astronomy \& Astrophysics, University of Toronto, 50 St. George Street, Toronto, ON, M5S 3H4, Canada} \\
\textsuperscript{45}{David A. Dunlap Department of Astronomy \& Astrophysics, University of Toronto, 50 St. George Street, Toronto, ON, M5S 3H4, Canada} \\
\textsuperscript{46}{NSF-Simons AI Institute for the Sky (SkAI), 172 E. Chestnut St., Chicago, IL 60611, USA} \\
\textsuperscript{47}{Vera C. Rubin Observatory Project Office, 950 N Cherry Ave, Tucson, AZ, 85719, USA} \\
\textsuperscript{48}{Center for Astrophysics \textbar{} Harvard \& Smithsonian, 60 Garden Street, Cambridge, MA, 02138, USA} \\
\textsuperscript{49}{Max-Planck-Institut f\"{u}r Radioastronomie, Auf dem H\"{u}gel 69, 53121 Bonn, Germany} \\
}
\date{%
Received 3 June 2025 / Accepted 14 November 2025%
\thanks{This version incorporates corrections published in the associated corrigendum, A\&A, 713, C1 (2026), DOI: \href{https://doi.org/10.1051/0004-6361/202662263e}{10.1051/0004-6361/202662263e}.
}%
}

\abstract 
{The South Pole Telescope third-generation camera (SPT-3G) has observed over $10\,000$ square degrees of sky at 95, 150, and 220 GHz (3.3, 2.0, 1.4 mm, respectively) and will significantly overlap the ongoing $14\,000$ square-degree \textit{Euclid} Wide Survey. The \textit{Euclid} collaboration recently released \textit{Euclid} Deep Field South (EDF-S) observations of 23 square degrees at wide field depths in the first quick data release (Q1).}
{With the goal of releasing complementary millimeter-wave data and encouraging legacy science, we performed dedicated observations of a \pixelmaskarea{}-square-degree field overlapping the EDF-S.}
{The observing time totaled 20 days, and we reached noise depths of 4.3, \mapdepthonefive{}, and \mapdepthtwotwo{} $\mu$K-arcmin at 95, 150, and 220 GHz, respectively.}
{In this work we present the temperature maps and two catalogs constructed from these data. The emissive source catalog contains \numsrcs{} objects (\numedfs{} inside EDF-S) with 54\% synchrotron-dominated sources and 46\% thermal dust emission-dominated sources. The 5$\sigma$ detection thresholds are \Sminnine{}, \Sminonefive{}, and \Smintwotwo{} mJy in the three bands. The cluster catalog contains \numclusterstot{} cluster candidates (\numclustersedfs{} inside EDF-S) with median mass \medmass{} and median redshift $z=\medclusterz{}$, corresponding to an order-of-magnitude improvement in cluster density over previous tSZ-selected catalogs in this region (3.81 clusters per square degree).}
{The overlap between SPT and \textit{Euclid} data will enable a range of multiwavelength studies of the aforementioned source populations. This work serves as the first step toward joint projects between SPT and \textit{Euclid} and provides a rich dataset containing information on galaxies, clusters, and their environments.}

\keywords{catalogs -- surveys -- galaxies: clusters: general -- galaxies: high-redshift -- cosmology: observations -- submillimeter: galaxies}

\maketitle

\section{Introduction} \label{sec:introduction}
Many future astronomical discoveries are expected to come from leveraging multiwavelength datasets to improve the reach and impact of next-generation surveys. Two projects contributing to such multiwavelength insights are the South Pole Telescope \citep[SPT;][]{carlstrom11} and \textit{Euclid} \citep{mellier25},  respectively at millimeter and optical/infrared wavelengths.

The SPT began observations in 2007, and the current camera, SPT-3G \citep{benson14, sobrin22}, began operating in 2017. The SPT and its cameras were designed for low-noise, high-resolution (1 arcmin) observations of the cosmic microwave background (CMB), and the total sky area observed with the SPT-3G camera through 2024 is over $10\,000$ square degrees \citep{prabhu24}. The deep arcminute-resolution millimeter-wave data enable precision cosmology with primary and secondary CMB anisotropies \citep[e.g.,][]{dutcher21, reichardt21, balkenhol21, balkenhol23}, CMB lensing \citep[e.g.,][]{pan23, ge24}, and galaxy clusters selected through the thermal Sunyaev-Zel'dovich (tSZ) effect (e.g., \citealt{bocquet24b}) as well as astrophysical studies using the kinematic SZ effect \citep[kSZ; e.g.,][]{raghunathan24}, the tSZ effect \citep[e.g.,][]{bleem22}, galactic and extragalactic transients \citep{whitehorn16, guns21, tandoi24}, and high-redshift star-burst galaxies, which inform studies of galaxy evolution in the early Universe \citep[e.g.,][]{vieira13, reuter20}.

\textit{Euclid} launched in 2023. Over its six-year mission, it will observe $14\,000$ square degrees in the ``Wide Survey'' \citep{scaramella22, mellier25}. The \textit{Euclid} VISual instrument (VIS) observes from 500-900 nm and the Near-Infrared Spectrometer and Photometer (NISP) observes 900-$2\,000$ nm in the Y, J, and H passbands. In addition to the Wide Survey, 12\% of the total \textit{Euclid} observing time is dedicated to observing three deep fields, with goals of calibrating the wide field data and enabling legacy science. The three deep fields have four-band \emph{Spitzer}/IRAC coverage \citep{moneti22}, and the \textit{Euclid} Deep Field South (EDF-S) in particular overlaps with two Legacy Survey of Space and Time (LSST) deep-drilling fields, Rubin-a and Rubin-b \citep{mellier25}. \textit{Euclid} catalog parameter estimates such as photometric redshifts and stellar masses will be informed by joint efforts between LSST \citep{rhodes17, mitra24} and other multiwavelength observatories. In March 2025, the first of many scheduled Quick Data Releases, Q1, was released and included deep field data at the depths of the planned \textit{Euclid} Wide Survey \citep{pettorino24}.

The data release described in this work includes SPT data products derived from or supporting observations overlapping the EDF-S, including maps at 95, 150, and 220 GHz and source catalogs that complement the optical/infrared EDF-S data taken by \textit{Euclid}. The map data products described in this work are similar to those in \citet{schaffer11}, which detailed products for a previous-generation SPT camera. Two of the three main populations of objects presented in this work, high-redshift dusty galaxies and galaxy clusters, have properties that make them detectable with a particularly clean selection by the SPT. As both of these samples are drawn from observations at depths approximating that of their larger area surveys (i.e., the \textit{Euclid} Wide Field and the SPT-3G $1\,500$-square-degree main survey), analyses of data products from this first joint survey field will provide important milestones regarding the exploitation of the full combined dataset. The products are publicly available online.\footnote{\url{https://pole.uchicago.edu/public/data/edfs25/}}

The typical millimeter-wave emissive source population at the flux levels probed in this work is composed primarily of synchrotron-emitting active galactic nuclei (AGNs) and dust-enshrouded star-forming galaxies emitting quasi-thermally \citep{everett20}. The two populations can be distinguished by their SPT spectral index. The AGN population probed at millimeter wavelengths (compared to typical radio-selected samples) is dominated by flat spectrum radio quasars \citep[FSRQs;][]{condon92, padovani17}. The dusty SPT sources are a combination of low-redshift luminous and ultraluminous infrared galaxies (LIRGs and ULIRGs) and distant dusty star-forming galaxies \citep[DSFGs;\footnote{DSFGs may also be referred to as ``submillimeter galaxies'' or SMGs due to typical strong submillimeter emission as described in e.g., \citealt{smail02, casey14, hodge20}.}][]{everett20}, including some galaxy protoclusters containing numerous DSFGs in a single SPT beam \citep[e.g.,][]{overzier16, casey16, miller18, wang21a}. Thought to be progenitors of present-day elliptical galaxies \citep{blain04, casey14}, DSFGs are of broad interest and relevance to the community, as the number density of such sources are not yet well understood \citep[see e.g.,][]{hayward21}, and they are integral to our picture of the cosmic star formation history of the Universe \citep{madau14, koprowski17}. These sources are efficiently detectable in the millimeter and submillimeter bands, as they are dust-enshrouded to the point of being obscured at optical wavelengths \citep{blain93, hughes98, blain02, casey14}, but a strong negative $K$ correction enables their detection in millimeter wavelengths almost independent of redshift \citep[$1 \lesssim z \lesssim 10$;][]{blain93, blain02}.

The millimeter regime is similarly well-suited to detecting clusters of galaxies. The hot dense gas in galaxy clusters interacts with CMB photons via the tSZ effect \citep{sunyaev72}. The tSZ surface brightness is redshift-independent, meaning that for a CMB instrument with angular resolution well-matched to the arcminute-scale signal, galaxy clusters can be detected to arbitrarily high redshift, in contrast to methods that rely on cluster emission (such as optical/infrared and X-ray), which suffer from cosmological dimming. The tSZ surface brightness is also a probe of the total cluster thermal energy, providing a low-scatter mass proxy. These properties lead to clean, approximately mass-limited cluster samples that are of great value for both cosmological and astrophysical studies \citep[e.g.,][]{carlstrom02, allen11, kravtsov12, bleem15b, bocquet24b}. This new joint dataset offers the possibility of impactful studies related to understanding the role of environment in the formation and evolution of galaxies, especially in the high-redshift universe at a crucial era ($1<z<2$), where there is evidence of strong changes in quenching efficiency and AGN activity in dense environments \citep[see e.g., the recent review by][and references therein]{alberts22}.   

This paper is organized as follows: In Sect. \ref{sec:instrument} we describe the SPT, SPT-3G camera, and SPT-3G EDF-S observations. In Sect. \ref{sec:mapmaking} we describe how we process the telescope data into maps. In Sect. \ref{sec:extractsignals} we outline the pipelines used for generating the emissive source and cluster catalogs. Section \ref{sec:resultsdataprod} presents the data products included in the release, such as maps, masks, and filter components. In Sect. \ref{sec:pointsourcecatalog} we describe the contents of the emissive source catalog, including the nature of the objects and associations with external datasets, and Sect. \ref{sec:clustercatalog} relates the results for the cluster catalog. We summarize our work in Sect. \ref{sec:conclusion}.

\paragraph{Conventions.} Cluster masses are reported in $M_{500c}$, defined as the mass enclosed in a radius, $r_{500c}$, at which the average enclosed density is 500$\times$ the critical density at the cluster redshift. In determining cluster masses we used the fiducial Lambda cold dark matter ($\Lambda$CDM) cosmology assumed in \cite{bleem15b} with $\sigma_8=0.80$, $\Omega_b = 0.046$, $\Omega_m = 0.30$, $h = 0.70$, $n_s(k_s=0.002) = 0.97$, and $\Sigma m_\nu=0.06$ eV. In accordance with the convention of CMB experiments, we report map noise depth in units of $\mu$K-arcmin, which expresses sky intensity as the equivalent CMB temperature fluctuation while referring to the typical noise level of a 1 square arcmin patch of map. This convention allows for a standardized noise comparison regardless of the choice of map pixel size when the noise is uncorrelated between pixels.

\section{Instrument and observations} \label{sec:instrument}
In this section we describe the characteristics of the telescope and camera. We also present the field observations and discuss them in the broader context of the SPT-3G surveys.

\subsection{The SPT and SPT-3G instruments} \label{sec:spt3gcamera}
The capabilities of SPT and the SPT-3G camera are described at length in \cite{carlstrom11} and \cite{sobrin22}, respectively, but we highlight relevant specifications in this section. The SPT has a 10-meter primary mirror and is located at the Amundsen-Scott South Pole Station. The site has the lowest annual median precipitable water vapor of any developed observatory on Earth, making it particularly suitable for ground-based millimeter observations. The SPT is sensitive to the arcminute angular scales required for tSZ surveys and has produced the deepest millimeter-wave wide surveys of the high-redshift galaxy population.

The SPT-3G camera has an improved detector count, field of view, and mapping speed than previous cameras on the SPT, resulting in significantly deeper maps over its predecessors. Though we do not include polarization data in the data release, the detectors are also sensitive to linear polarization in all three bands (95, 150, and 220 GHz), enabling studies of CMB polarization \citep{dutcher21, balkenhol21, balkenhol23, ge24}, polarized atmosphere \citep{coerver24}, and polarized transients \citep[e.g.,][]{guns21}.

\subsection{Euclid Deep Field South observations} \label{sec:observations}
\begin{figure}[h]
                \centering
                        \includegraphics[width=1\columnwidth]{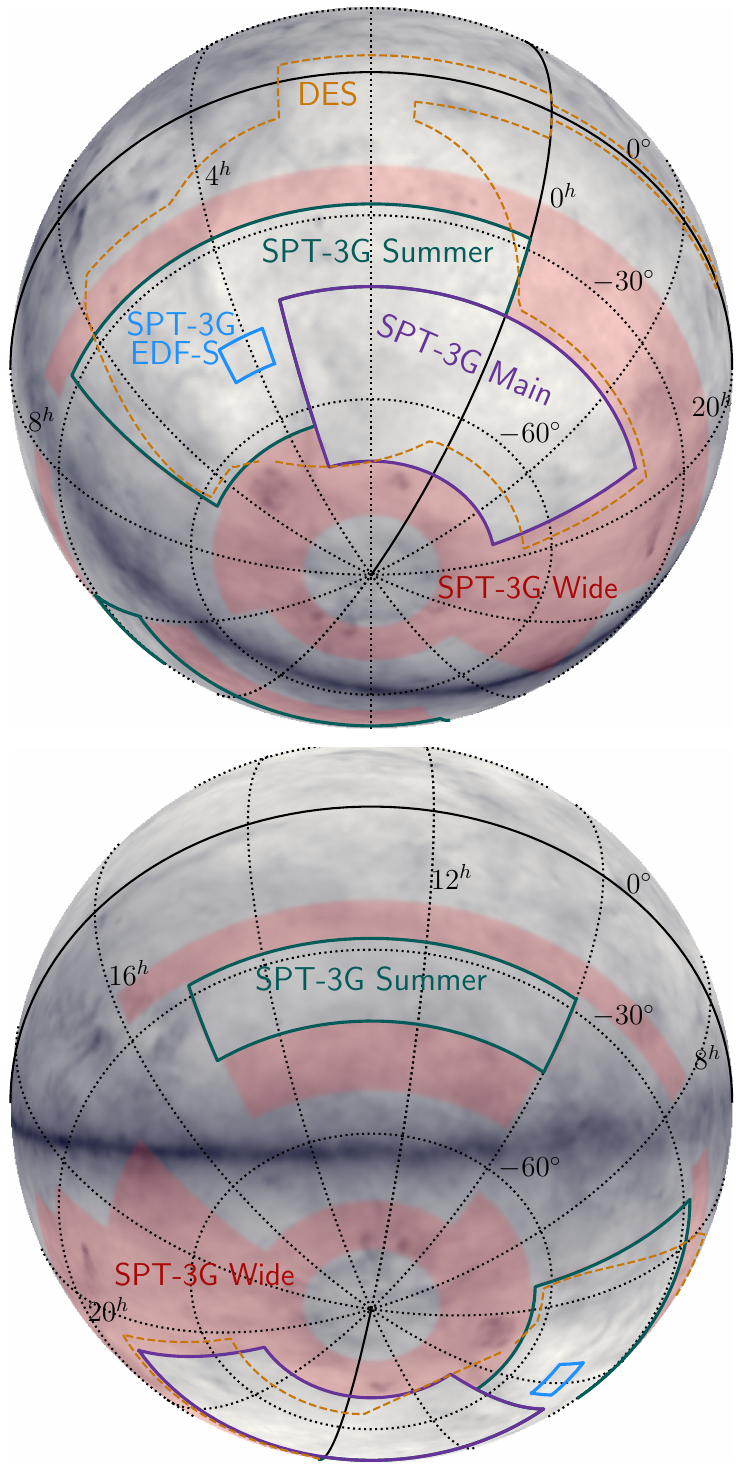}
                        \caption{Dedicated SPT-3G observations of the EDF-S (blue outline) were taken from October to December 2024. SPT-3G observations of the Main (purple outline), Summer (green outline), and Wide (red shade) fields total approximately $10\,000$ square degrees of sky coverage. The background image is a dust map from \textit{Planck} \citep{planck15-10}, and the Dark Energy Survey (indicated as DES) footprint is also shown (orange dotted outline).}
                \label{fig:allsptfields}
        \end{figure}

Since 2019, the SPT-3G camera has been used to observe the ``SPT-3G Main'' field during most austral winters, ``SPT-3G Summer'' fields during most austral summers, and the ``SPT-3G Wide'' field during the 2024 observing season. These fields and the \pixelmaskarea{}-square-degree ``SPT-3G EDF-S'' overlapping the 30-square-degree EDF-S\footnote{Though the Q1 area reported in \cite{aussel25} is 28.1 square degrees, the version of the mask we use is equivalent to 30 square degrees.} are shown in Fig. \ref{fig:allsptfields}.

We observed the EDF-S from 6 October 2024 to 3 December 2024 in \numobs{} non-consecutive individual observations (we observed the SPT-3G Main field when we were not observing the EDF-S). The SPT's sky scanning strategy is a result of its location on Earth at the geographic South Pole and a requirement for constant-elevation-scans. The telescope scans at constant declination, with constant speed in right ascension (R.A., 1.0 deg/s), across the field, reverses direction to scan back across the field, then takes a 12.5 arcmin step in declination, and this sequence repeats until the full declination range of the field has been scanned. Therefore, to observe the 30-square-degree EDF-S region, the SPT observed a 69-square-degree box centered at R.A.=61.25 deg (04$h$04$m$) and declination (Dec.)=$-$48.75 deg. The 69-square-degree area is reduced to \pixelmaskarea{} square degrees in the analysis due to necessary edge apodization. At approximately 2.5 hours per observation and 197 individual observations, we observed the SPT-3G EDF-S for approximately 20 days. Though we previously observed the EDF-S during SPT-3G Summer observations, the dedicated EDF-S observations dominate the map depth over that region. Therefore, we opted not to include the observations from previous years in this analysis.

\begin{figure*}[h]
                \centering
                        \includegraphics[width=1.0\textwidth]{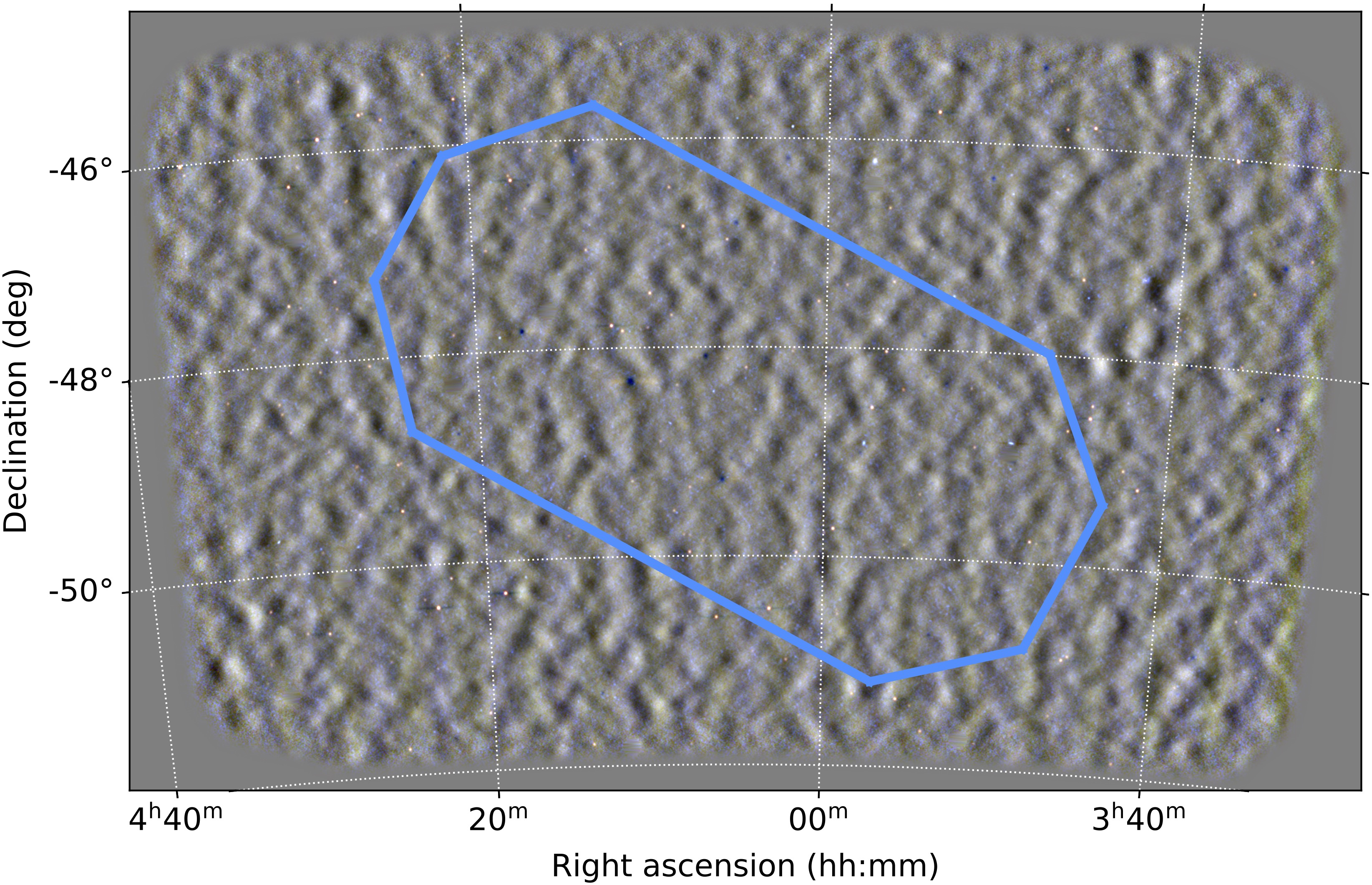}
                        \caption{Approximate full-depth EDF-S sky footprint (shown in a blue outline) overlaid on top of the \pixelmaskarea{}-square-degree SPT-3G coadded temperature maps in the ZEA projection. In this RGB image, the 95 GHz data are shown in the red channel, 150 GHz data are green, and 220 GHz data are blue. The image was generated using the \texttt{stiff} \citep{bertin12} command \texttt{stiff coadd\_map\_ZEA\_95GHz.fits coadd\_map\_ZEA\_150GHz.fits coadd\_map\_ZEA\_220GHz.fits -MIN\_TYPE MANUAL -MAX\_TYPE MANUAL -MIN\_LEVEL -140 -MAX\_LEVEL 140 -GAMMA}. Notable features of the map include the small-angular-scale CMB temperature fluctuations, emissive point sources which appear as red and blue bright dots, and tSZ decrements from galaxy clusters which appear as compact dark spots. Artifacts from interpolated bright sources appear as discs of which a close-up is shown in Fig. \ref{fig:mapprocess150}. The map border is an apodized edge (apodization is described in Sect. \ref{sec:masks}).}
                \label{fig:edfs_region}
        \end{figure*}

The field characteristics are summarized in Table \ref{tab:field}. The EDF-S boundary overlaid onto the SPT maps made from the inverse-variance-weighted average of all individual observations (also called ``coadds'') is shown in Fig. \ref{fig:edfs_region}.

    \begin{table}[h]
    \caption{Information about the observing footprint used in this work.}
    \centering
        \begin{tabular}{lc}
            \hline
            \textbf{Quantity} & \textbf{Value} \\
            SPT field name & \textsc{ra4h04dec-48} \\
            R.A. center & 61.25 deg (04$h$04$m$) \\
            Dec. center & $-48.75$ deg \\
            $\Delta$ R.A. & $11.5$ deg \\
            $\Delta$ Dec. & $6$ deg \\
            Effective area & \pixelmaskarea{} square degrees \\
            Observing dates & \startobs{}--\lastobs{} \\
            Time on field & 20 days \\
            \hline
        \end{tabular}
        \tablefoot{We calculated an apodization mask (see Sect. \ref{sec:masks}) based on the weights of the map pixels that down-weights the noisy edge regions of the map. The effective area of the field is determined by the mask and is the region over which we search for point sources and clusters.}
        \label{tab:field}
    \end{table}

\section{Map making} \label{sec:mapmaking}
The first step in our analysis is to process the camera data into maps (``map making''), which are used by both the point-source and cluster-finding algorithms. The SPT-3G map-making procedure is detailed in \cite{dutcher21}, with changes and choices specific to our analysis described here.

\subsection{Timestream processing} \label{sec:timestream}
The SPT-3G observations are recorded as time-ordered data (TOD), or timestreams.\footnote{The SPT-3G public software repository is available on GitHub \citep{spt3gsoftware}.} We filtered the TOD and binned data into pixels to create temperature maps. Different science analyses in the SPT collaboration necessitate different TOD filtering and map binning choices, such as the angular scales to high- and low-pass filter, treatment of bright sources, and map resolution and pixelization. In this work, we optimized the TOD processing and map pixelization for small-angular-scale science, namely detections of point sources and galaxy clusters.

The timestreams were high-pass filtered along the scan direction below $\ell_\textrm{HP}$=500 (using the relation $\theta = \pi/\ell$, this corresponds to 21.6 arcmin in angular units; the high-pass cutoff was chosen to remove as much atmospheric signal as possible without significantly impacting science signal of interest) and low-pass filtered above $\ell_{\mathrm{LP}}$=$20\,000$ (corresponding to 0.54 arcmin) to reduce aliasing when the data were binned into map pixels. We note that, because the scan speed is constant in R.A. (see Sect. \ref{sec:observations}), the on-sky speed, and hence the translation between scan-direction multipole $\ell_x$ and temporal frequency $f$, is a function of declination. For example, the high-pass cutoff $\ell_\mathrm{HP} = 500$ varies in temporal frequency from $f = 0.86$ Hz to $f = 0.96$ Hz across the declination extent of the field. We applied a polynomial filter by fitting and subtracting a 9th-order Legendre polynomial from the timestreams, which has a similar effect as the Fourier-space high-pass filter. Finally, a ``common-mode'' filter was applied, which averages signals in each SPT-3G camera wafer and band and subtracts the common signal (typically caused by the atmosphere) from the TOD.

The filters are represented by a series of linear operations on the TOD \citep{dutcher21} and are well approximated by analytic expressions in the Fourier domain. The effects of the TOD filtering can be modeled in map space in the Fourier domain by Eqs. (\ref{eqn:filters}) and (\ref{eqn:commonmode}). The functional forms for our low-pass, $H_{\mathrm{LP}}$, and high-pass, $H_{\mathrm{HP}}$, filters as scan-synchronous exponentials are

    \begin{equation} 
        \begin{split}
            H_{\mathrm{LP}} = e^{-\left(\frac{\ell_x}{\ell_\mathrm{LP}}\right)^6}
            \\
            H_{\mathrm{HP}} = e^{-\left(\frac{\ell_{\mathrm{HP}}}{\ell_x}\right)^6}
        \end{split}
    \label{eqn:filters},
    \end{equation}

\noindent and the common-mode filter, $H_{\mathrm{CM}}$, is

    \begin{align}
        H_{\mathrm{CM}} = 
        \begin{split}
            \begin{cases} 
            e^{-\frac{(\ell-\ell_{0})^2}{2\sigma^2}} & \ell \leq \ell_0 \\
            1 & \ell > \ell_0
            \end{cases},
            \\
        \end{split} \label{eqn:commonmode}
    \end{align}

\noindent where the cutoff scale ($\ell_0$) is 745 and the width ($\sigma$) is 240. Taken together, we analytically described the TOD filtering using the transfer function, $H$, which was used to extract point sources and clusters (see Sect. \ref{sec:extractsignals}). 

\subsection{Map pixelization} \label{sec:pixelization}
The maps have 0.25 arcmin pixels, which is sufficient to Nyquist sample our smallest beam at 220 GHz. We present temperature maps in both the oblique Lambert zenithal equal-area \citep[ZEA;][]{calabretta02} and Sanson-Flamsteed projections \citep[SFL;][]{calabretta02}, which are used in the point-source-finding and galaxy-cluster-finding pipelines, respectively. We further note that since the high- and low-pass filtering is along the scan direction, i.e., along lines of constant declination, the analytic forms presented in Eqs. (\ref{eqn:filters}) and (\ref{eqn:commonmode}) can only be applied to maps in which the $x$-direction of the map is also at constant declination. This is satisfied by the SFL maps at the expense of shape distortions at the map edges, but not the ZEA maps.

The SFL projection was used in previous SPT cluster analyses \citep[e.g.,][]{bleem15b, kornoelje25} because of the ability to conveniently filter the map. The ZEA projection was used in the most recent SPT point source analysis \citep{everett20}. The pipelines were independently developed and optimized according to the SPT-3G data landscape, taking into account all the sky areas shown in Fig. \ref{fig:allsptfields}.

\subsection{Deconvolving detector time constants}
Unlike previous SPT-3G analyses, we measured, then corrected for, two independent timing effects of our system: detector time constants, $\tau$, and a constant timing offset between detector and telescope pointing data. A detector time constant measures how long it takes for a detector to respond to a changing input signal. The value of $\tau$ is affected by the atmospheric loading on the detectors and is therefore different for every observation and individual bolometer.

The time constants were measured and calculated using methods presented in \cite{pan18}. First, we recorded timestreams while the detectors were illuminated by a chopped thermal calibration source, acquiring data at several frequencies of the chopper. We then fit the amplitude of each detector's fundamental mode response, as a function of frequency, to the Fourier transform of an exponential decay model in which the detector time constant, $\tau$, was the free parameter. We deconvolved the measured time constants from the timestreams in Fourier space before binning them into map pixels, multiplying by the deconvolution expression

    \begin{equation}
        G(\nu) = 1 + 2 \pi i \nu \tau \mathrm{.}
        \label{eqn:decontau}
    \end{equation}
    
\noindent We deconvolved the median $\tau$ value per detector across all measurements because an individual detector's behavior is not likely to change significantly between observations. Furthermore, the time constants depend on the telescope's elevation and season of observing, so we restricted the $\tau$ measurements to calibrator frequency sweeps performed between observations of the EDF-S.

The median $\tau$ for all time constants for all detectors per band is 5.6, 5.5, and 3.3 ms for 95, 150, and 220 GHz, respectively. The distributions of $\tau$ are skewed, with a tail extending to large values of $\tau$. We therefore report the median absolute deviation, which is more robust to outliers than standard deviation, as 1.9, 1.9, and 1.4 ms for the bands.

The time offset is understood as a disagreement between the recorded detector time stamps and telescope pointing time stamps. We used scan-direction differenced observations of the high signal-to-noise H\textsc{ii} region MAT 5A to measure the offset. In each observation, the telescope slewed in a right-going and a left-going direction over the same sky, and the data were all added together in the final maps. For this measurement, we created two separate maps consisting of data from either all left-going or all right-going scans with the detector time constants deconvolved. We then subtracted the two maps, and the resulting difference map contained a dipole structure due to the difference in apparent position of MAT 5A between the two maps. We measured the width of the dipole in the difference map and converted the width to time using the telescope's scan speed. The time offset is constant in time and was measured to be $-4.6 \pm 0.4$ ms for all frequencies. We similarly deconvolved this global time offset from the timestreams before map making.

\subsection{Calibration and astrometry} \label{sec:astrometry}
For the EDF-S observations, we followed the same field calibration procedures as described in \cite{sobrin22}. Regular observations of the galactic H\textsc{ii} region RCW 38, which has a precisely known location in the sky and reference flux that was calibrated to \textit{Planck} in \cite{mocanu19}, were taken throughout the EDF-S observation period. The primary purpose of observing RCW 38 is to calibrate each SPT-3G bolometer based on the measured temperature and the known value.

We improved the telescope's pointing accuracy by comparing source positions in the single observation maps to The Australia Telescope 20 GHz Survey (AT20G; \citealt{murphy10}), which has positional uncertainties of less than an arcsecond, and by remaking the single observation maps with the pointing corrections applied \citep{chichura25}. To quantify the accuracy of the final positions, we again compared the catalog positions of 18 sources from the coadded map to those in AT20G and found rms positional offsets between sources of \rmsRAoffset{} and \rmsDECoffset{} arcsec and mean offsets of \meanRAoffset{} and \meanDECoffset{} arcsec in R.A. and Dec., respectively.

\subsection{Bright source treatment} \label{sec:brightsrcs}
As a result of the TOD processing, especially bright point sources in the maps will have significant ``ringing'' features around them that extend along R.A. if not addressed at the timestream processing stage. Because our maps are low-noise, the ringing features can be significant and difficult to remove or ignore after maps are made. Other unwanted effects include hiding point sources and clusters beneath the ringing artifacts and slowing down the source-finding pipelines.

To resolve these issues at the timestream level, we interpolated over \numinterpsrcs{} point sources that had approximate flux $>$30 mJy at 95 GHz using a 3 arcmin radius circle. To make the bright source list, we created preliminary maps and a preliminary source catalog from which we identified sources above the threshold. We then created a final set of maps, interpolating over the sources that were above the threshold. A total of \interparea{} square degrees of map area is lost to interpolation (which is \percentinterparea{}\% of the \pixelmaskarea{}-square-degree SPT-3G EDF-S footprint). The source-finding pipelines did not search these regions for point sources or clusters, and a single point source was assumed to lie in each bright source disc. The interpolated regions appear as discs in the coadded maps shown in Fig. \ref{fig:edfs_region} with a close-up in Fig. \ref{fig:mapprocess150}.

Because we removed the sources from the maps, but include them in the emissive source catalog, we provide separate files of the bright source maps as 30$\times$30 arcmin thumbnails in the data release. The thumbnail maps have the same filtering settings described throughout this section as the larger field maps. We also provide a mask (described in Sect. \ref{sec:masks}) with the interpolation regions zeroed out.

In the emissive source catalog, bright source thumbnails are indicated by a string with the source name (if \texttt{True}) in the \texttt{thumb} column and all other sources are 0 (for \texttt{False}); see Tables \ref{tab:catalog_flags} and \ref{tab:columnlabels} for an example and more details.

\subsection{Beam convolution} \label{sec:beam}
Characterizing the point spread function, or telescope ``beam,'' is necessary to recover as much sky signal as possible and calculate accurate fluxes of astrophysical sources. Releasing the complete SPT-3G beam is outside the scope of this work, and therefore, we simplified the beam and map products. (The empirical beam is the focus of a forthcoming publication by the SPT-3G collaboration.)

To simplify the maps, we first deconvolved the best-fit empirical beam model from them, then convolved them with two-dimensional Gaussians (see Sect. \ref{sec:beamproducts}) that were not significantly wider than the empirical beams. There are six fewer, five more, and one more detections in the 95, 150, and 220 GHz Gaussian beam maps when compared to using our empirical beam measurement, which is consistent with scattering across the detection threshold. There is a 1-2\% difference in measured fluxes in each band, so we determined that the impact of this choice on the final catalog is negligible.

\subsection{Data quality}
The 197 individual observations of the EDF-S were combined using a weighted average to produce a coadded map. Every map pixel in every individual observation has an associated weight; individual-observation pixels that are especially noisy (for example, due to poor weather or unusual bolometer behavior) are down-weighted in the final coadd \citep{dutcher21}.

As long as the weights accurately describe the pixel variance, coadding maps in the manner previously described will produce map pixel values that are unbiased with minimum variance, even if some individual observations are especially noisy. We computed some basic statistics on each map, such as pixel mean and variance, to confirm that we only included maps whose weights accurately describe the noise. The map statistics were uniform enough that no cuts were warranted. Therefore, the final coadd includes all individual observations that the SPT took of the EDF-S.

The maps included in our release are not intended for CMB power spectrum or lensing analyses, mainly owing to the bandpass filtering choices which remove large angular scale features, where the relevant cosmological signals are most significant, while preserving small angular scale features that would contaminate power spectra. Maps and data products optimized for analyzing CMB lensing and cross-correlations with \textit{Euclid} data are intended for a future study and data release.

To compute the white noise levels reported in Table \ref{tab:mapnoise}, we subtracted one half of the individual maps from the other in 25 different random groups of maps, then computed the power spectrum of the difference maps in the range $5\,500 < \ell < 6\,500$ and calculated the median noise value. The final map depths are 4.3, \mapdepthonefive{}, and \mapdepthtwotwo{} $\mu$K-arcmin at 95, 150, and 220 GHz; for comparison, the field depths for four observing seasons of the SPT-3G Main field, which covers approximately $1\,500$ square degrees, are 3.2, 2.6, and 9.0 $\mu$K-arcmin at 95, 150, and 220 GHz \citep{kornoelje25}. The 5$\sigma$ detection threshold flux values for the emissive source catalog are also reported in Table \ref{tab:mapnoise}. The process to compute flux from CMB temperature, and the relevant conversion factors for the point source-processed maps, are described in Sect. \ref{sec:psfinding}.

    \begin{table}[h]
    \caption{Map noise levels of the coadded maps.}
        \begin{tabular}{lccc}
            \hline
            \textbf{Nominal band (GHz)} & \textbf{95} & \textbf{150} & \textbf{220} \\
            Nominal band (mm) & 3.3 & 2.0 & 1.4 \\
            Map noise ($\mu$K$_{\mathrm{CMB}}$-arcmin) & 4.3 & \mapdepthonefive{} & \mapdepthtwotwo{} \\
            5$\sigma$ point source (mJy) & \Sminnine{} & \Sminonefive{} & \Smintwotwo{} \\
            \hline
        \end{tabular}
        \tablefoot{Noise levels are provided in CMB units of $\mu$K$_{\mathrm{CMB}}$-arcmin and flux thresholds of a 5$\sigma$ point source. The sky area was observed in such a way to achieve nearly equal depth across the maps with the exception of the top approximately 0.6 degrees, which has about 20\% higher noise than the rest of the observed area; see Sect. \ref{sec:masks} for a detailed explanation.}
        \label{tab:mapnoise}
    \end{table}

\section{Signal extraction and catalog generation} \label{sec:extractsignals}
Both the point-source and galaxy-cluster-detection pipelines use the coadded maps described in Sect. \ref{sec:mapmaking} and have significant similarities in the methods used to detect and characterize signals of interest. We describe these commonalities in this section and expand upon the specific choices for point-source and cluster-detection in Sects. \ref{sec:psfinding} and \ref{sec:clusterfinding}, respectively. 

The SPT maps contain significant scale-dependent contributions from different astrophysical origins. We modeled the temperature fluctuations in the maps as a function of spatial position ($\theta$) and frequency ($\nu_i$) by  
 
    \begin{equation}
        \begin{split}
            T(\vec{\theta},\nu_i) = B(\vec{\theta},\nu_i) * [\Delta T(\vec{\theta},\nu_i) + N_{\mathrm{astro}}(\vec{\theta},\nu_i)] + N_{\mathrm{instr}}(\vec{\theta},\nu_i)
        \end{split} 
    ,\end{equation}

\noindent where the sky signals were convolved ($*$) with the telescope beam and the transfer function (which encodes the impact of our timestream filtering; see Sect. \ref{sec:timestream}), with $\Delta T$ representing the sky signal of interest (e.g., tSZ), $B$ representing the beam and transfer function, $N_{\mathrm{instr}}$ representing the instrumental and residual atmospheric noise, and $N_{\mathrm{astro}}$ representing all undesired astrophysical power, including primary CMB, tSZ, kSZ, and extragalactic emissive sources below the SPT-3G detection threshold. The simulated power for each of the astrophysical noise quantities comes from \cite{reichardt21}. 

Following common practice in millimeter-wave surveys \citep[e.g.,][]{tegmark98, melin06}, we used matched filtering techniques to optimize sensitivity to desired signals in the presence of these contaminants. The optimal filter applied to the maps takes the following form for frequency, $\nu_i$:

    \begin{equation}
        \psi(\ell,\nu_i) = \sigma_{\psi}^2\sum_jN^{-1}_{ij}(\ell)f(\nu_j)S_{\mathrm{filt}}(\ell,\nu_j) \mathrm{,}
        \label{eqn:mvmf}
    \end{equation}

\noindent where $N_{ij}(\ell)$ is the noise covariance matrix, $S_{\mathrm{filt}}(\ell,\nu_j)$ is the spatial profile of the targeted signal, and $f(\nu_j)$ is its spectral dependence. The noise covariance matrix combines the measured instrumental and atmospheric noise residuals (Sect. \ref{sec:noiseasds}) with the models of astrophysical noise from \citet{reichardt21}. The summation, $j$, occurs over either a single or all SPT frequencies for point sources and clusters, respectively. Finally, the variance in the optimal filter is given by 

    \begin{equation}
        \sigma_\psi^{-2} = \int d^2\ell \sum_{i,j}f(\nu_i)S_{\textrm{filt}}(\ell,\nu_i)N_{ij}^{-1}(\ell)f(\nu_j)S_{\textrm{filt}}(\ell,v_j)
        \label{eqn:varmvmf}
    ,\end{equation}

\noindent which is used to normalize the filter such that the response to our desired signal is unity. We made a number of different choices for the modeling of these components in the point source and cluster analyses.

\subsection{Point source filtering}
\paragraph{Spatial template.} The source profile adopted for the point source analysis is a $\delta$ function. Emissive extended sources are flagged according to Sect. \ref{sec:flags}.

\paragraph{Spectral template.} In this analysis, we did not assume a frequency dependence for the point source emission a priori. As such, each frequency map was filtered independently and the single-band source lists were combined.

\paragraph{Instrumental noise.} For instrumental noise, we used the map white noise levels in Table \ref{tab:mapnoise}. At low-$\ell$ the CMB signal dominates over the instrument noise, so approximating the instrument noise as white in the construction of the point source filter is a reasonable approximation.

\paragraph{Map projection.} We used both the ZEA and Plate Carr\'ee \citep[CAR,][]{calabretta02} map projections in the application of the point source filter. Starting in the area-preserving ZEA projection, we applied the beam-only component from $S_{\mathrm{filt}}$ and the noise components, $N^{-1}_{ij}$, in the Fourier domain. Back in map space, we converted the partially filtered ZEA map to CAR and Fourier transformed it again. In the CAR projection, strips of declination are also strips of $x$, so we then applied the transfer function component of $S_{\mathrm{filt}}$ in one dimension. Finally, we transformed the fully filtered map back into real space and into the ZEA projection.

\subsection{Galaxy cluster filtering}
\paragraph{Spatial template.} Galaxy clusters are extended structures in the SPT maps. As the physical scale of the tSZ signal depends on a cluster's mass and redshift, we made use of a range of spatial templates in our blind cluster search by modeling the spatial profile of the tSZ decrement signal assuming an isothermal projected $\beta$-model as in \cite{cavaliere76}: 

    \begin{equation}
        S = \Delta T_0 \left(1 + \frac{\theta^2}{\theta_c^2}\right)^{-\frac{3}{2}\beta+\frac{1}{2}} \mathrm{,}
        \label{eqn:beta}
    \end{equation}
    
\noindent with normalization $\Delta T_0$, core radii ($\theta_c$) ranging from 0.25 to 3 arcmin in increments of 0.25 arcmin, and a fixed $\beta = 1$. The choice of $\beta$-model as a spatial template is consistent with previous SPT analyses and has negligible impact when compared to other common profiles \citep{vanderlinde10}. For the cases of common candidate detections between filter scales, the core size that maximizes the significance is included in the final catalog.

\paragraph{Spectral template.} The tSZ signature is a spectral distortion caused by the inverse Compton scattering of CMB photons off of high-energy electrons within the intracluster medium (ICM). The spectral distortion can be represented as temperature fluctuations dependent on the electron number density, $n_e$; temperature, $T_e$; and Thomson scattering cross section, $\sigma_{\mathrm{T}}$:

    \begin{equation}
        \begin{split}
        \Delta T_{\mathrm{SZ}} &= T_{\mathrm{CMB}}f_{\mathrm{SZ}}(x)\int n_e \frac{k_{\mathrm{B}}T_e}{m_ec^2}\sigma_{\mathrm{T}}d \ell \\
        &= T_{\mathrm{CMB}}f_{\mathrm{SZ}}(x)y_{\mathrm{SZ}} \mathrm{,}
        \end{split} \label{eqn:tsz}
    \end{equation}
    
\noindent where we take the Compton $y$ parameter, $y_{\mathrm{SZ}}$, to be the total thermal energy of the electron gas integrated along the line of sight \citep[LOS;][]{sunyaev72}. Here, $f_{\mathrm{SZ}}(x)$ is the frequency dependence of the tSZ effect,

    \begin{equation}
        f_{\mathrm{SZ}}(x) = \left( x\frac{e^x+1}{e^x-1} \right)(1+\delta_{\mathrm{rc}}) \mathrm{,}
        \label{eqn:fsz}
    \end{equation}
    
\noindent where \mbox{$x \equiv h\nu$/$k_\mathrm{B} T_{\mathrm{CMB}}$}, $h$ is Planck's constant, and $k_\mathrm{B}$ is the Boltzmann constant. We used the effective frequencies for the tSZ signal in each band, where $\nu$ is 95.7, 148.9, and 220.2 GHz for nominal bands 95, 150, and 220 GHz. The term ``effective frequency'' refers to the frequency at which the instrument response to a $\delta$ function would be equivalent to that of a source of a specified spectrum after integrating over the real bandpass and taking into account the beam dependence on frequency. $\delta_{\mathrm{rc}}$ represents the relativistic correction to the electron gas's energy spectrum \citep{erler18}, which is assumed to be negligible for this analysis.

\paragraph{Instrumental noise.} We empirically estimated the instrumental and residual atmospheric noise by creating ``signal-free'' coadded maps by randomly multiplying half of the observations by $-$1 to remove sky signals. The average of the Fourier transform of 25 such realizations was used to estimate the $N_{\mathrm{instr}}$ term in the noise covariance matrix used in constructing the optimal filters.

\paragraph{Map projection.} We used the SFL projection in the entire cluster-finding procedure, which aligns the rows of map pixels with the telescope's scan direction. This projection makes it convenient to correct for effects from our constant-declination-scan filtering using the transfer function described in Sect. \ref{sec:timestream} at the expense of small shape distortions at the field edges. 

\subsection{Point source catalog generation} \label{sec:psfinding}
\begin{figure}[h]
                \centering
                        \includegraphics[width=1\columnwidth]{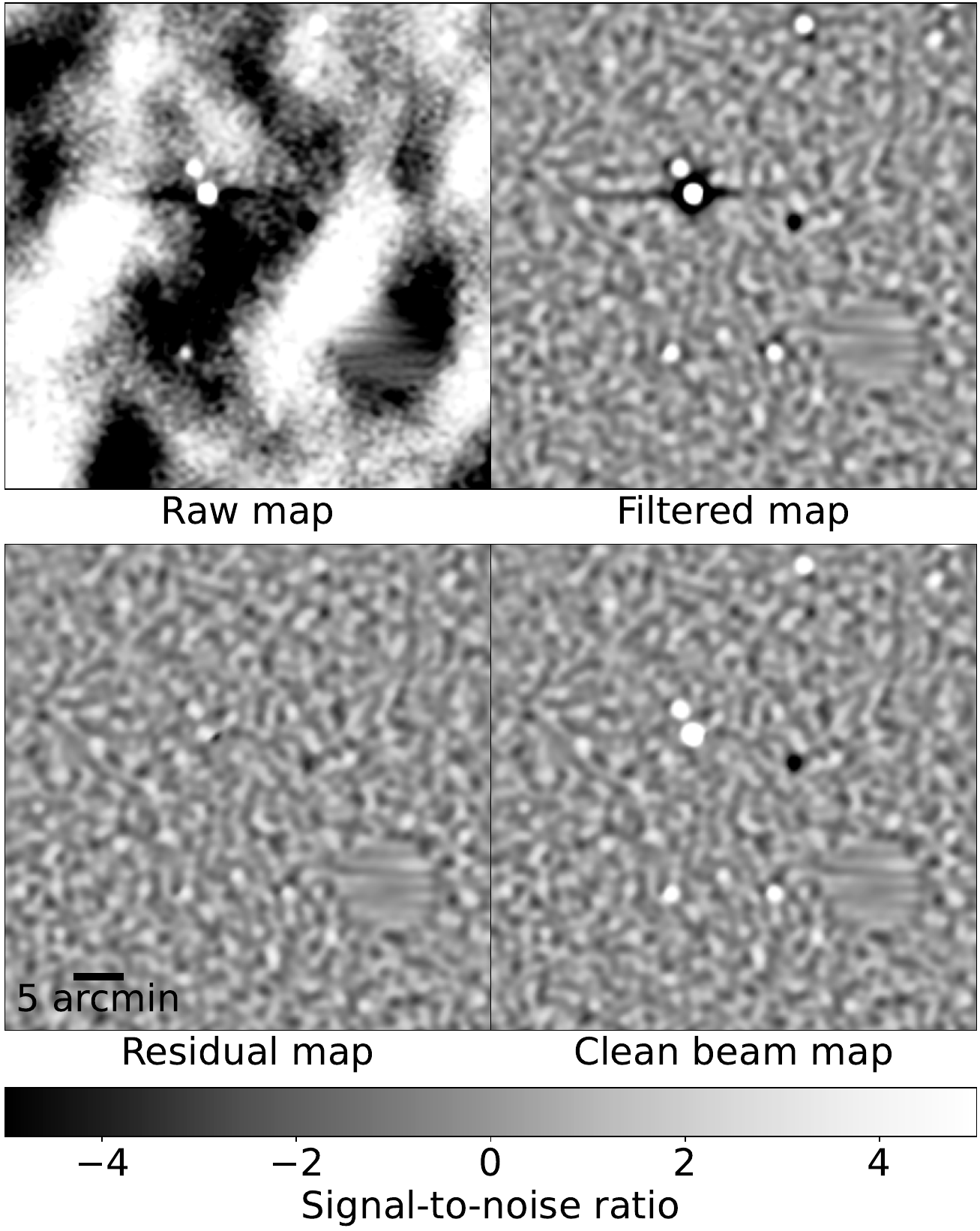}
                        \caption{Each panel shows a major map processing step in the point-source-finding procedure. For demonstration purposes, the 150 GHz map is shown; however, the coadd of each band undergoes source finding independently. \textit{Top left:} Coadded temperature map before any processing. On the same color scale as all other panels, we observe the large CMB fluctuations dominating the emissive sources and the galaxy cluster. The first step is to remove the CMB. \textit{Top right:} Map after an optimal filter has been applied and suppressed most of the CMB modes. The emissive galaxies are positive points with negative filtering wings around them. The galaxy cluster is a decrement. The second step is to remove the filtering effects around sources. \textit{Bottom left:} Residual map after applying a CLEAN algorithm, which iteratively removes ``dirty beams'' at the location of sources. The pixels left over in their place are just below the 5$\sigma$ threshold. \textit{Bottom right:} Map after putting back in the ``clean beam'' at source locations without the filtering artifacts. This clean beam map is the image to which a simple pixel grouping algorithm was applied to generate the emissive source catalog. Note that while the cluster in this example is affected by the procedure, it is not optimal to find and characterize clusters using these methods, and therefore a separate pipeline was used. The disk feature in the lower right of all panels is an interpolated bright source (see Sect. \ref{sec:brightsrcs}).}
                \label{fig:mapprocess150}
        \end{figure}

The point-source-finding procedure can be described in three coarse steps: optimally filtering the raw coadd, processing the filtered map, and pixel-grouping the processed map. Once the pixels are grouped into sources, the CMB brightness temperature fluctuations from the sources are converted to flux density. The specific details of these steps will be described in a forthcoming publication by the SPT-3G collaboration, though an overview is provided in this section. A zoomed-in portion of the 150 GHz map undergoing the point-source-finding procedure is shown in Fig. \ref{fig:mapprocess150} as a reference for the following steps.

\paragraph{CLEAN.} We first applied the point source filter described in the previous section to the individual frequency maps. The filter was applied in Fourier space and normalized such that the resulting map is in CMB temperature units (top-right panel of Fig. \ref{fig:mapprocess150}).

Next, an implementation of the CLEAN algorithm \citep{hogbom74} was applied to the optimally filtered map to remove filtering artifacts present around sources. In Sect. \ref{sec:brightsrcs} we discuss the \numinterpsrcs{} sources $>$30 mJy at 95 GHz that were interpolated over as the solution to ringing features from particularly bright sources so that we did not have to CLEAN them; one of these is visible in Fig. \ref{fig:mapprocess150}. In the top-right panel, the emissive sources are the white dots and the filtering artifacts are the dark features encircling the white dots. The inverse of these colors towards the middle of the panel is a galaxy cluster (which is a decrement at 150 GHz).

As in \cite{everett20}, CLEAN was implemented where the brightest pixel in the map was found, and a fraction of a source template (called the ``loop gain,'' which we set at 0.1) was removed from that location, gradually decreasing the maximum map amplitude. The iterations proceeded until a 5$\sigma$ signal-to-noise threshold was reached, where the noise was estimated in 0.5 degree-tall declination-dependent strips. The map left over after the first part of CLEAN is called the ``residual map'' (lower left panel of Fig. \ref{fig:mapprocess150}). The reason one can still see the remnants of the sources, and the galaxy cluster in particular, is because these features are just below the threshold at which CLEAN stopped.

During the CLEAN iterations, the source locations and pixel amplitudes were recorded, then a new source template---the ``clean beam'' or point source template without the filtering artifacts---was iteratively returned to the map with the appropriate amplitude. The final result is a ``clean beam map'' (lower right panel of Fig. \ref{fig:mapprocess150}) which has the following key characteristics: large-scale modes such as CMB and atmosphere have been suppressed, and unresolved point sources appear as simple telescope beams. To generate the catalog from this map, a simple pixel grouping algorithm (\texttt{photutils}, \citealt{bradley23}) was applied to the clean beam map, which extracts groups of pixels above 5$\sigma$ most likely to be distinct sources and deblends sources near each other. The peak temperature map value of a source was recorded from the centroid location found by \texttt{photutils}.

\paragraph{SPT band cross-matching.} The aforementioned map processing steps were applied to each frequency map individually to create three separate source catalogs. The three catalogs were unified using a simple radial association radius of \internalrassoc{} arcsec between bands. This value comes from 3$\times$1.2 arcmin/5$\sigma$ \citep{ivison07}, where 1.2 arcmin is the 150 GHz Gaussian beam FWHM, 5$\sigma$ is the lowest possible detection significance for a source, and multiplying by three is a discretionary choice that minimizes adverse effects when using the emissive source catalog for source subtraction during cluster finding (see Sect. \ref{sec:clusterfinding}). We opted to use the 150 GHz beam size rather than the other bands because it is between the other sizes, and sources are more likely to be detected at 150 GHz plus another band than any other combination of bands. The location of the highest signal-to-noise detection in any map is recorded as the source's R.A. and Dec. in the catalog. If a source was not detected above 5$\sigma$ in a band's map, we used forced photometry in the map where the source was below the threshold and recorded the signal-to-noise and forced-photometry temperature value in the catalog. 

The raw coadded temperature maps and the final clean beam maps in the ZEA map projection are included in the released products. One could construct the intermediate maps using the ancillary data products described in Sect. \ref{sec:resultsdataprod}.

\paragraph{Flux calculation.} The conversion between the CMB temperature and flux density was computed from the derivative of the Planck blackbody function and evaluated at a fiducial frequency and temperature, as follows,

    \begin{equation}
        \begin{split}
            S(\mathrm{Jy}) = T_{\mathrm{peak}} \cdot \Delta\Omega_{\mathrm{f}} \cdot 10^{26} \cdot \frac{2k_{\mathrm{B}}}{c^2} 
            \cdot \left( \frac{k_{\mathrm{B}} T_{\mathrm{CMB}}}{h} \right)^2  \frac{x^4 e^x}{(e^x - 1)^2}
        \end{split},
        \label{eqn:fluxconversion}
    \end{equation}

\noindent where $x \equiv h\nu/(k_{\mathrm{B}}T_{\mathrm{CMB}})$ and $\Delta \Omega_{\mathrm{f}}$ is the solid angle of the optimal filter, $\psi$ \citep{everett20},

    \begin{equation}
            \Delta \Omega_{\mathrm{f}} = \left[ \int d^2 \ell \ \psi (\ell) \ B (\ell) \right]^{-1} \mathrm{,}
        \label{eqn:effectivearea}
    \end{equation}

\noindent and $B$ is the transfer function combined with the beam. Though we could have made any choice of fiducial frequency at which to calculate and report source fluxes, we chose 94.2, 147.8, and 220.7 GHz for the nominal 95, 150, and 220 GHz bands, which correspond to the effective band centers for a flat spectrum, point-like source. The fiducial frequencies depend on the real SPT-3G bandpasses \citep{sobrin22}, taking into account how the telescope beam behaves with frequency. At this choice of fiducial frequencies, the conversion factors between CMB temperature and flux are 0.07513, 0.05856, and 0.05068 mJy/$\mu$K for nominal bands 95, 150, and 220 GHz, respectively.

Unlike previous SPT point source analyses \citep{vieira10, mocanu13, everett20}, we did not account for certain biases in the flux calculation that are caused by selecting peaks in a Gaussian noise field; this accounting is known as ``flux de-boosting'' \citep{coppin05}. On average, the boosting results in fluxes that are overestimated by $11\pm3$\%, $15\pm9$\%, and $23\pm10$\% at 95, 150, and 220 GHz, respectively, for sources detected between $5$ and $5.5\sigma$. Above $5.5\sigma$, the effect from boosting is smaller than the uncertainty on the flux measurement. It should also be noted that the bias only affects sources detected above $5\sigma$---for sources with forced photometry measurements in non-detection bands, we do not expect the flux to be overestimated. One should take this potential bias into account if using the reported fluxes to construct spectral energy distributions based on photometry.

We made this choice because the main goal of the emissive catalog release is to study the individual sources themselves and not number counts for the types of millimeter sources represented, which have already been well characterized in previous analyses in other parts of the sky \citep[e.g.,][]{everett20, vargas23}. The practical effect on source classification of not accounting for flux boosting is to shift the classification of 3\% of sources (based on preliminary measurements) from ``dusty'' to ``synchrotron'' or vice versa (source classification is discussed in detail in Sect. \ref{sec:classification}). However, the affected sources have other indications of synchrotron emission, such as radio counterparts and dipping or peaking spectral behavior, so not knowing their de-boosted flux values does not significantly hinder one's ability to understand the source's nature.

\subsection{Galaxy cluster catalog generation} \label{sec:clusterfinding}
Galaxy clusters were identified as peaks in the frequency-combined minimum variance tSZ maps filtered by the optimal $\beta$-model filters discussed in the beginning of Sect. \ref{sec:extractsignals}. Before applying these filters, we first removed signals from emissive sources to mitigate spurious contamination from these sources in our cluster candidate list. 

\paragraph{Emissive source subtraction and masking.} Similar to emissive source detection, cluster detection is sensitive to the ringing wings caused by the timestream filtering procedure. The wings have an opposite sign to the central source and can hold a significant portion of the source's amplitude, which can lead to spurious false detections that impact the purity of the final cluster sample \citep[see e.g., discussion in ][]{kornoelje25}. \citet{bleem24} introduced a treatment of these sources through ``source subtraction.'' Similar to the previous subsection, we adopted a model of the beam convolved with the transfer function as our model for the spatial profile of emissive sources. We subtracted templates scaled by measured source amplitudes from emissive source locations in the coadded temperature maps for all three frequencies for sources detected at 95 GHz at a signal-to-noise greater than five---essentially acting as a CLEAN iteration with a loop gain of 1 (see Sect. \ref{sec:psfinding}).

We additionally adopted a masking procedure similar to \cite{reichardt13} and SPT cluster analyses thereafter by masking a 4 arcmin radius region around bright source locations interpolated over during the map-making procedure. Following an initial cluster detection run, we visually inspected\footnote{Using the inspection tool \texttt{Image Marker} \citep{walker25}.} all tSZ cluster candidates and flagged an additional 10 regions for masking in the construction of the final sample owing to poor source subtraction. The problematic regions are at the locations of bright (signal-to-noise $>50$ at 95 GHz) or extended sources \citep[such as NGC 1493 and NGC 1494;][]{sulentic73}.

\paragraph{Cluster detection observable, $\xi$.} We define the cluster significance, $\xi$, as the signal-to-noise ratio maximized over 12 spatial filter scales. The numerator for $\xi$ is directly returned by the matched filter at every location in the map. We form the basis for the denominator by estimating the noise in our minimum-variance matched-filtered maps.

We estimated the noise by fitting a Gaussian to the distribution, in a series of declination-dependent strips, of pixel values that fall within 5$\sigma$ of the mean of that strip. The averaging of pixel values occurs in strips that are 1.5 degrees tall to capture declination-dependent noise that stems from unequal area coverage and variations in atmospheric loading. We set the minimum significance threshold for detection at $\xi_{\mathrm{min}}=4$, balancing purity with the size of the sample. The purity of the sample degrades as $\xi$ decreases due to the noise fluctuations, the implications of which are discussed further in Sect. \ref{sec:completeness}.

Once tSZ candidates were identified by the preceding filtering scheme, we estimated redshifts and masses for the sample, and we compared to optical and infrared data. The methods are described in the following subsections, and the results are discussed further in Sect. \ref{sec:clustercatalog}.

\subsubsection{Photometric redshift estimation} \label{sec:redshifts}
As the tSZ effect is redshift-independent, we require external survey data to confirm our tSZ candidates as galaxy clusters and to provide redshift estimates. We provide a brief summary of the multicomponent matched filter method \citep[MCMF;][]{klein18, klein19, klein24a} that we used to probabilistically assign optical and infrared galaxy cluster counterparts to the tSZ detections and estimate their redshifts. In this work, we used data from the DR10 release of the DECam Legacy Survey \citep[DECaLS;][]{dey19}, which combines data from the DES \citep{flaugher15} and the Wide-field Infrared Survey Explorer \citep[WISE;][]{wright10} in the SPT-3G EDF-S region. Readers are referred to preceding publications for further details on the implementation and performance of MCMF for tSZ galaxy cluster redshifts. In total, we estimated redshifts for \numclusterswithredshifts{} clusters.

The MCMF identifies counterparts by identifying excesses of bright red-sequence \citep[shown to be excellent tracers of optical galaxy clusters, see e.g.,][]{gladders00, rykoff14} or infrared galaxies along the LOS at SPT candidate locations. Similar estimates along random sight lines were used to assess the probability of falsely associating tSZ candidates with optical overdensities as a function of redshift, $z$, and cluster galaxy richness, $\lambda$ \citep[the latter being a weighted form of cluster galaxy count, see][]{klein18}. We quantified the false association probability through the statistic $f_\textrm{cont}$ given by

    \begin{equation}
        f_\textrm{cont}(\lambda_i, z_i) = \frac{ \int_{\lambda_i}^{\infty}f_\textrm{rand}(\lambda, z_i)d\lambda}{ \int_{\lambda_i}^{\infty}f_\textrm{SZ}(\lambda, z_i)d\lambda}
        \label{eqn:fcont},
    \end{equation}

\noindent where $f_\textrm{rand}$ traces the galaxy richness distributions along random lines of sight and $f_\textrm{SZ}$ is the richness distribution along lines of sight from tSZ candidates. We considered a candidate confirmed as a galaxy cluster if $f_\textrm{cont}<0.2$. For candidates in which multiple optical or infrared overdensities along the line of sight have $f_\textrm{cont}<0.2$, we adopted the association with the smallest contamination fraction as the primary association (and assigned the cluster this redshift and richness) but also provide optical properties for these additional structures.

When combined with the intrinsic purity, $p(\xi>\xi_\textrm{min})$, of the tSZ candidate sample (see Sect. \ref{sec:completeness}), the overall contamination for the optically confirmed sample is 

    \begin{equation}\label{equation:opticalpurity}
        \textrm{contamination}= f_\textrm{cont}^{\textrm{max}}\times[1-p(\xi>\xi_\textrm{min})],
    \end{equation} 

\noindent or approximately 3\% at $\xi>4$ for $f_\textrm{cont}^{\textrm{max}}=0.2$ given the high purity, \clusterpurity{}, of the tSZ candidate list. In the catalog, we provide redshifts, galaxy richnesses, and contamination fractions for confirmed clusters. The exact column labels and details are provided in Table \ref{tab:columnlabels}.

\subsubsection{Spectroscopic redshift assignment}
We followed our previous work on optical follow-up \citep{klein23, klein24b} and obtained spectroscopic redshifts for clusters using three methods. The first method involved cross-matching confirmed clusters with previously published clusters that have known spectroscopic redshifts, using a maximum separation of 2 arcmin from the cluster center. In the second method, we used public spectroscopic surveys to search for multiple galaxies with consistent redshifts within a 2 Mpc radius around the cluster center. In the third method, we searched for spectroscopic redshifts of the brightest cluster galaxy (BCG) in the literature.

In total, we assigned spectroscopic redshifts to nine clusters. Eight of these were obtained by matching to clusters with existing spectroscopic measurements \citep{depropris02, williamson11, bayliss16, tempel16, hilton21, xu22}. For the remaining cluster, SPT-CL\,J0357$-$4935, we adopted the spectroscopic redshift of the BCG identified in the 2dF Galaxy Redshift Survey \citep[2dFGRS;][]{colless01}.

\subsubsection{Mass estimation} \label{sec:masses}
Since the magnitude of the tSZ effect is dependent on the electron pressure in the ICM, the significance of the cluster has been shown to have a strong correlation to the integrated mass of the system \citep{dehaan16, bocquet24b}. We estimated the masses of clusters using the significance-mass relation from \cite{benson13},

    \begin{equation}
        \langle \mathrm{ln} \zeta \rangle = \mathrm{ln} \left[A_{\mathrm{SZ}}\left(\frac{M_{500}}{3 \times 10^{14}M_{\odot}h^{-1}}\right)^{B_{\mathrm{SZ}}}\left( \frac{E(z)}{E(0.6)}\right)^{C_{\mathrm{SZ}}}\right] \mathrm{,}
        \label{eqn:mz}
    \end{equation}

\noindent where the parameters $A_{\mathrm{SZ}}$, $B_{\mathrm{SZ}}$, and $C_{\mathrm{SZ}}$ characterize the normalization, mass slope, and redshift evolution of the significance-mass relation, respectively, and $E(z) \equiv H(z)/H_0$. As $\xi$ is a biased tracer of detection significance due to its maximization over preferred position and filter scale, we introduce the unbiased estimator of our detection significance, $\zeta$. Following \citet{vanderlinde10}, $\zeta$ is related to $\xi$ by 

    \begin{equation}
        \zeta \equiv \sqrt{\langle \xi \rangle^2-3},
    \end{equation}

\noindent which holds for $\xi>2$. We assumed a unit-width Gaussian scatter on $\xi$ and a log-normal scatter on $\zeta$ of 0.2. 

The significance-mass relationship is dependent on the level of noise in the map. To remove this field-level noise dependence, as in previous SPT publications, $A_{\mathrm{SZ}}$ was rescaled as $\gamma A_{\mathrm{SZ}}$, where $\gamma$ parametrizes the noise level of the field. Given the SPT-3G EDF-S field only spans \pixelmaskarea{} square degrees, fitting for the parameters of Eq. (\ref{eqn:mz}) by using cluster abundances of the EDF-S at our fixed fiducial cosmology \citep[as was done in, e.g., ][for the SPT-SZ and SPTpol surveys]{bleem15b, bleem24} leads to poor constraints on the scaling relation parameters. Thus, we adopted the best-fit $A_{\mathrm{SZ}}$, $B_{\mathrm{SZ}}$, and $C_{\mathrm{SZ}}$ from \citet{bleem24} to estimate $\gamma$ by comparing the masses of cross-matched clusters between the EDF-S and the SPT-SZ survey, allowing $\gamma$ to vary until the median ratio of the masses of these clusters reached unity. We found a value of 4.05 for the SPT-3G EDF-S; this is approximately three times the $\gamma$ value for the SPT-SZ survey in this region, meaning that the average significance of clusters in common between the SPT-SZ and SPT-3G EDF-S fields is roughly a factor of three greater in the new SPT-3G EDF-S data.

\section{Temperature maps and data products} \label{sec:resultsdataprod}
We provide the coadded temperature maps at 95, 150, and 220~GHz in both the ZEA and SFL projections in the data release. Figure \ref{fig:edfs_region} shows an RGB image with the 95, 150, and 220 GHz maps, respectively, featuring the large spatial scale CMB fluctuations, bright individual galaxies, and tSZ decrements. The data products released with this paper include the entire \pixelmaskarea{} square degrees, though it is noted in the catalogs whether the object is strictly inside the EDF-S boundaries. We do not include polarization or lensing maps in this release.

In the following subsections we describe the ancillary data products we provide that are necessary components to reproduce the main analysis results. The specific details of how the data products are used are described in Sect. \ref{sec:extractsignals}.

\subsection{Masks} \label{sec:masks}
The maps are saved with an apodized edge for ease of use in Fourier transform operations. The edge apodization masks were created to gently roll off the map edges based on the values of the pixel weights; specifically, to roll off the pixels whose weights were $<$0.5 of the median weight value on the left and right edges. We decreased the threshold to $<$0.1 of the median weight value, thereby including slightly noisier map area, for the top and bottom of the field to fully encompass the EDF-S field.\footnote{We designed the SPT-3G observations based on the science area of \textit{Euclid} Q1 data. However, the mask provided by the \textit{Euclid} collaboration includes extra area required to achieve uniform coverage in the intended science region; further, the \textit{Euclid} Q1 science analyses apply different area selections depending on the analysis. The SPT-3G data we present includes the possible area imaged by \textit{Euclid} Q1, but consequently our observations increase in noise by about 20\% in the top approximately 0.6 degrees.} The apodization mask data products are map-shaped arrays, where the values at the edges slowly transition from 0 to 1 in the valid portions of the map. We provide these edge apodization masks in both map projections.

Both the point-source and cluster-finding pipelines use an additional binary ``pixel mask'' that was used to discard objects found in problematic areas of the map: without modifying the map itself, if a source was found in a 0 region of the pixel mask, it was removed from the catalog. We set the edges of the pixel mask equal to 1 where the apodization masks were $>0.999$, ensuring that source-finding was restricted to the uniform coverage region of the maps. In the pixel masks, we also zeroed out the discs that were left over from the bright source interpolation procedure (Sect. \ref{sec:brightsrcs}). We provide the pixel masks used to generate the catalogs.

To flag detections as ``inside EDF-S,'' we used a mask provided by the \textit{Euclid} collaboration (private communication). The binary mask encompasses 30 square degrees. We include the pixel mask we created from the \textit{Euclid}-provided mask in our data release.

\subsection{Telescope beams} \label{sec:beamproducts}
As described in Sect. \ref{sec:beam}, our maps are convolved with two-dimensional Gaussians rather than the best-fit empirical SPT-3G beams. We supply the $B_{\ell}$ file for the Gaussian beams, which were generated using \texttt{healpy} \citep{zonca19}. The FWHMs are 1.8, 1.2, and 1.0 arcmin for 95, 150, and 220 GHz, respectively. Though these beam FWHMs are similar to the published values in \cite{sobrin22}, they differ slightly and were chosen to ensure the Gaussian beams have 0 power at the same angular scales as the empirical beams. The Gaussian beams have nonzero values to arbitrarily high multipoles, and therefore we set the values of $B_{\ell}$ to 0 when the amplitudes reach 0.005. When using the beams to create the Fourier-space matched filters for finding point sources and clusters (Sect. \ref{sec:extractsignals}), we assumed azimuthal symmetry to create two-dimensional versions.

\subsection{Noise amplitude spectral densities} \label{sec:noiseasds}
As detailed in Sect. \ref{sec:extractsignals}, we constructed empirical noise estimates from our observations by making many coadded difference maps of our field. We used the Fourier transform of these noise maps to create noise amplitude spectral densities (ASDs). We provide the ASDs used in the cluster-detection procedure in the SFL projection. 

\subsection{Transfer function} \label{sec:transfer}
We aggressively filtered the TOD to make temperature maps (see Sect. \ref{sec:mapmaking}). For this analysis, we applied low-pass, high-pass, polynomial, and common mode filters to the TOD and analytically described the resulting transfer function using Eqs. (\ref{eqn:filters}) and (\ref{eqn:commonmode}). We used the analytic transfer function to create the matched filters in Sect. \ref{sec:extractsignals}, as well as to create ``source templates'' for removing sources from the data during cluster source subtraction and point source CLEAN. Therefore, we provide the analytic transfer function in the two-dimensional Fourier domain as part of the set of data products. It should be noted that our map making filters along the scan-direction of the telescope, so this analytic form for the transfer function only applies to maps in the SFL projection. We used a one-dimensional, scan-direction only version of the same transfer function in the ZEA projection for point source finding.

\section{Results: Emissive point source catalog} \label{sec:pointsourcecatalog}
\begin{figure*}[h]
                \centering
                        \includegraphics[width=1\textwidth]{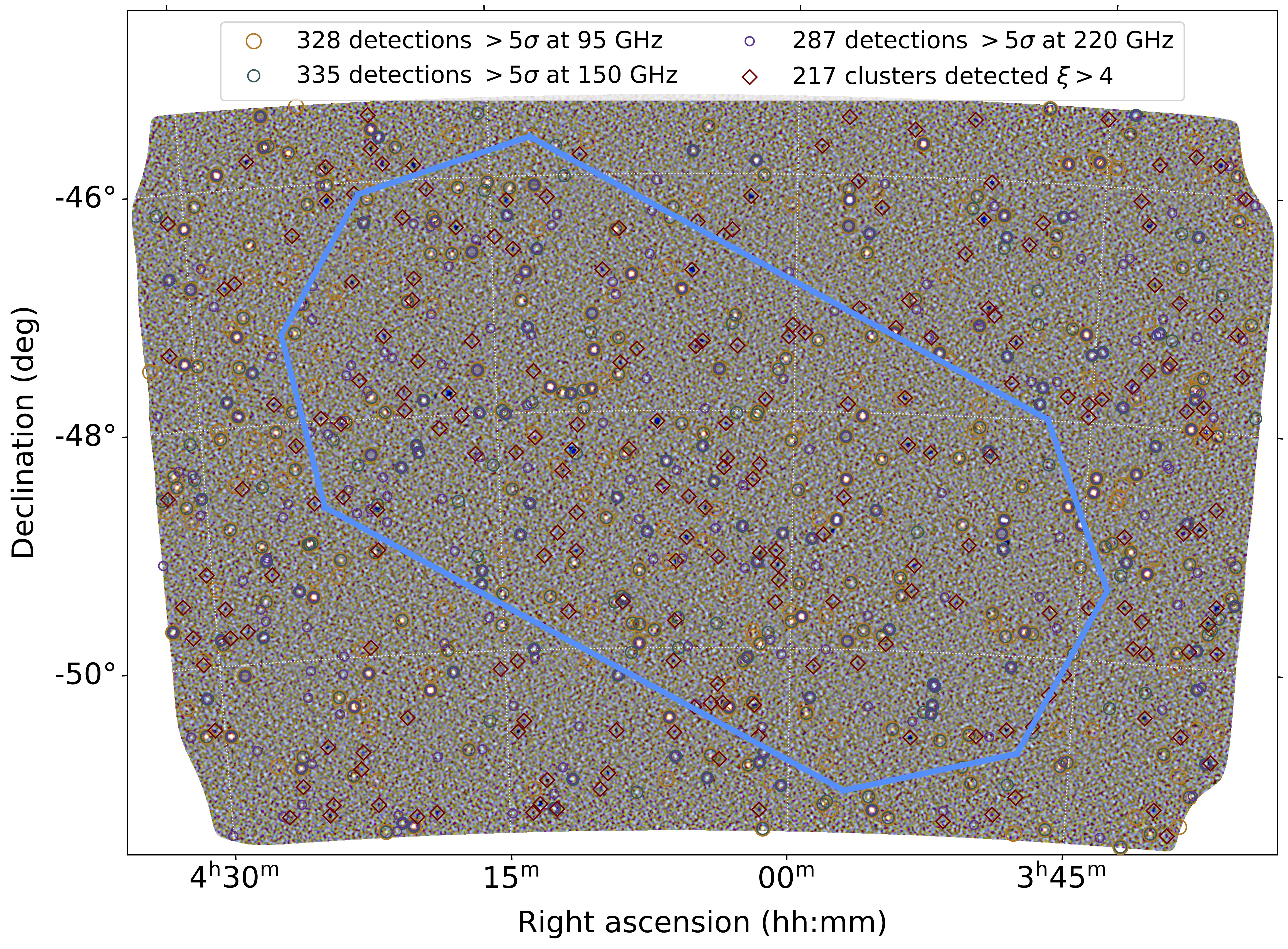}
                        \caption{Maps of the SPT-3G \pixelmaskarea{}-square-degree clean beam in RGB color channels, with R=95 GHz, G=150 GHz, and B=220 GHz. The image was generated using the \texttt{stiff} \citep{bertin12} command \texttt{stiff clean\_beam\_map\_ZEA\_mJy\_95GHz.fits clean\_beam\_map\_ZEA\_mJy\_150GHz.fits clean\_beam\_map\_ZEA\_mJy\_220GHz.fits -MIN\_TYPE QUANTILE -MIN\_LEVEL 0.001,0.03,0.15 -MAX\_TYPE QUANTILE -MAX\_LEVEL 0.999,0.999,0.99 -GAMMA\_LOG YES}. Emissive source locations are indicated by colored circles (95, 150, and 220 GHz detections in orange, green, and purple, respectively) and \numclusterstot{} galaxy clusters are indicated by red diamonds. Bright spots are individual galaxies in the emissive source catalog and dark spots are tSZ galaxy clusters in the cluster catalog. The blue border indicates the EDF-S sky area observed by \textit{Euclid}.}
                \label{fig:indicate_srcs}
        \end{figure*}

We present two catalogs of astrophysical objects: emissive point sources in this section and galaxy cluster candidates from tSZ decrements in Sect. \ref{sec:clustercatalog}. Detections from both catalogs are shown on the 150 GHz clean beam map in Fig. \ref{fig:indicate_srcs}, where the dominant signals are from individual point sources (bright dots) and galaxy clusters (dark spots). Comparing the clean beam maps in Fig. \ref{fig:indicate_srcs} to the raw temperature maps in Fig. \ref{fig:edfs_region}, we see how the source-finding procedure suppresses CMB signal to leave only point-like features. The census of detections and source designations in the emissive source catalog are provided in Table \ref{tab:census}.

    \begin{table*}[h]
    \caption{Census of the emissive point source catalog.}
                \centering
                \begin{tabular}{l c c}
            \hline
            \textbf{Criterion} & \textbf{Number of sources (inside EDF-S)} & \textbf{\% of sources (inside EDF-S)} \\
            Sources detected only at 95 GHz & \numdetonlynine{} (\edfsnumdetonlynine{}) & \fracdetonlynine{} (\edfsfracdetonlynine{}) \\
            Sources detected only at 150 GHz & \numdetonlyonefive{} (\edfsnumdetonlyonefive{}) & \fracdetonlyonefive{} (\edfsfracdetonlyonefive{}) \\
            Sources detected only at 220 GHz & \numdetonlytwotwo{} (\edfsnumdetonlytwotwo{}) & \fracdetonlytwotwo{} (\edfsfracdetonlytwotwo{}) \\
            Sources detected only at 95, 150 GHz & \numdetonlynineandonefive{} (\edfsnumdetonlynineandonefive{}) & \fracdetonlynineandonefive{} (\edfsfracdetonlynineandonefive{}) \\
            Sources detected only at 150, 220 GHz & \numdetonlyonefiveandtwotwo{} (\edfsnumdetonlyonefiveandtwotwo{}) & \fracdetonlyonefiveandtwotwo{} (\edfsfracdetonlyonefiveandtwotwo{}) \\
            Sources detected only at 95, 220 GHz & \numdetonlynineandtwotwo{} (\edfsnumdetonlynineandtwotwo{}) & \fracdetonlynineandtwotwo{} (\edfsfracdetonlynineandtwotwo{}) \\
            Sources detected in all three bands & \numdetonlyallbands{} (\edfsnumdetonlyallbands{}) & \fracdetonlyallbands{} (\edfsfracdetonlyallbands{}) \\
            Total number of sources & \numsrcs{} (\numedfs{}) & 100 (\fracedfs{}) \\
            \hline
            Synchrotron-dominated & \numsynch{} (\edfsnumsynch{}) & \fracsynch{} (\edfsfracsynch{}) \\
            Dust-dominated & \numdusty{} (\edfsnumdusty{}) & \fracdusty{} (\edfsfracdusty{}) \\
            DSFG candidate & \numdsfgcand{} (\edfsnumdsfgcand{}) & \fracdsfgcand{} (\edfsfracdsfgcand{}) \\
            Extended & \numextended{} (\edfsnumextended{}) & \fracextended{} (\edfsfracextended{}) \\
            \hline
                \end{tabular}
                \tablefoot{``Sources detected only at'' refers to unique detections, and therefore a source in the catalog is only represented once in these rows. The numbers and percent shares inside the \textit{Euclid} EDF-S region are in parentheses. The criteria used to classify sources as synchrotron- or dust-dominated are described in Sect. \ref{sec:classification} and the criteria to flag sources as DSFG candidates and extended objects are described in Sect. \ref{sec:flags}.}
                \label{tab:census}
        \end{table*}

Point sources detected at millimeter wavelengths can broadly be categorized into groups of synchrotron-emitting, typically radio-loud AGNs and thermally emitting dusty galaxies at low and high redshift \citep{everett20, vargas23}.\footnote{We searched for stars in the point source catalog using methods described in \cite{everett20} and \cite{tandoi24} and determined there are no star candidates.} Synchrotron emission is associated with a falling or flat source spectrum as frequency increases while thermal emission is associated with a rising spectrum (we do not consider the case of optically thick synchrotron emission, which may appear similar to a thermal source spectrum). The point source catalog comprises \numsrcs{} (\numedfs{}) total sources (inside EDF-S), \numsynch{} (\edfsnumsynch{}) of which are synchrotron-dominated and \numdusty{} (\edfsnumdusty{}) of which are dust-dominated. The multiwavelength counterparts, classifications as synchrotron or dusty, and characterizations of objects in the emissive source catalog are described in the following subsections. 

\subsection{Type classification} \label{sec:classification}
\begin{figure*}[h]
                \centering
                        \includegraphics[width=1\textwidth]{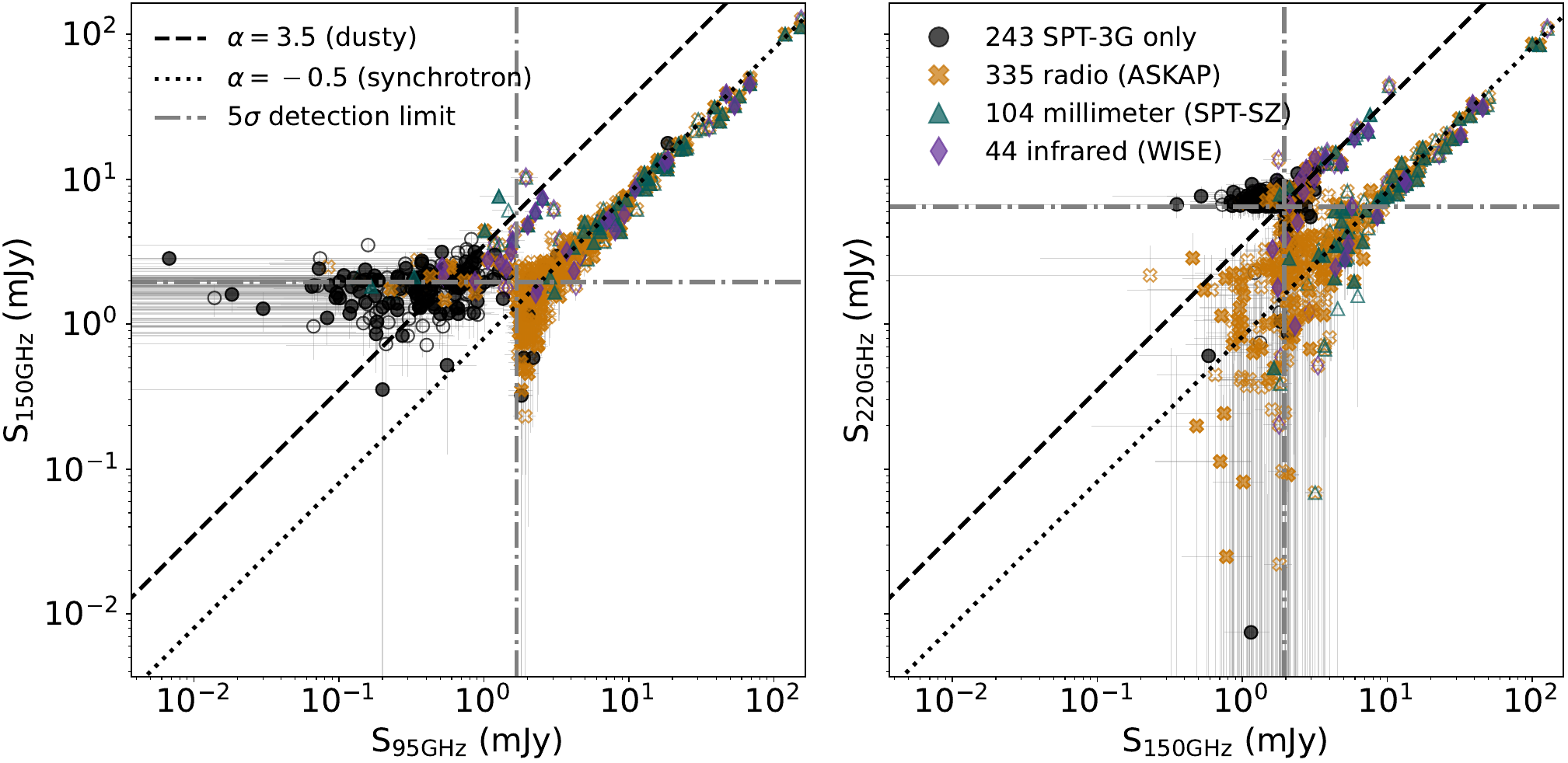}
                        \caption{Comparison of the 150 GHz and 95 GHz fluxes (left panel) and the 220 GHz and 150 GHz fluxes (right panel) of entries in the emissive source catalog. The external counterparts of the sources are marked in black, orange, purple, and green to indicate no association, radio (ASKAP), infrared (WISE), and millimeter (SPT-SZ) associations, respectively. Sources inside the EDF-S footprint are indicated with filled-in symbols, and sources outside the footprint are left unfilled. We note that 37 sources are omitted from the left panel and 43 from the right because their flux values are negative in one of the bands and thus cannot be log scaled. The negative fluxes are a result of noise fluctuations from forced photometry in a source's non-detection band. Of the 80 omitted sources, 37 have no associations, 42 have radio counterparts, and one has an infrared counterpart. We also indicate typical spectral indices for synchrotron ($\alpha=-0.5$) and dusty ($\alpha=3.5$) sources as dotted and dashed lines, respectively. Finally, we show the 5$\sigma$ detection thresholds for catalog admission as dot-dash lines. In general, the brightest objects are synchrotron-dominated AGNs and the dimmest objects are dust-dominated DSFGs.}
                \label{fig:fluxvsflux}
        \end{figure*}

The catalog contains fluxes measured in three bands and signal-to-noise ratios for each source. Fluxes are shown in Fig. \ref{fig:fluxvsflux} and marked by external catalog associations, which are discussed in Sect. \ref{sec:crossmatch}. Two populations of sources, synchrotron and dusty, become obvious in the flux versus flux plots.

We used the logarithm of the ratio of fluxes between our bands, otherwise known as the spectral index, to classify sources as synchrotron-dominated or dust-dominated. The spectral index, $\alpha$, is

    \begin{align}
        \begin{split}
           \alpha^{1}_{2} &= \frac{\log (\frac{\mathrm{S}_1}{\mathrm{S}_2}) }{\log (\frac{\nu_1}{\nu_2})},
        \end{split} \label{eqn:alpha}
    \end{align}

\noindent where the subscripts 1 and 2 refer to the frequency bands, such as 95 and 150 GHz or 150 and 220 GHz. We used the fiducial frequencies given in Sect. \ref{sec:psfinding}, though the effect of choice of $\nu$ on the resulting $\alpha$ and flux values is small.

The turnover point between the source populations is not a strictly flat spectrum ($\alpha=0$) because the distributions overlap somewhat; we used the \cite{everett20} value $\alpha^{150}_{220}<\alphacut{}$ to classify sources as synchrotron-dominated and $>\alphacut{}$ as dust-dominated. Synchrotron sources make up most of the 95 GHz detections, and to a lesser extent the 150 GHz detections, while dusty sources make up most of the 220 GHz detections.

\begin{figure*}[h]
                \centering
                        \includegraphics[width=1\textwidth]{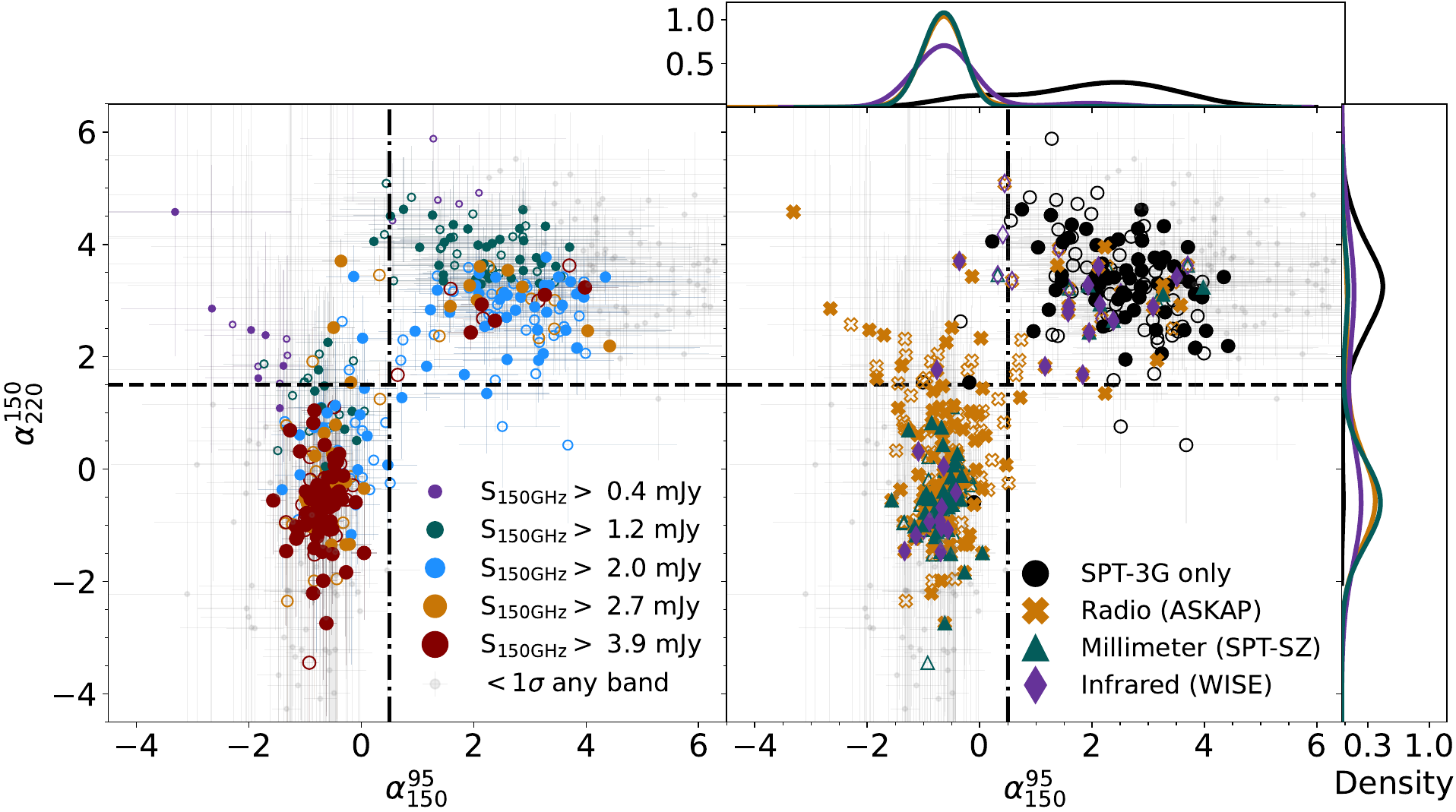}
                        \caption{Source spectral indices between 150 and 220 GHz versus the spectral indices between 95 and 150 GHz. Note that \numnans{} sources are omitted from the plots because their fluxes are below 0 and filled-in symbols indicate sources inside the EDF-S. \textit{Left panel:} Spectral indices that are color-coded according to their 150 GHz flux values: 0.4, 1.2, 2.0, 2.7, and 3.9 mJy are roughly equivalent to 1, 3, 5, 7, and 10$\sigma$ detections. This view highlights the population separation between high signal-to-noise sources, which typically have synchrotron-dominated spectra, and dim sources which comprise the dusty population. \textit{Right panel:} Spectral indices that are represented by colors and symbols according to their external catalog associations (see the caption of Fig. \ref{fig:fluxvsflux}). In this view, we emphasize that sources with dust-dominated spectra are uniquely discovered by low-noise millimeter-wavelength surveys such as those conducted by the SPT. The top and right panels are the kernel density estimations of each $\alpha$ axis, showing the bimodal distributions in both $\alpha^{95}_{150}$ and $\alpha^{150}_{220}$ but especially how $\alpha^{150}_{220}$ can be used to assign a spectral type of synchrotron or dusty; in this analysis, we adopted the cutoff value of $\alpha^{150}_{220}=1.5$ (marked by dotted lines).}
                \label{fig:alphavsalpha}
        \end{figure*}

The median spectral index between 150 and 220 GHz fluxes is $-0.5$ and 3.5 for synchrotron-dominated and dust-dominated sources, respectively. We highlight that the synchrotron sources in Fig. \ref{fig:fluxvsflux} consistently share a spectral index of about $-0.5$ from both 95 to 150 GHz and from 150 to 220 GHz; however, the dusty population has far less emission at 95 GHz overall. For this reason, we used the prescription adopted in \cite{everett20} for source classification, where the $\alpha^{150}_{220}$ quantity determines the classification because the forced photometry in the 95 GHz band for dusty sources is often not significant. The distribution of $\alpha^{150}_{220}$ versus $\alpha^{95}_{150}$ is shown in Fig. \ref{fig:alphavsalpha}, where the separation between the two populations are most distinct.

\paragraph{Synchrotron-dominated sources.} Blazars are AGNs with relativistic jets pointed towards the observer. They are further broken into the subclasses FSRQs and BL Lac objects, where FSRQs tend to be brighter and may exhibit emission lines that BL Lac objects lack \citep{urry95}. The sample of SPT-detected synchrotron-dominated sources are primarily FSRQs. These \numsynch{} sources (\fracsynch{}\% of the catalog) are located in the lower left quadrant of Fig. \ref{fig:alphavsalpha}, where the brightest sources in the catalog have fairly flat spectral indices (both $\alpha^{150}_{220}$ and $\alpha^{95}_{150}$ are close to 0).

Because blazars are typically bright across a wide range of wavelengths, it is expected that the majority of synchrotron sources the SPT detects will have multiwavelength, and particularly radio, counterparts. Indeed, 95\% of the SPT-3G sources classified as synchrotron-dominated have counterparts in the radio catalog with which we compared (there are more details on cross-matching with external datasets in Sect. \ref{sec:crossmatch}). In the spectral energy distribution of an AGN, synchrotron radiation, which is caused by the acceleration of relativistic charged particles, is observed from radio to ultraviolet wavelengths \citep{dutka17}, and therefore it is likely that the SPT is probing the same mechanism as the radio regime.

\paragraph{Dust-dominated sources.} The other \fracdusty{}\% of the catalog consists of \numdusty{} dust-dominated sources, in which the millimeter emission is primarily reprocessed starlight emitted quasi-thermally by dust grains. A subset of this population of particular astrophysical and cosmological interest are high-redshift DSFGs, some of which are expected to be strongly gravitationally lensed \citep{vieira13, giulietti24}. We selected high-redshift DSFG candidates by removing from the dust-dominated sample any source that was detected strongly in The AllWISE Data Release catalog \citep{cutri14}, which are predominantly at low redshift ($z \lesssim 0.1$), and any source that exhibited unusual or flat spectral behavior not indicative of a high-redshift DSFG. Quantitative details of the DSFG flagging procedure are described in Sect. \ref{sec:flags}. By these criteria, almost three in four dusty sources in the SPT-3G sample are high-redshift DSFG candidates and do not have counterparts at other wavelengths (emphasized in the right panel of Fig. \ref{fig:alphavsalpha}).

SPT-selected DSFGs have been measured to have a median redshift of $z \approx 4$ \citep{reuter20} and are important laboratories for studying the extreme end of galaxy formation and evolution. Some objects in the DSFG count may also be galaxy protoclusters \citep{miller18}. While protoclusters can be discovered in millimeter-wavelength surveys and selected by searching for DSFG characteristics, high-resolution imaging is needed to confirm them and differentiate the two populations \citep{wang21a}. According to dusty source models (such as \citealt{negrello07}) that are able to reproduce the strongly lensed DSFG sample discovered with the first-generation SPT camera, SPT-SZ \citep{vieira13, everett20}, we would expect approximately 25 strongly lensed and 23 unlensed DSFGs in the EDF-S (blue outline in Fig. \ref{fig:indicate_srcs}). Total source counts in this field are approximately $1 \sigma$ higher than those in \cite{everett20}; that we find \edfsnumdsfgcand{} DSFG candidates (i.e., and not 48) is under investigation. While some of the discrepancy can be attributed to flux boosting (see Sect. \ref{sec:psfinding}), it does not account for the entirety of the difference, and thus we attribute this mostly to random statistical fluctuation. 

\paragraph{Extended sources.} We find \numextended{} extended sources in the catalog (\edfsnumextended{} inside EDF-S) based on the procedure described in Sect. \ref{sec:flags}. These sources span emission type---14 of 19 synchrotron-dominated extended sources have a radio association, indicating that these may be AGNs with resolved radio lobes \citep[e.g.,][]{davies11, mahony11}. Four dust-dominated extended objects are local galaxies with infrared counterparts, and 13 of the 19 dust-dominated extended sources are flagged as DSFGs with few external counterparts; these sources will be studied further for galaxy protocluster candidacy. The last two non-DSFG, dusty, extended sources with no counterparts have slightly negative 95 GHz forced photometry.

\subsection{Purity of emissive detections} \label{sec:emissivepurity}
We evaluated the purity of the 220 GHz detections by applying the source-finding procedure to the negative 220 GHz map, assuming that no peak in the negative map at 220 GHz was a genuine astrophysical source. This assessment is most important for the DSFG candidates, which are most strongly detected at 220 GHz. We were unable to do the same test at 95 and 150 GHz because of the true tSZ decrements in these maps that were detected by the point-source-finding pipeline. There are no 220 GHz decrements above 5$\sigma$ and just three detections above 4.5$\sigma$. The emissive source catalog we provide has a 5$\sigma$ threshold, so the lack of 220 GHz decrements indicates high purity of the catalog. A full completeness and purity assessment of the SPT-3G emissive source-finding procedure is the subject of a forthcoming work.

\subsection{Flags in the emissive source catalog} \label{sec:flags}
We provide a few useful binary flags (1=\texttt{True} and 0=\texttt{False}) for each entry in the emissive source catalog: \texttt{edfs\_flag}, which indicates whether the object lies in the footprint observed by \textit{Euclid}; \texttt{extended}, which indicates whether the source is resolved in the SPT maps; and \texttt{dsfg\_flag}, which indicates whether the source is a high-redshift DSFG candidate. The criteria for the flags are described in this section. 

\paragraph{\texttt{edfs\_flag.}} We used a mask provided by the \textit{Euclid} collaboration to assess whether objects in the SPT-3G catalog lie in \textit{Euclid's} EDF-S observational area. \fracedfs{}\% of the total sources in the sample lie in the EDF-S footprint. Throughout this work, we provide numbers on both the EDF-S-only sources and the total sample. 

\paragraph{\texttt{extended.}} To determine source extension, we made a small cutout of the clean beam map (Fig. \ref{fig:indicate_srcs}) around every source, fit a two-dimensional Gaussian to the cutout, and calculated the major and minor axis FWHM values. A source was flagged as extended in the catalog if one of its axes had a FWHM that was $>2.5\sigma$ from the median FWHM value in any band. Using these criteria, \numextended{} (\edfsnumextended{}) sources are flagged as extended and \numFalse{} (\edfsnumFalse{}) are not extended (inside EDF-S).

\paragraph{\texttt{dsfg\_flag.}} We further scrutinized the dust-dominated group to select DSFG candidates that we then flagged in the catalog. Following the selection criteria in \cite{everett20}, we restricted the DSFG candidate sample to sources with strictly rising spectra ($\alpha^{95}_{150}>0.5$ in addition to $\alpha^{150}_{220} > \alphacut{}$, which included forced photometry values for sources below the detection threshold) and no infrared counterpart. When applied to the SPT-SZ catalog, these criteria yielded 506 DSFG candidates (out of 4,845 total for approximately 10\% of the sample). We find \numdsfgcand{} sources out of \numsrcs{} (\fracdsfgcand{}\% of the sample) in this catalog are DSFG candidates by the same standards, and \edfsnumdsfgcand{} of these objects lie in the EDF-S region observed by \textit{Euclid}. The reason for the drastic increase in the fraction of DSFG candidates is the improved map depths of the SPT-3G EDF-S observations that reveal more sources along the steep luminosity function of dusty sources. These cuts are agnostic towards lensed and unlensed DSFGs because the differentiation between them cannot be constrained with the arcminute-resolution SPT data alone.

\subsection{Cross-matching with external catalogs} \label{sec:crossmatch}
To assist in characterizing the properties of the objects, we associated the emissive source catalog with multiwavelength datasets using external radio, millimeter, and infrared catalogs. For cross-matching to the millimeter catalog, we associated sources inside radial distances calculated using

    \begin{align}
        \begin{split}
           r_{\mathrm{assoc}} = \sqrt{ 2 \left[ (\Delta \mathrm{SPT})^2 + \left( \frac{\Gamma_{\mathrm{SPT}}}{5} \right) ^2 \right]},
        \end{split} \label{eqn:crossmatch}
    \end{align}

\noindent where $\Delta\mathrm{SPT}$ is the positional uncertainty for SPT (rounded up to 5 arcsec), $\Gamma_{\mathrm{SPT}}$ is the beam FWHM for SPT at 150 GHz (1.2 arcmin), and the denominator 5 is the signal-to-noise of the detection (set at 5 for all sources, which is the lowest possible detection value); the result is 43.2 arcsec. For the radio and infrared catalogs, which have significantly greater source densities and smaller beam sizes, we used a 30 arcsec association radius. These choices result in probabilities of random association, $P(\mathrm{random}) = \Sigma \cdot \pi \cdot r_{\mathrm{assoc}}^2$, where $\Sigma$ is catalog source density, that are $\leq$ 2\% for all catalogs.
    
Though we experimented with varying the distance threshold depending on the SPT signal-to-noise of the detection, we found that it did not significantly improve or affect the results. We cross-matched the following external datasets whose relevant characteristics are outlined in Table \ref{tab:crossmatch}.

        \begin{table*}[h]
    \caption{Characteristics of the external catalogs with which the emissive point source catalog is cross-matched and results of the cross-matching.}
                \centering
                \begin{tabular}{c c c c c c c}
                        \hline
            \textbf{Survey} & \textbf{Band} & \textbf{Beam FWHM} & \textbf{$\Sigma$ (deg$^{-2}$)} & \textbf{$r_{\mathrm{assoc}}$ (arcsec)} & \textbf{$P$(random)} (\%) & \textbf{N. sources} \\
                        ASKAP low & 887.5 MHz & 25 arcsec & \srcdensityASKAPlow{} & \rassocASKAPlow{} & \PrandomASKAPlow{} & \nummatchesASKAPlow{} \\
            ASKAP mid & 1367.5 MHz & 25 arcsec & \srcdensityASKAPmid{} & \rassocASKAPmid{} & \PrandomASKAPmid{} & \nummatchesASKAPmid{} \\
            ASKAP high & 1655.5 MHz & 25 arcsec & \srcdensityASKAPhigh{} & \rassocASKAPhigh{} & \PrandomASKAPhigh{} & \nummatchesASKAPhigh{} \\
                        AT20G & 20 GHz (1.5 cm) & 4.6 arcsec & \srcdensityATtwentyG{} & \rassocATtwentyG{} & \PrandomATtwentyG{} & \nummatchesATtwentyG{} \\
            SPT-SZ & 95, 150, 220 GHz & 1.7, 1.2, 1.0 arcmin & \srcdensitySZ{} & \rassocSZ{} & \PrandomSZ{} & \nummatchesSZ{} \\
            WISE & 22 $\mu$m & 12 arcsec & \srcdensityWISE{} & \rassocWISE{} & \PrandomWISE{} & \nummatchesWISE{} \\
            \hline
            SPT-3G & 95, 150, 220 GHz & 1.8, 1.2, 1.0 arcmin & \srcdensitySPT{} & \internalrassoc{} & \PrandomSPT{} & \numnocps{}\tablefootmark{a} \\
                \hline  
        \end{tabular}
                \tablefoot{We took into consideration the relative beam sizes and source densities when determining the association radii, ensuring $P$(random) is $\leq2$\%. The single-band SPT-3G source densities are \srcdensitySPTnine{}, \srcdensitySPTonefive{}, and \srcdensitySPTtwotwo{} sources per square degree for 95, 150, and 220 GHz, respectively.\\ 
        \tablefoottext{a}{This number reflects sources without counterparts in any of the listed surveys.}
        }
                \label{tab:crossmatch}
        \end{table*}

\paragraph{ASKAP.} The Rapid ASKAP Continuum Survey (RACS) from the Australian Square Kilometre Array Pathfinder (ASKAP; RACS-low at 887.5 MHz, \citealt{hale21}; RACS-mid at 1367.5 MHz, \citealt{duchesne24}; and RACS-high at 1655.5 MHz, \citealt{duchesne25}) has \nummatchesASKAP{} total matches. We separately cross-matched the emissive source catalog to each of the ASKAP bands because the catalogs were treated as independent entities in their release. In the SPT-3G catalog file, we indicate only if the SPT detection was associated with a source in any of the ASKAP bands (see the ``Counterparts'' column of Table \ref{tab:catalog_flags}). As previously mentioned, 95\% of the synchrotron sources have a radio counterpart in ASKAP, which is consistent with our understanding of the AGN emission mechanism from radio to millimeter wavelengths. Future and ongoing radio surveys in the Southern Hemisphere with the Square Kilometre Array mid-frequency array and its precursors, MeerKAT \citep{jonas16} and ASKAP, will provide further constraints and discriminatory power for the SPT-3G emissive sources.

\paragraph{SPT-SZ.} The SPT-SZ survey at 95, 150, and 220 GHz \citep{everett20} has \nummatchesSZ{} matches. The SPT-SZ 4.5$\sigma$ detection limits are 9.8, 5.8, and 20.4 mJy in 95, 150, and 220 GHz, respectively (or factors of 7, 3.6, and 3.9 higher than the 5$\sigma$ thresholds presented here). The first-generation SPT survey completely covers the EDF-S, enabling some direct comparisons: one strongly lensed DSFG, SPT-S J041839-4752.0, is in the EDF-S, and has a spectroscopic redshift of 4.2 from the Atacama Large Millimeter/submillimeter Array \citep[ALMA,][]{reuter20} and characterization from \textit{James Webb} Space Telescope (JWST) observations \citep{rigby25}. As a comparison, we also checked the Atacama Cosmology Telescope Southern Surveys point source catalog at 150, 220, and 280 GHz (ACT, \citealt{vargas23}) and confirmed all SPT-3G matches in ACT were also matched to SPT-SZ. Only 5\% of sources classified as dusty in SPT-3G are associated with SPT-SZ sources, meaning that the vast majority of dusty sources in this work are new discoveries.

\paragraph{WISE.} WISE at 22 $\mu$m (band 4) only \citep{cutri14} has \nummatchesWISE{} matches. We used counterparts from WISE to flag low-redshift infrared galaxies. In \cite{everett20}, the Faint Source Survey (FSS) from the Infrared Astronomical Satellite at 12, 25, 60, and 100 $\mu$m (IRAS, \citealt{moshir90}) was used for this purpose. We did not expect the WISE maps to be sensitive to the (largely obscured) stellar emission of DSFGs, whereas low-redshift infrared galaxies are bright at 22 $\mu$m. We emphasize that we only associated with sources brighter than 8.05 magnitudes (5.0 mJy) in WISE band 4, which is approximately 5$\sigma$, rather than the entire WISE catalog, which would result in numerous false associations. Fewer than 10\% of dusty SPT-3G sources have a WISE counterpart, indicating that a small fraction of SPT-3G dusty sources are at low redshift or are rare, mid-infrared-bright strongly lensed galaxies at $z\sim1.5-2$ \citep{diazsanchez17, diazsanchez21}. As a check, we also cross-matched with IRAS and confirmed that SPT-3G matches in IRAS are all included in the WISE matches.

There are \numnocps{} sources in the emissive sample with no multiwavelength counterparts. We show the fluxes and spectral indices for sources in the catalog indicated by their external associations in Figs. \ref{fig:fluxvsflux} and \ref{fig:alphavsalpha}; in general, synchrotron sources have a radio counterpart while dusty sources are more likely to be unassociated. Both synchrotron and dusty sources have infrared and millimeter counterparts. Over 90\% of the sources without counterparts are classified as dust-dominated, highlighting the distinctive ability of millimeter surveys to discover sources unable to be detected at other wavelengths.

\subsection{Cross-matching with \textit{Euclid} Q1 catalogs} \label{sec:euclidmatch}
As discussed in the previous sections, millimeter-selected DSFGs have been found to also be members of rare subpopulations: strongly gravitationally lensed systems and galaxy protoclusters. We would like to investigate the comparative strengths and weaknesses of the SPT and \textit{Euclid} for the discovery and characterization of such sources. It is known from the previous SPT emissive source catalog \citep{vieira13} that a significant fraction of SPT-selected sources are strongly gravitationally lensed. Though the background sources are expected to be obscured in the optical bands, using \textit{Euclid} data to characterize some lensing systems would increase efficiency over individual-source follow-up of SPT DSFGs. SPT-selected protoclusters may also be constrained by characteristics in the \textit{Euclid} data that the SPT cannot observe. The \textit{Euclid} Q1 release included a strong lens catalog \citep{walmsley25} and the \textit{Euclid} view of \textit{Planck} protoclusters \citep{dusserre25} with which we compared the SPT sample. These comparisons are not meant to be exhaustive, but rather provide a first look at SPT-selected DSFGs in \textit{Euclid}, with some deeper insight on the classifications of the sources as strong lens and protocluster candidates.

We used a 43.2 arcsec radial association threshold to match to the strong lens catalog and a 5 arcmin threshold to match to the protoclusters, as was used in \cite{dusserre25}. The results of the cross-matching are in Tables \ref{tab:catalog_flux} and \ref{tab:catalog_flags}, showing seven matches to strong lenses (out of over 1\,000 in the EDF-S) and five matches to \textit{Planck} protoclusters (out of seven in the EDF-S). There are no matches to grade A strong lenses, one match to grade B, and six matches to grade C lenses. Just two of the lens matches are flagged as DSFG candidates; this low number may indicate that the types of strong lens systems that SPT and \textit{Euclid} respectively probe are different.

\cite{dusserre25} presents seven \textit{Planck} protoclusters in the EDF-S for comparison and recovers two \textit{Euclid} counterparts. We associate five SPT sources from the list of seven, including the two that \cite{dusserre25} recovers (SPT-S J040259-4711.6 and SPT-S J041604-4926.0); of the five, four are flagged as DSFGs. While further study is required to fully design a joint protocluster search using SPT and \textit{Euclid} data, it appears promising that millimeter-wave surveys can flag possible candidates while \textit{Euclid} can help distinguish them from DSFGs.

\begin{figure*}[h]
                \centering
                        \includegraphics[width=1\textwidth]{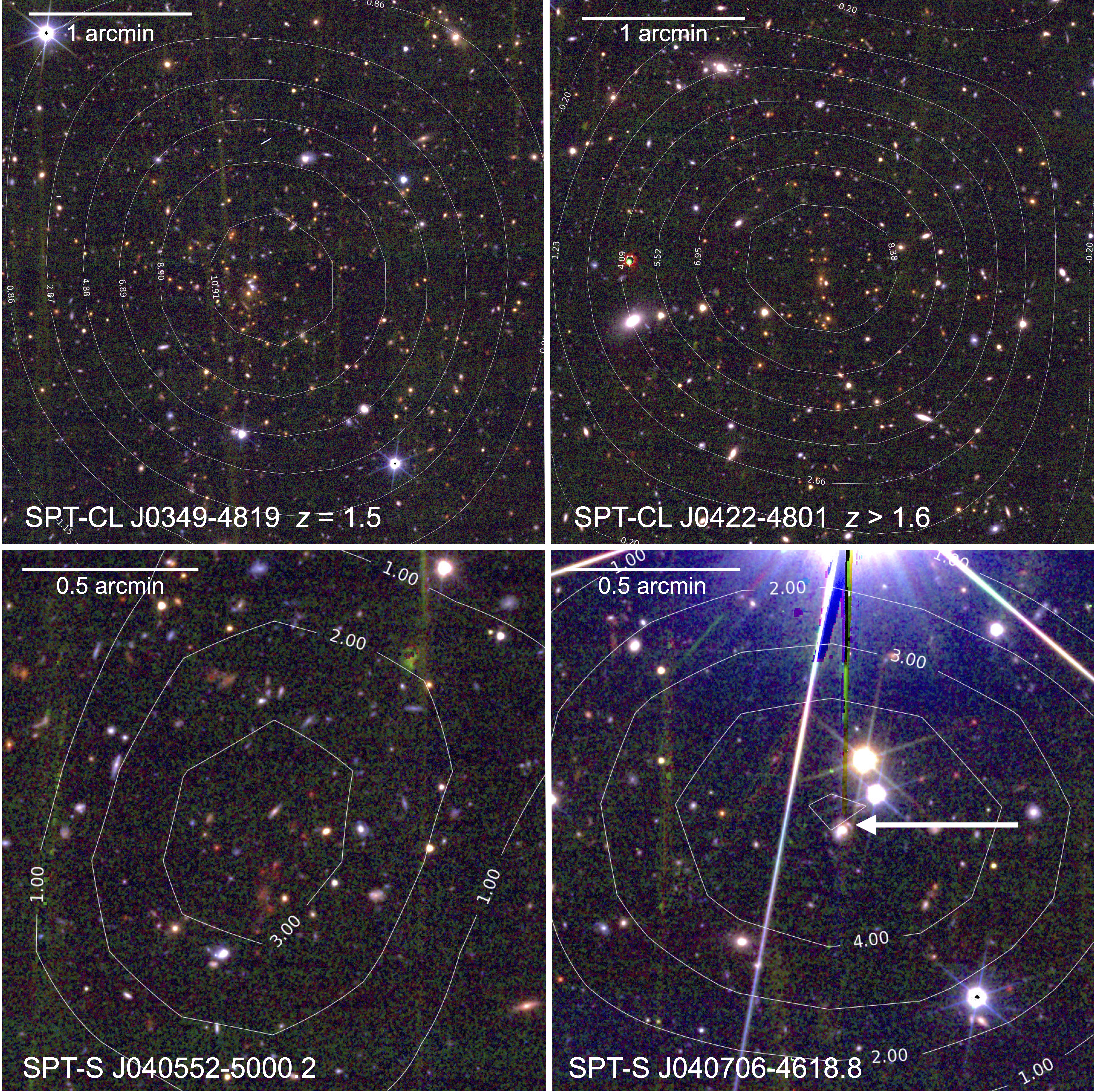}
                        \caption{Detection contours of SPT overlaid on \textit{Euclid} (RGB) H, Y, and VIS imaging from the Q1 data release for four detections in our catalogs. \textit{Top row:} Two tSZ identified clusters from the SPT-3G EDF-S survey (Sect. \ref{sec:clustercatalog}). These images highlight the sensitivity of both SPT and \textit{Euclid} to high-redshift galaxy clusters. \textit{Bottom row:} Two examples of DSFGs in the emissive SPT-3G source catalog (Sect. \ref{sec:pointsourcecatalog}). In the bottom-left, we show a protocluster candidate characterized as a DSFG and associated with the approximately dozen red dots centered at the SPT location. In the bottom right, we show a strongly lensed DSFG candidate, identified by the ``halo'' feature at the SPT location indicated by an arrow. The visible arc may be an unobscured component of the background DSFG. This strong lens candidate is not identified in the \textit{Euclid} Q1 strong lens catalog, possibly pointing to differences in strong lens selections between \textit{Euclid} and SPT.}
                \label{fig:euclidimages} 
        \end{figure*}

In Fig. \ref{fig:euclidimages} we show two DSFG-flagged sources from the emissive source catalog (lower panels) as seen in \textit{Euclid} data (H, Y, and VIS as RGB colors). Here, we highlight an SPT strong lens candidate in the lower right panel that is not associated with an object in the \cite{walmsley25} strong lens catalog. We show a protocluster candidate in \textit{Euclid} data in the lower left panel of Fig. \ref{fig:euclidimages}. It is not one of the \textit{Planck} protocluster matches---its protocluster candidacy is based on its SPT characteristics and visual identification of an apparent overdensity of red galaxies at the SPT location.

\section{Results: Galaxy cluster catalog} \label{sec:clustercatalog}
We present a tSZ-selected galaxy cluster catalog over the \pixelmaskarea{}-square-degree SPT-3G EDF-S patch. The catalog contains \numclusterstot{} candidates detected at $\xi > 4$, with \numclustersedfs{} located in the 30-square-degree EDF-S patch. This is a cluster density of 3.81 clusters per square degree, which is an order of magnitude improvement over previous tSZ-selected cluster catalogs in this region. We provide the equatorial coordinates, detection significance, core size, and estimates of the redshift and mass for each optically confirmed (see Sect. \ref{sec:redshifts}) candidate in the catalog, which is accessible online with catalog quantities described in Table \ref{tab:columnlabels}. In the top panels of Fig. \ref{fig:euclidimages} we show two examples of high-redshift clusters found by the SPT in this sample; the tSZ detection contours are overlaid onto imaging data from the \textit{Euclid} Q1 release. 

In this section we analyze the purity and completeness of the cluster sample and draw comparisons to other ICM-selected cluster samples in the SPT-3G EDF-S region. Drawing comparison statistics to other tSZ-selected ICM catalogs shows the advances that have been made in the past several years in obtaining higher cluster counts. Comparing to X-ray catalogs, on the other hand, shows the power of the redshift independent tSZ selection and the different mass regimes that these two methods probe.

\begin{figure}[h]
                \centering
                \includegraphics[width=1\columnwidth]{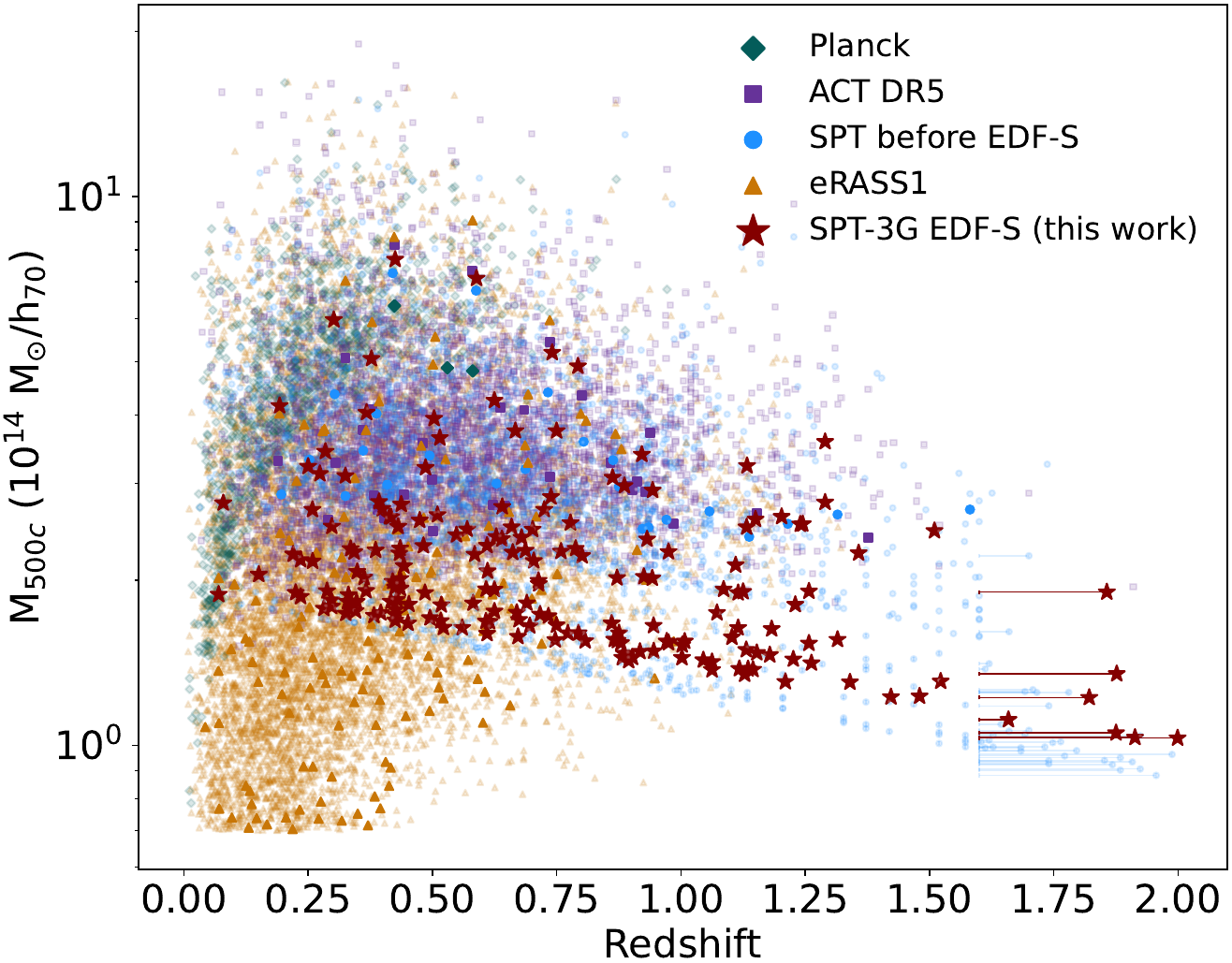}
                        \caption{Mass versus redshift for the SPT-3G EDF-S cluster catalog compared to samples from other ICM-based wide field cluster surveys. In pale shading, we include all clusters from published catalogs and, in bold, clusters in the SPT-3G EDF-S footprint. We plot sources from this work (red stars), \textit{Planck} \citep[green diamonds,][]{planck15-27},  ACT \citep[purple squares,][]{hilton21}, eROSITA \citep[orange triangles,][]{bulbul24}, and prior SPT samples \citep[blue circles,][]{bleem20, klein24a, bleem24, kornoelje25}. The SPT-3G sample extends to significantly lower masses at high redshift than previously published works in the EDF-S region. Finally, note that as we only have lower redshift limits for SPT clusters at $z>1.6$ (see Sect. \ref{sec:redshifts}), clusters at higher redshifts are plotted with redshifts drawn from a mass function for our fiducial $\Lambda$CDM cosmology.} 
                \label{fig:massvsz} 
        \end{figure}

Using the optical confirmation criterion described in Sect. \ref{sec:redshifts}, we report \numclusterswithredshifts{} clusters with redshifts (\fracclusterswithredshifts{} of the catalog). We find the median redshift of the cluster sample to be $z=$ \medclusterz{} and the total redshift range spanned is \minclusterz{} $<z \lesssim$ \maxclusterz{} with \numclustersaboveone \ clusters found above $z>1$. We are conservative with our high-redshift estimates, as there are many uncertainties at $z>1.6$. We refer the reader to the discussion of these uncertainties in \cite{kornoelje25, bleem24} and report all redshifts $z>1.6$ in a lower bound $z=1.6+$ bin. The median mass of the sample is \medmass{}  with a minimum of \minmass{}. MCMF (described in Sect. \ref{sec:redshifts}) provides excellent redshift precision ($\sigma_z/(1+z)$=0.005) at $z<1.1$, with degraded precision and confirmation power at higher redshifts ($\sigma_z/(1+z)$=0.03), owing to the limitations of the WISE data. Data from \textit{Euclid} is expected to significantly improve both cluster confirmation and redshift precision for systems at higher redshift. In Fig. \ref{fig:massvsz} we show the mass-redshift distribution of the SPT-3G EDF-S sample as compared to wide field samples from \textit{Planck} \citep{planck15-27}, eROSITA \citep{bulbul24}, ACT \citep{hilton21}, and SPT \citep[][combined]{bleem20, klein24a, bleem24, kornoelje25}, marking in bold those clusters in the SPT-3G EDF-S footprint. As can be seen here and in Fig. \ref{fig:clusterredshifts}, the SPT sample greatly expands the population of high-redshift ICM-selected clusters in the EDF-S region. 

\subsection{Completeness and purity of the cluster sample} \label{sec:completeness}
We estimated the purity and completeness of the SPT-3G EDF-S cluster catalog. Both quantities are a function of significance that worsen near the detection threshold ($\xi=4.0$).

\begin{figure*}[h]
        \centering
        \includegraphics[width=1.0\textwidth]{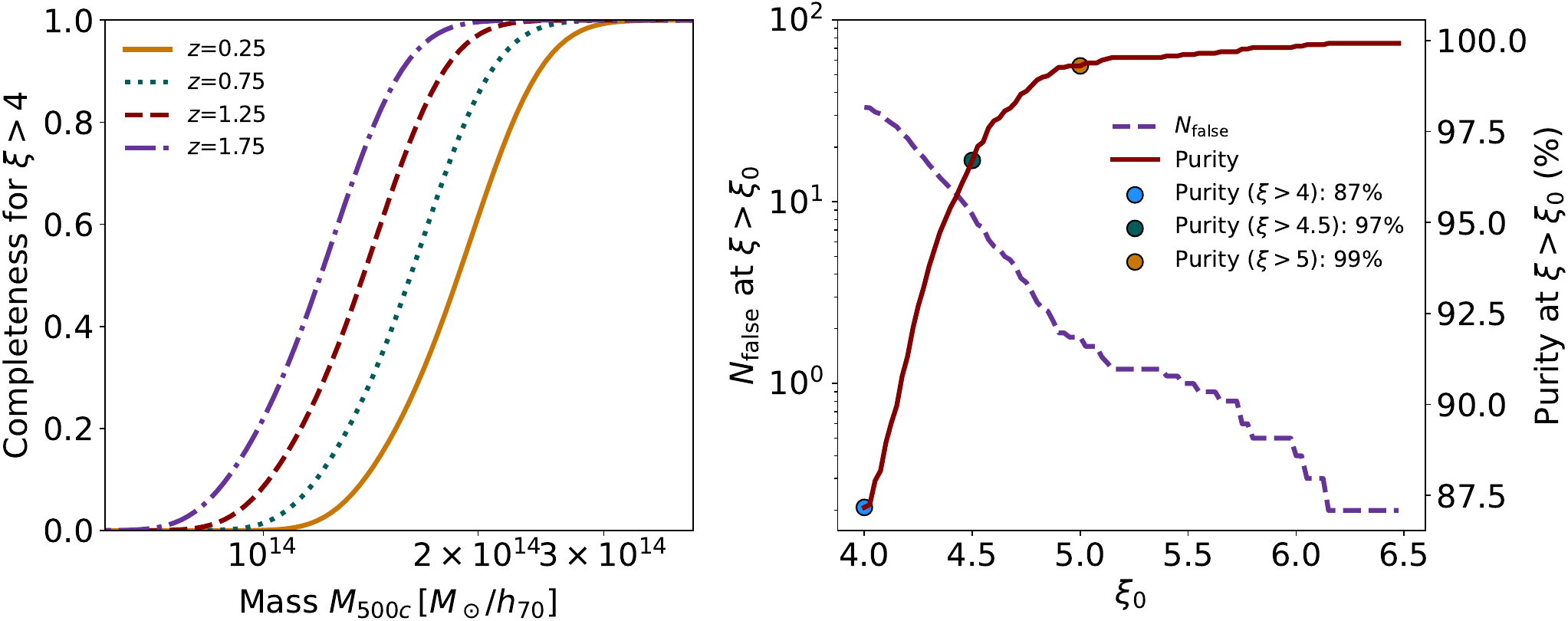}
        \caption{\textit{Left panel:} Completeness of the SPT-3G EDF-S cluster catalog as a function of mass at four different slices in redshift, $z =0.25, 0.75, 1.25,$ and $1.75$. The sample is expected to be more than $90\%$ complete above $2.5 \times 10^{14} M_{\odot}/h_{70}$ ~at $z > 0.25$. \textit{Right panel:} Number of false detections and purity of the cluster catalog as a function of minimum detection significance. We find a purity of $99\%$ at $\xi > 5$, which was also found in previous SPT publications, and a purity of \clusterpurity{} above the minimum significance threshold of $\xi = 4$, corresponding to about 30 predicted false detections in the catalog.}
        \label{fig:completeness_purity}
    \end{figure*}

We modeled the completeness as the probability that a cluster of a given mass and redshift is detected in the SPT-3G EDF-S field. The significance of the completeness was modeled as a Heaviside function of the form $\Theta(\xi - 4.0)$, which we translated into completeness in mass and redshift using Eq. (\ref{eqn:mz}). The conversion considers both the intrinsic scatter from the $\zeta$-mass relationship and the observational scatter on $\xi$, as discussed in Sect. \ref{sec:masses}. We find that the SPT-3G EDF-S cluster catalog is more than $90\%$ complete at masses above $2.5 \times 10^{14} M_{\odot}/h_{70}$ at $z > 0.25$, shown in the left panel of Fig. \ref{fig:completeness_purity}.

We defined the purity as the percent of candidate detections that are true galaxy clusters at a given detection significance. We quantified the purity of the cluster sample by detecting clusters in a set of 20 simulated SPT-3G EDF-S-like maps. The construction of the simulated temperature maps follows methods in \cite{kornoelje25}. We summarize the process as follows: first, we constructed a set of Gaussian realizations of the CMB from the best-fit lensed \textit{Planck} 2018 \(\Lambda\)CDM primary CMB parameters \citep{planck15-27}. We did the same for the kSZ and a background of faint dusty sources, which are constructed from the best-fit spatial and spectral values from \citet{reichardt21}. 

The remaining simulated map components---the tSZ signal, radio galaxies, and instrumental noise---are not Gaussian realizations of an input power spectrum. The simulated tSZ maps were made by pasting tSZ profiles at the locations of massive halos in a lightcone from the Outer Rim simulation \citep{heitmann19}; the maps produced from this process were rescaled such that the amplitude of the tSZ power spectrum matched that measured in \citet{reichardt21} at $\ell = 3\,000$. Radio galaxies were populated in the simulated maps using the point source flux distribution from \citet{lagache20}. We assumed a spectral index of $-$0.7 and $-$0.9 with a scatter of 0.3 to rescale the point source maps made at 150 GHz to 95 GHz and 220 GHz, respectively. We finally used instrumental noise maps described in Sect. \ref{sec:extractsignals}. 

We applied the cluster-finding procedure described in Sect. \ref{sec:clusterfinding} to the simulated maps to best mimic the detection and catalog generation pipeline. Following \citet{bleem24}, we associated candidate tSZ detections with halos from the Outer Rim simulation to probabilistically estimate the number of false detections in the maps. The final purity estimates are shown in the right panel of Fig. \ref{fig:completeness_purity} as a function of detection significance. We find that the SPT-3G EDF-S cluster sample is \clusterpurity{} pure above the detection threshold $\xi = 4$, which correspondingly predicts about 30 false detections, consistent with the number of candidates without optical confirmations. When considering the subsample of clusters that have optical confirmations, the purity is 97\%.

\subsection{Comparisons to other ICM-selected cluster catalogs}\label{sec:sampleresults}
\begin{figure}[h]
                \centering
                        \includegraphics[width=1\columnwidth]{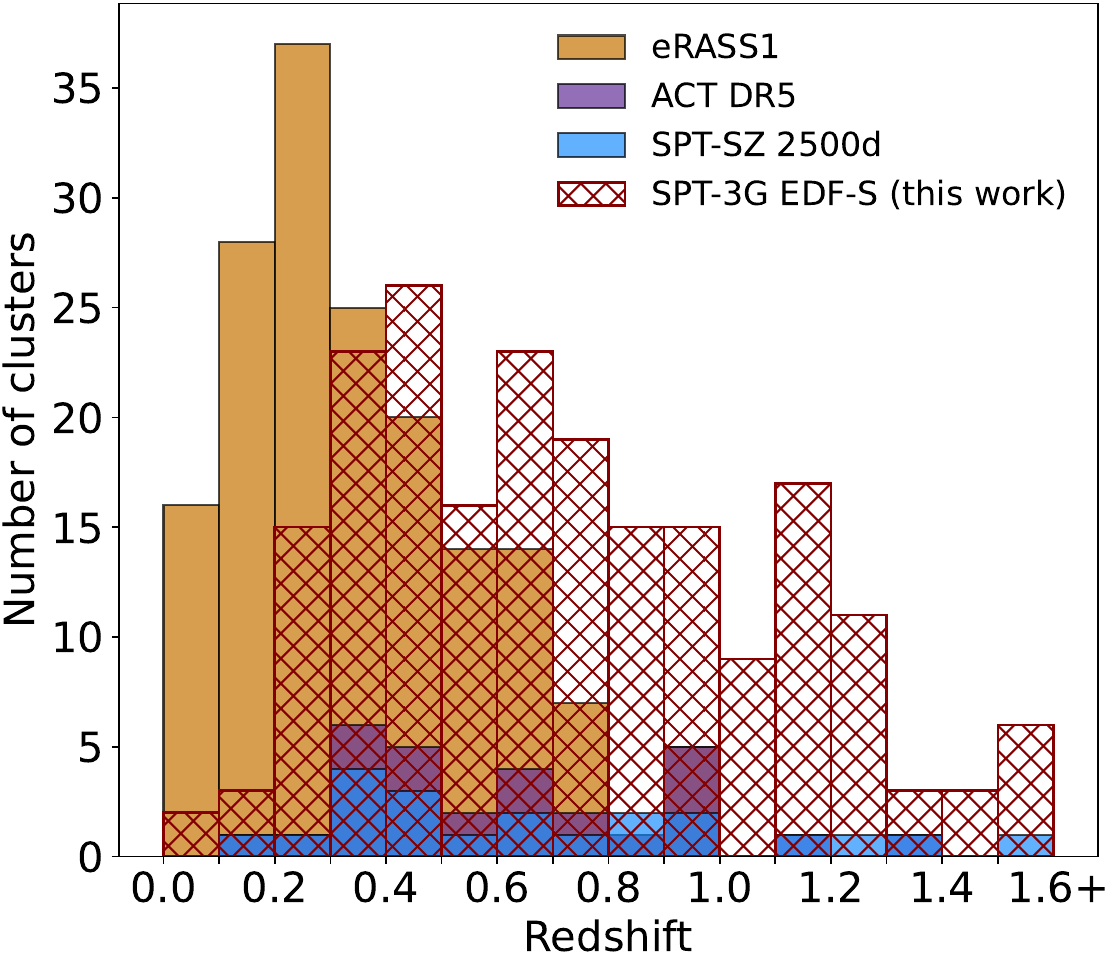}
                        \caption{Redshift distribution of the SPT-3G EDF-S clusters and other ICM-selected clusters in the SPT-3G EDF-S footprint. The SPT-3G EDF-S cluster sample (red hatch pattern) has a median redshift of $z=$ \medclusterz{} and spans the range $\minclusterz{} <z\lesssim \maxclusterz{}$. The eRASS1 (X-ray, yellow) subsample has 167 clusters, a median redshift of $z=0.32$, and spans the range $0.04<z<0.99$ \citep{kluge24}. The 29 ACT DR5 clusters have median redshift $z=0.58$ and their redshifts span the range $0.19<z<1.38$ \citep[purple,][]{hilton21}. We also compared to the SPT-SZ subsample of 21 clusters \citep[blue,][]{bleem15b, klein24a}. This subsample has a median redshift of $z=0.58$ and spans the range $0.24<z<1.38$. The SPT-3G EDF-S cluster sample significantly increases the number of ICM-selected clusters in this patch, especially at high redshift ($z>1$).}
                \label{fig:clusterredshifts}
        \end{figure}

We compared the cluster sample to subsamples of other ICM-selected galaxy cluster catalogs. In general, the SPT-3G EDF-S cluster sample spans a wider redshift range and significantly increases the number of confirmed clusters when compared to other samples over the same sky area. We show the redshift distribution of multiple samples overlapping the SPT-3G EDF-S footprint in Fig. \ref{fig:clusterredshifts}. We also cross-matched our catalog with these external catalogs with a 1.5 arcmin matching radius. We calculated the median angular separation and mass ratios for the matches, with uncertainties on the medians estimated by bootstrapping the respective data vectors for angular separation and mass ratio. More details are provided on the cross-comparisons between SPT and other surveys in this section. 

\paragraph{eRASS1.} The SRG/eROSITA All-Sky Survey (eRASS1) cataloged $12\,000$ galaxy clusters over $13\,116$ square degrees of the Western Galactic hemisphere by identifying hot overdensities in the ICM from X-ray emission over the energy range 0.2$-$10 keV, with the majority of the sensitivity to hot gas in the ICM coming from the ``soft X-ray'' energy range \citep[0.2$-$2.3 keV,][]{bulbul24}. In the SPT-3G EDF-S patch of sky, the eRASS1 catalog contains the largest subsample of galaxy clusters with 167 X-ray-selected candidates, which have a median redshift of $z=0.32$ and span the redshift range $0.04<z<0.99$. The eRASS1 subsample has a median mass of \mass$=1.3 \times 10^{14}$\msun \ with a minimum of \mass$=0.083 \times 10^{14}$\msun, which is over an order of magnitude lower than the SPT-3G minimum mass. 

The eRASS1 catalog is complementary to our work as it has comparable cluster counts, but has a lower median redshift. Of the 167 eRASS1 clusters, we cross-match 61 to the SPT-3G EDF-S cluster catalog. The median angular separation between the SPT-3G (millimeter) and eRASS1 (X-ray) is 0.31$\pm$0.11 arcmin and the samples have a median mass ratio, $M_{\mathrm{eRASS1}}/M_{\mathrm{SPT}}$, of 1.20$\pm$0.10. We recover 73\% of the eRASS1 clusters that are located within the SPT-3G EDF-S patch with masses greater than $2\times 10^{14}$\msun. We recover 90\% of the clusters with masses greater than $3\times 10^{14}$\msun. Three of the 12 unmatched X-ray clusters with masses greater than $2\times 10^{14}$\msun \ fall inside the bright point source mask holes described in Sect. \ref{sec:brightsrcs}. We do not expect to detect the majority of X-ray clusters with masses less than $2\times 10^{14}$\msun \ and at low redshift ($z<0.25$) due to our atmospheric filtering suppressing an already weak tSZ signal (see Fig. \ref{fig:completeness_purity}, left panel).

\paragraph{ACT DR5.} The fifth data release of the Atacama Cosmology Telescope \citep[ACT DR5,][]{hilton21} included a tSZ-selected galaxy cluster catalog using the 98 and 150 GHz channels over $13\,211$ square degrees of sky. From this sample, $4\,200$ clusters at significance $>4$ were optically confirmed. We find a subsample of 29 ACT clusters in the SPT-3G EDF-S footprint with a median redshift of $z=0.58$ that span the redshift range $0.19<z<1.38$. The subsample has a median mass of \mass$=3.05\times 10^{14}$\msun \ with a minimum of \mass$=2.39\times 10^{14}$\msun. 

We match 25 ACT DR5 clusters in the SPT-3G EDF-S field. Of the remaining four unmatched clusters, two line up with bright source mask holes. The median angular separation between the matched clusters from SPT-3G and ACT is 0.31$\pm$0.14 arcmin, which is similar to the matching to eRASS1. The median mass ratio is 0.99 $\pm$ 0.14, showing the mass estimation between these two tSZ-selected catalogs is in agreement.

\paragraph{SPT-SZ 2500d.} The SPT-SZ cluster catalog over $2\,500$ square degrees contains 811 tSZ-selected clusters with $\xi>4.0$ \citep{bleem15b, klein24a}. We find 21 of these clusters in the SPT-3G EDF-S footprint with a median redshift of $z=0.62$ and spanning the redshift range $0.2<z<1.6$. The subsample has a median mass of \mass$=3.5\times 10^{14}$\msun \ and a minimum mass of \mass$=2.8\times 10^{14}$\msun. We match 19 SPT-SZ clusters in the SPT-3G EDF-S sample with a median angular separation of 0.24 $\pm$ 0.07 arcmin and a median mass ratio of 1.0 $\pm$ 0.05 (by construction; see Sect. \ref{sec:masses}), showing these catalogs are also in agreement as expected.

\paragraph{}
This work shows an increase in cluster counts of nearly an order of magnitude over previous tSZ-selected cluster catalogs in the same region. Our catalog also offers a complementary higher median redshift sample to eRASS1. \numclusterswithhighredshifts{} of the SPT-3G EDF-S clusters are found at high redshift ($z>1)$, which is more than 10 and 30 times more high-redshift clusters than SPT-SZ and ACT DR5, respectively. Our catalog maintains the tSZ-selected cluster paradigm of a sample with high median redshift ($z$=\medclusterz{}), which will be matched or improved upon by future SPT-3G catalogs.

\subsection{Comparisons to \textit{Euclid} Q1 data}
The photometric redshifts we report in this work were derived from a combination of DES and WISE data. As a consistency check, we also used an early adaptation of MCMF that uses \textit{Euclid} Q1 data \citep{klein25}. It uses the \textit{Euclid} Q1 photometric dataset \citep{aussel25} that incorporates ground-based DES photometry alongside the \textit{Euclid} data. Despite the fact that this is a preliminary version lacking calibration data due to the small Q1 field size, we find good photometric redshift agreement with a median $\Delta_z/(1+z) = 0.0002$ and a median absolute deviation of 0.011 based on 107 clusters. As outlined in \citet{klein25}, the Q1 dataset suffers from some photometric issues for bright cluster galaxies, impacting the performance of the richness measurements at low and intermediate redshifts. We therefore leave the incorporation of \textit{Euclid}-based redshift and richness measurements to future studies using the larger \textit{Euclid}-DR1 dataset.

As a second point of comparison, \citet{bhargava25} presented an optical and infrared-selected cluster sample of 426 galaxy clusters detected at high signal-to-noise using both the \texttt{AMICO} \citep{maturi18} and \texttt{PZWav} \citep{thongkham24} algorithms in the full 63-square-degree Q1 data. The primary sample spanned redshift $0.2\leq z \leq 1.5$ and an additional 15 systems at $z>1.5$ were also identified. Of this 426 member sample, 35 of the highest signal-to-noise detections in the Q1 fields were publicly released as of this publication.

We first compared the SPT-3G EDF-S cluster candidate list to the 18 clusters released so far in the EDF-S Q1 field at $z < 1.5$; this \textit{Euclid} subsample has a median redshift of $z = 0.56$. For these 18 systems, we find matches for 13 clusters within a 1.5 arcmin matching radius. The median spatial separation is 0.39 arcmin and maximum 0.75 arcmin. The median difference in redshift (adopting the \texttt{AMICO} redshifts for reference) is 0.018/(1+$z$) with a maximum absolute difference of 0.04/(1+$z$), showing excellent agreement between our DES and WISE estimated redshifts and the \textit{Euclid} derived redshifts.

Investigating the five systems without a match in the SPT catalog, we do not find any with tSZ signal in the SPT maps higher than 2.6$\sigma$ at a 0.25 arcmin spatial filter scale. We also do not find any of these unmatched clusters aligned with bright source mask holes. \citet{bhargava25} report cluster richnesses which gives an indication of cluster mass; of the five unmatched clusters, four of them have low richnesses. Therefore, not detecting them in SPT is consistent with our expectations of completeness at low mass (see the left panel of Fig. \ref{fig:completeness_purity}).

Comparing the five $z>1.5$ clusters released in EDF-S, we find a spatial match between our full candidate list with one system (EUCL-Q1-CL J034533.23$-$500806.7) at 0.3 arcmin separation, but at highly discrepant redshifts ($z_\textrm{AMICO}=1.78$ versus $z_\textrm{SPT}=1.18$; however, we note the MCMF run on \textit{Euclid} data does identify a counterpart at $z=1.7$, so the DES and WISE data might be too shallow in this case). While this is just a single cluster, exploration of such discrepancies on larger joint samples using upcoming data from the respective experiments' wide field surveys will be valuable for understanding the construction of well-characterized cluster samples. 

The analysis of \textit{Euclid} data will be informed by tSZ cluster catalogs sampled over a large redshift range, while SPT-3G cluster science will benefit from deriving redshift information from \textit{Euclid's} state-of-the-art photometric data. The collaboration between SPT-3G and \textit{Euclid} will produce galaxy cluster samples with significant power for constraining cosmology and has implications for studying the formation history of clusters through targeted follow-up observations. These advancements in analysis of galaxy clusters would not be possible with either SPT-3G or \textit{Euclid} data alone.

\section{Conclusion} \label{sec:conclusion}
In this work, we have presented millimeter-wave maps and source catalogs from SPT-3G observations of the \textit{Euclid} Deep Field South. Though the EDF-S is 30 square degrees, we included all the data from our \pixelmaskarea{}-square-degree observations. The data correspond to 20 days of on-source observation time with SPT-3G, where we reached map noise depths of \mapdepthnine{}, \mapdepthonefive{}, and \mapdepthtwotwo{} $\mu$K-arcmin at 95, 150, and 220~GHz, respectively. We processed the data to optimize source detection by retaining the small-angular-scale features of the maps. We used two distinct but related, pipelines to generate an emissive source catalog and a galaxy cluster catalog, and we showed how millimeter selection of the various source populations, especially DSFGs and galaxy clusters, has unique benefits over selections at other wavelengths.

The emissive source catalog comprises \numsrcs{} objects in total (\numedfs{} inside EDF-S). Of the emissive sources, \fracsynch{}\% are synchrotron-dominated AGNs, the majority of which have radio counterparts. The other \fracdusty{}\% are dust-dominated galaxies, where approximately one in three lacks a counterpart in external datasets and has spectral behavior characteristic of high-redshift DSFGs. The population of high-redshift DSFGs is of current interest in studies of galaxy evolution, as their observed physical properties are not fully understood with current models of mass assembly in the early Universe. 

The galaxy cluster catalog includes \numclusterstot{} total clusters (\numclustersedfs{} inside EDF-S), with an expected purity of \clusterpurity. We used previously demonstrated methods to obtain redshifts for \numclusterswithredshifts{} clusters. The median redshift of the sample is \medclusterz{}, with \numclustersaboveone{} clusters above $z>1$. The purity of the optically confirmed subsample rises to 97\%. The masses for the EDF-S catalog span \minmass \ to \maxmass, with a median of \medmass. The sample presented in this work increases the number of tSZ-selected clusters in the SPT-3G EDF-S patch by nearly an order of magnitude, with a high median redshift that is complementary to X-ray-selected clusters, such as those detected by eROSITA.

The SPT-3G data and catalogs described in this work, when combined with the deep \textit{Euclid} optical and infrared data in EDF-S, are designed to enable joint projects on a range of topics in astrophysics and cosmology. These topics include studies of strongly lensed galaxies and protoclusters, the physical properties of the most luminous unlensed DSFGs, redshift and weak lensing mass estimations of galaxy clusters, and cluster galaxy populations. Looking further ahead, the SPT-3G collaboration intends to perform cosmological studies on CMB lensing maps constructed from the EDF-S observations discussed in this work as well as cross-correlation studies of galaxy counts with CMB lensing and galaxy-shape derived weak lensing with the tSZ effect. The observations and data products presented here provide a framework for future collaborative efforts of multiwavelength astrophysical and cosmological analyses from the \textit{Euclid} and SPT-3G wide field surveys.

\section*{Data availability} \label{sec:dataavailability}
Data products described in Sect. \ref{sec:resultsdataprod}, including catalogs, sky maps, and ancillary products, are available for download at \url{https://pole.uchicago.edu/public/data/edfs25/}, the NASA LAMBDA archive at \url{https://lambda.gsfc.nasa.gov/product/spt/spt3g_euclid_deep_field_south_info.html}, and the CDS via anonymous ftp to cdsarc.u-strasbg.fr (130.79.128.5) or via \url{http://cdsweb.u-strasbg.fr/cgi-bin/qcat?J/A+A/}.

\begin{acknowledgements}
The South Pole Telescope program is supported by the National Science Foundation (NSF) through awards OPP-1852617 and OPP-2332483. Partial support is also provided by the Kavli Institute of Cosmological Physics at the University of Chicago. 

Argonne National Laboratory’s work was supported by the U.S. Department of Energy, Office of High Energy Physics, under contract DE-AC02-06CH11357. 

The UC Davis group acknowledges support from Michael and Ester Vaida. 

Work at the Fermi National Accelerator Laboratory (Fermilab), a U.S. Department of Energy, Office of Science, Office of High Energy Physics HEP User Facility, is managed by Fermi Forward Discovery Group, LLC, acting under Contract No. 89243024CSC000002.

The Melbourne authors acknowledge support from the Australian Research Council’s Discovery Project scheme (No. DP210102386). 

The Paris group has received funding from the European Research Council (ERC) under the European Union’s Horizon 2020 research and innovation program (grant agreement No 101001897), and funding from the Centre National d’Etudes Spatiales. 

The SLAC group is supported in part by the Department of Energy at SLAC National Accelerator Laboratory, under contract DE-AC02-76SF00515.

M. Aravena gratefully acknowledges support from ANID Basal Project FB210003 and ANID MILENIO NCN2024\_112.

RPD acknowledges funding from the South African Radio Astronomy Observatory (SARAO), which is a facility of the National Research Foundation (NRF), an agency of the Department of Science, Technology, and Innovation (DSI).

The Cosmic Dawn Center (DAWN) is funded by the Danish National Research Foundation under grant No. 140. TRG is grateful for support from the Carlsberg Foundation via grant No. CF20-0534.

This research was done using services provided by the OSG Consortium \citep{osg07, osg09, osg06, osg15}, which is supported by the National Science Foundation awards \#2030508 and \#2323298.

Some results in this paper have been derived using the \texttt{healpy} and \texttt{HEALPix} packages.

This publication makes use of data products from the Wide-field Infrared Survey Explorer, which is a joint project of the University of California, Los Angeles, and the Jet Propulsion Laboratory/California Institute of Technology, and NEOWISE, which is a project of the Jet Propulsion Laboratory/California Institute of Technology. WISE and NEOWISE are funded by the National Aeronautics and Space Administration.

This scientific work uses data obtained from Inyarrimanha Ilgari Bundara, the CSIRO Murchison Radio-astronomy Observatory. We acknowledge the Wajarri Yamaji People as the Traditional Owners and native titleholders of the Observatory site. CSIRO’s ASKAP radio telescope is part of the Australia Telescope National Facility (https://ror.org/05qajvd42). Operation of ASKAP is funded by the Australian Government with support from the National Collaborative Research Infrastructure Strategy. ASKAP uses the resources of the Pawsey Supercomputing Research Centre. Establishment of ASKAP, Inyarrimanha Ilgari Bundara, the CSIRO Murchison Radio-astronomy Observatory and the Pawsey Supercomputing Research Centre are initiatives of the Australian Government, with support from the Government of Western Australia and the Science and Industry Endowment Fund. This work includes archived data obtained through the CSIRO ASKAP Science Data Archive, CASDA (http://data.csiro.au).

The Legacy Surveys consist of three individual and complementary projects: the Dark Energy Camera Legacy Survey (DECaLS; Proposal ID \#2014B-0404; PIs: David Schlegel and Arjun Dey), the Beijing-Arizona Sky Survey (BASS; NOAO Prop. ID \#2015A-0801; PIs: Zhou Xu and Xiaohui Fan), and the Mayall z-band Legacy Survey (MzLS; Prop. ID \#2016A-0453; PI: Arjun Dey). DECaLS, BASS and MzLS together include data obtained, respectively, at the Blanco telescope, Cerro Tololo Inter-American Observatory, NSF’s NOIRLab; the Bok telescope, Steward Observatory, University of Arizona; and the Mayall telescope, Kitt Peak National Observatory, NOIRLab. Pipeline processing and analyses of the data were supported by NOIRLab and the Lawrence Berkeley National Laboratory (LBNL). The Legacy Surveys project is honored to be permitted to conduct astronomical research on Iolkam Du’ag (Kitt Peak), a mountain with particular significance to the Tohono O’odham Nation.

NOIRLab is operated by the Association of Universities for Research in Astronomy (AURA) under a cooperative agreement with the National Science Foundation. LBNL is managed by the Regents of the University of California under contract to the U.S. Department of Energy.

This project used data obtained with the Dark Energy Camera (DECam), which was constructed by the Dark Energy Survey (DES) collaboration. Funding for the DES Projects has been provided by the U.S. Department of Energy, the U.S. National Science Foundation, the Ministry of Science and Education of Spain, the Science and Technology Facilities Council of the United Kingdom, the Higher Education Funding Council for England, the National Center for Supercomputing Applications at the University of Illinois, Urbana-Champaign, the Kavli Institute of Cosmological Physics at the University of Chicago, Center for Cosmology and Astro-Particle Physics at the Ohio State University, the Mitchell Institute for Fundamental Physics and Astronomy at Texas A\&M University, Financiadora de Estudos e Projetos, Fundacao Carlos Chagas Filho de Amparo, Financiadora de Estudos e Projetos, Fundacao Carlos Chagas Filho de Amparo a Pesquisa do Estado do Rio de Janeiro, Conselho Nacional de Desenvolvimento Cientifico e Tecnologico and the Ministerio da Ciencia, Tecnologia e Inovacao, the Deutsche Forschungsgemeinschaft and the Collaborating Institutions in the Dark Energy Survey. The Collaborating Institutions are Argonne National Laboratory, the University of California at Santa Cruz, the University of Cambridge, Centro de Investigaciones Energeticas, Medioambientales y Tecnologicas-Madrid, the University of Chicago, University College London, the DES-Brazil Consortium, the University of Edinburgh, the Eidgenossische Technische Hochschule (ETH) Zurich, Fermi National Accelerator Laboratory, the University of Illinois, Urbana-Champaign, the Institut de Ciencies de l’Espai (IEEC/CSIC), the Institut de Fisica d’Altes Energies, Lawrence Berkeley National Laboratory, the Ludwig Maximilians Universitat Munchen and the associated Excellence Cluster Universe, the University of Michigan, NSF’s NOIRLab, the University of Nottingham, the Ohio State University, the University of Pennsylvania, the University of Portsmouth, SLAC National Accelerator Laboratory, Stanford University, the University of Sussex, and Texas A\&M University.

The Legacy Surveys imaging of the DESI footprint is supported by the Director, Office of Science, Office of High Energy Physics of the U.S. Department of Energy under Contract No. DE-AC02-05CH1123, by the National Energy Research Scientific Computing Center, a DOE Office of Science User Facility under the same contract; and by the U.S. National Science Foundation, Division of Astronomical Sciences under Contract No. AST-0950945 to NOAO.
\end{acknowledgements}

\bibliography{spt}

@ARTICLE{alberts22,
       author = {{Alberts}, Stacey and {Noble}, Allison},
        title = "{From Clusters to Proto-Clusters: The Infrared Perspective on Environmental Galaxy Evolution}",
      journal = {Universe},
         year = 2022,
        month = oct,
       volume = {8},
       number = {11},
          eid = {554},
        pages = {554},
          doi = {10.3390/universe8110554},
archivePrefix = {arXiv},
       eprint = {2209.02781},
 primaryClass = {astro-ph.GA},
       adsurl = {https://ui.adsabs.harvard.edu/abs/2022Univ....8..554A}
}

@ARTICLE{allen11,
   author = {{Allen}, S.~W. and {Evrard}, A.~E. and {Mantz}, A.~B.},
    title = "{Cosmological Parameters from Observations of Galaxy Clusters}",
  journal = {\araa},
archivePrefix = "arXiv",
   eprint = {1103.4829},
 primaryClass = "astro-ph.CO",
     year = 2011,
    month = sep,
   volume = 49,
    pages = {409-470},
      doi = {10.1146/annurev-astro-081710-102514},
   adsurl = {http://adsabs.harvard.edu/abs/2011ARA%26A..49..409A}
}

@ARTICLE{aussel25,
       author = {{Euclid Collaboration} and {Aussel}, H. and {Tereno}, I. and {Schirmer}, M. and {Alguero}, G. and {Altieri}, B. and {Balbinot}, E. and {de Boer}, T. and {Casenove}, P. and {Corcho-Caballero}, P. and {Furusawa}, H. and {Furusawa}, J. and {Hudson}, M.~J. and {Jahnke}, K. and {Libet}, G. and {Macias-Perez}, J. and {Masoumzadeh}, N. and {Mohr}, J.~J. and {Odier}, J. and {Scott}, D. and {Vassallo}, T. and {Verdoes Kleijn}, G. and {Zacchei}, A. and {Aghanim}, N. and {Amara}, A. and {Andreon}, S. and {Auricchio}, N. and {Awan}, S. and {Azzollini}, R. and {Baccigalupi}, C. and {Baldi}, M. and {Balestra}, A. and {Bardelli}, S. and {Basset}, A. and {Battaglia}, P. and {Belikov}, A.~N. and {Bender}, R. and {Biviano}, A. and {Bonchi}, A. and {Bonino}, D. and {Branchini}, E. and {Brescia}, M. and {Brinchmann}, J. and {Camera}, S. and {Ca{\~n}as-Herrera}, G. and {Capobianco}, V. and {Carbone}, C. and {Cardone}, V.~F. and {Carretero}, J. and {Casas}, S. and {Castander}, F.~J. and {Castellano}, M. and {Castignani}, G. and {Cavuoti}, S. and {Chambers}, K.~C. and {Cimatti}, A. and {Colodro-Conde}, C. and {Congedo}, G. and {Conselice}, C.~J. and {Conversi}, L. and {Copin}, Y. and {Courbin}, F. and {Courtois}, H.~M. and {Cropper}, M. and {Cuby}, J. -G. and {Da Silva}, A. and {da Silva}, R. and {Degaudenzi}, H. and {de Jong}, J.~T.~A. and {De Lucia}, G. and {Di Giorgio}, A.~M. and {Dinis}, J. and {Dolding}, C. and {Dole}, H. and {Douspis}, M. and {Dubath}, F. and {Duncan}, C.~A.~J. and {Dupac}, X. and {Dusini}, S. and {Ealet}, A. and {Escoffier}, S. and {Fabricius}, M. and {Farina}, M. and {Farinelli}, R. and {Faustini}, F. and {Ferriol}, S. and {Fotopoulou}, S. and {Fourmanoit}, N. and {Frailis}, M. and {Franceschi}, E. and {Franzetti}, P. and {Galeotta}, S. and {George}, K. and {Gillard}, W. and {Gillis}, B. and {Giocoli}, C. and {G{\'o}mez-Alvarez}, P. and {Gracia-Carpio}, J. and {Granett}, B.~R. and {Grazian}, A. and {Grupp}, F. and {Guzzo}, L. and {Gwyn}, S. and {Haugan}, S.~V.~H. and {Herent}, O. and {Hoar}, J. and {Hoekstra}, H. and {Holliman}, M.~S. and {Holmes}, W. and {Hook}, I.~M. and {Hormuth}, F. and {Hornstrup}, A. and {Hudelot}, P. and {Ili{\'c}}, S. and {Jhabvala}, M. and {Joachimi}, B. and {Keih{\"a}nen}, E. and {Kermiche}, S. and {Kiessling}, A. and {Kubik}, B. and {Kuijken}, K. and {K{\"u}mmel}, M. and {Kunz}, M. and {Kurki-Suonio}, H. and {Lahav}, O. and {Le Boulc'h}, Q. and {Le Brun}, A.~M.~C. and {Le Mignant}, D. and {Liebing}, P. and {Ligori}, S. and {Lilje}, P.~B. and {Lindholm}, V. and {Lloro}, I. and {Mainetti}, G. and {Maino}, D. and {Maiorano}, E. and {Mansutti}, O. and {Marcin}, S. and {Marggraf}, O. and {Markovic}, K. and {Martinelli}, M. and {Martinet}, N. and {Marulli}, F. and {Massey}, R. and {Maurogordato}, S. and {McCracken}, H.~J. and {Medinaceli}, E. and {Mei}, S. and {Melchior}, M. and {Mellier}, Y. and {Meneghetti}, M. and {Merlin}, E. and {Meylan}, G. and {Mora}, A. and {Moresco}, M. and {Morris}, P.~W. and {Moscardini}, L. and {Mourre}, S. and {Nakajima}, R. and {Neissner}, C. and {Nichol}, R.~C. and {Niemi}, S. -M. and {Nightingale}, J.~W. and {Nutma}, T. and {Padilla}, C. and {Paltani}, S. and {Pasian}, F. and {Peacock}, J.~A. and {Pedersen}, K. and {Percival}, W.~J. and {Pettorino}, V. and {Pires}, S. and {Polenta}, G. and {Pollack}, J.~E. and {Poncet}, M. and {Popa}, L.~A. and {Pozzetti}, L. and {Racca}, G.~D. and {Raison}, F. and {Rebolo}, R. and {Renzi}, A. and {Rhodes}, J. and {Riccio}, G. and {Rix}, H. -W. and {Romelli}, E. and {Roncarelli}, M. and {Rossetti}, E. and {Rusholme}, B. and {Saglia}, R. and {Sakr}, Z. and {S{\'a}nchez}, A.~G. and {Sapone}, D. and {Sartoris}, B. and {Sauvage}, M. and {Schewtschenko}, J.~A. and {Schneider}, P. and {Scodeggio}, M. and {Secroun}, A. and {Sefusatti}, E. and {Seidel}, G.},
        title = "{Euclid Quick Data Release (Q1) -- Data release overview}",
      journal = {\aap, submitted},
         year = 2025,
        month = mar,
          eid = {arXiv:2503.15302},
        pages = {arXiv:2503.15302},
          doi = {10.48550/arXiv.2503.15302},
archivePrefix = {arXiv},
       eprint = {2503.15302},
 primaryClass = {astro-ph.GA},
       adsurl = {https://ui.adsabs.harvard.edu/abs/2025arXiv250315302E}
}

@ARTICLE{balkenhol21,
       author = {{Balkenhol}, L. and {Dutcher}, D. and {Ade}, P.~A.~R. and {Ahmed}, Z. and {Anderes}, E. and {Anderson}, A.~J. and {Archipley}, M. and {Avva}, J.~S. and {Aylor}, K. and {Barry}, P.~S. and {Basu Thakur}, R. and {Benabed}, K. and {Bender}, A.~N. and {Benson}, B.~A. and {Bianchini}, F. and {Bleem}, L.~E. and {Bouchet}, F.~R. and {Bryant}, L. and {Byrum}, K. and {Carlstrom}, J.~E. and {Carter}, F.~W. and {Cecil}, T.~W. and {Chang}, C.~L. and {Chaubal}, P. and {Chen}, G. and {Cho}, H. -M. and {Chou}, T. -L. and {Cliche}, J. -F. and {Crawford}, T.~M. and {Cukierman}, A. and {Daley}, C. and {de Haan}, T. and {Denison}, E.~V. and {Dibert}, K. and {Ding}, J. and {Dobbs}, M.~A. and {Everett}, W. and {Feng}, C. and {Ferguson}, K.~R. and {Foster}, A. and {Fu}, J. and {Galli}, S. and {Gambrel}, A.~E. and {Gardner}, R.~W. and {Goeckner-Wald}, N. and {Gualtieri}, R. and {Guns}, S. and {Gupta}, N. and {Guyser}, R. and {Halverson}, N.~W. and {Harke-Hosemann}, A.~H. and {Harrington}, N.~L. and {Henning}, J.~W. and {Hilton}, G.~C. and {Hivon}, E. and {Holder}, G.~P. and {Holzapfel}, W.~L. and {Hood}, J.~C. and {Howe}, D. and {Huang}, N. and {Irwin}, K.~D. and {Jeong}, O.~B. and {Jonas}, M. and {Jones}, A. and {Khaire}, T.~S. and {Knox}, L. and {Kofman}, A.~M. and {Korman}, M. and {Kubik}, D.~L. and {Kuhlmann}, S. and {Kuo}, C. -L. and {Lee}, A.~T. and {Leitch}, E.~M. and {Lowitz}, A.~E. and {Lu}, C. and {Meyer}, S.~S. and {Michalik}, D. and {Millea}, M. and {Montgomery}, J. and {Nadolski}, A. and {Natoli}, T. and {Nguyen}, H. and {Noble}, G.~I. and {Novosad}, V. and {Omori}, Y. and {Padin}, S. and {Pan}, Z. and {Paschos}, P. and {Pearson}, J. and {Posada}, C.~M. and {Prabhu}, K. and {Quan}, W. and {Rahlin}, A. and {Reichardt}, C.~L. and {Riebel}, D. and {Riedel}, B. and {Rouble}, M. and {Ruhl}, J.~E. and {Sayre}, J.~T. and {Schiappucci}, E. and {Shirokoff}, E. and {Smecher}, G. and {Sobrin}, J.~A. and {Stark}, A.~A. and {Stephen}, J. and {Story}, K.~T. and {Suzuki}, A. and {Thompson}, K.~L. and {Thorne}, B. and {Tucker}, C. and {Umilta}, C. and {Vale}, L.~R. and {Vanderlinde}, K. and {Vieira}, J.~D. and {Wang}, G. and {Whitehorn}, N. and {Wu}, W.~L.~K. and {Yefremenko}, V. and {Yoon}, K.~W. and {Young}, M.~R. and {SPT-3G Collaboration}},
        title = "{Constraints on {\ensuremath{\Lambda}} CDM extensions from the SPT-3G 2018 E E and T E power spectra}",
      journal = {\prd},
         year = 2021,
        month = oct,
       volume = {104},
       number = {8},
          eid = {083509},
        pages = {083509},
          doi = {10.1103/PhysRevD.104.083509},
archivePrefix = {arXiv},
       eprint = {2103.13618},
 primaryClass = {astro-ph.CO},
       adsurl = {https://ui.adsabs.harvard.edu/abs/2021PhRvD.104h3509B}
}

@ARTICLE{balkenhol23,
       author = {{Balkenhol}, L. and {Dutcher}, D. and {Spurio Mancini}, A. and {Doussot}, A. and {Benabed}, K. and {Galli}, S. and {Ade}, P.~A.~R. and {Anderson}, A.~J. and {Ansarinejad}, B. and {Archipley}, M. and {Bender}, A.~N. and {Benson}, B.~A. and {Bianchini}, F. and {Bleem}, L.~E. and {Bouchet}, F.~R. and {Bryant}, L. and {Camphuis}, E. and {Carlstrom}, J.~E. and {Cecil}, T.~W. and {Chang}, C.~L. and {Chaubal}, P. and {Chichura}, P.~M. and {Chou}, T. -L. and {Coerver}, A. and {Crawford}, T.~M. and {Cukierman}, A. and {Daley}, C. and {de Haan}, T. and {Dibert}, K.~R. and {Dobbs}, M.~A. and {Everett}, W. and {Feng}, C. and {Ferguson}, K.~R. and {Foster}, A. and {Gambrel}, A.~E. and {Gardner}, R.~W. and {Goeckner-Wald}, N. and {Gualtieri}, R. and {Guidi}, F. and {Guns}, S. and {Halverson}, N.~W. and {Hivon}, E. and {Holder}, G.~P. and {Holzapfel}, W.~L. and {Hood}, J.~C. and {Huang}, N. and {Knox}, L. and {Korman}, M. and {Kuo}, C. -L. and {Lee}, A.~T. and {Lowitz}, A.~E. and {Lu}, C. and {Millea}, M. and {Montgomery}, J. and {Nakato}, Y. and {Natoli}, T. and {Noble}, G.~I. and {Novosad}, V. and {Omori}, Y. and {Padin}, S. and {Pan}, Z. and {Paschos}, P. and {Prabhu}, K. and {Quan}, W. and {Rahimi}, M. and {Rahlin}, A. and {Reichardt}, C.~L. and {Rouble}, M. and {Ruhl}, J.~E. and {Schiappucci}, E. and {Smecher}, G. and {Sobrin}, J.~A. and {Stark}, A.~A. and {Stephen}, J. and {Suzuki}, A. and {Tandoi}, C. and {Thompson}, K.~L. and {Thorne}, B. and {Tucker}, C. and {Umilta}, C. and {Vieira}, J.~D. and {Wang}, G. and {Whitehorn}, N. and {Wu}, W.~L.~K. and {Yefremenko}, V. and {Young}, M.~R. and {Zebrowski}, J.~A. and {SPT-3G Collaboration}},
        title = "{Measurement of the CMB temperature power spectrum and constraints on cosmology from the SPT-3G 2018 T T , T E , and E E dataset}",
      journal = {\prd},
         year = 2023,
        month = jul,
       volume = {108},
       number = {2},
          eid = {023510},
        pages = {023510},
          doi = {10.1103/PhysRevD.108.023510},
archivePrefix = {arXiv},
       eprint = {2212.05642},
 primaryClass = {astro-ph.CO},
       adsurl = {https://ui.adsabs.harvard.edu/abs/2023PhRvD.108b3510B}
}

@ARTICLE{bayliss16,
   author = {{Bayliss}, M.~B. and {Ruel}, J. and {Stubbs}, C.~W. and {Allen}, S.~W. and 
	{Applegate}, D.~E. and {Ashby}, M.~L.~N. and {Bautz}, M. and 
	{Benson}, B.~A. and {Bleem}, L.~E. and {Bocquet}, S. and {Brodwin}, M. and 
	{Capasso}, R. and {Carlstrom}, J.~E. and {Chang}, C.~L. and 
	{Chiu}, I. and {Cho}, H.-M. and {Clocchiatti}, A. and {Crawford}, T.~M. and 
	{Crites}, A.~T. and {de Haan}, T. and {Desai}, S. and {Dietrich}, J.~P. and 
	{Dobbs}, M.~A. and {Doucouliagos}, A.~N. and {Foley}, R.~J. and 
	{Forman}, W.~R. and {Garmire}, G.~P. and {George}, E.~M. and 
	{Gladders}, M.~D. and {Gonzalez}, A.~H. and {Gupta}, N. and 
	{Halverson}, N.~W. and {Hlavacek-Larrondo}, J. and {Hoekstra}, H. and 
	{Holder}, G.~P. and {Holzapfel}, W.~L. and {Hou}, Z. and {Hrubes}, J.~D. and 
	{Huang}, N. and {Jones}, C. and {Keisler}, R. and {Knox}, L. and 
	{Lee}, A.~T. and {Leitch}, E.~M. and {von der Linden}, A. and 
	{Luong-Van}, D. and {Mantz}, A. and {Marrone}, D.~P. and {McDonald}, M. and 
	{McMahon}, J.~J. and {Meyer}, S.~S. and {Mocanu}, L.~M. and 
	{Mohr}, J.~J. and {Murray}, S.~S. and {Padin}, S. and {Pryke}, C. and 
	{Rapetti}, D. and {Reichardt}, C.~L. and {Rest}, A. and {Ruhl}, J.~E. and 
	{Saliwanchik}, B.~R. and {Saro}, A. and {Sayre}, J.~T. and {Schaffer}, K.~K. and 
	{Schrabback}, T. and {Shirokoff}, E. and {Song}, J. and {Spieler}, H.~G. and 
	{Stalder}, B. and {Stanford}, S.~A. and {Staniszewski}, Z. and 
	{Stark}, A.~A. and {Story}, K.~T. and {Vanderlinde}, K. and 
	{Vieira}, J.~D. and {Vikhlinin}, A. and {Williamson}, R. and 
	{Zenteno}, A.},
    title = "{SPT-GMOS: A Gemini/GMOS-South Spectroscopic Survey of Galaxy Clusters in the SPT-SZ Survey}",
  journal = {\apjs},
archivePrefix = "arXiv",
   eprint = {1609.05211},
     year = 2016,
    month = nov,
   volume = 227,
      eid = {3},
    pages = {3},
      doi = {10.3847/0067-0049/227/1/3},
   adsurl = {http://adsabs.harvard.edu/abs/2016ApJS..227....3B}
}

@ARTICLE{benson13,
   author = {{Benson}, B.~A. and {de Haan}, T. and {Dudley}, J.~P. and {Reichardt}, C.~L. and 
	{Aird}, K.~A. and {Andersson}, K. and {Armstrong}, R. and {Ashby}, M.~L.~N. and 
	{Bautz}, M. and {Bayliss}, M. and {Bazin}, G. and {Bleem}, L.~E. and 
	{Brodwin}, M. and {Carlstrom}, J.~E. and {Chang}, C.~L. and 
	{Cho}, H.~M. and {Clocchiatti}, A. and {Crawford}, T.~M. and 
	{Crites}, A.~T. and {Desai}, S. and {Dobbs}, M.~A. and {Foley}, R.~J. and 
	{Forman}, W.~R. and {George}, E.~M. and {Gladders}, M.~D. and 
	{Gonzalez}, A.~H. and {Halverson}, N.~W. and {Harrington}, N. and 
	{High}, F.~W. and {Holder}, G.~P. and {Holzapfel}, W.~L. and 
	{Hoover}, S. and {Hrubes}, J.~D. and {Jones}, C. and {Joy}, M. and 
	{Keisler}, R. and {Knox}, L. and {Lee}, A.~T. and {Leitch}, E.~M. and 
	{Liu}, J. and {Lueker}, M. and {Luong-Van}, D. and {Mantz}, A. and 
	{Marrone}, D.~P. and {McDonald}, M. and {McMahon}, J.~J. and 
	{Mehl}, J. and {Meyer}, S.~S. and {Mocanu}, L. and {Mohr}, J.~J. and 
	{Montroy}, T.~E. and {Murray}, S.~S. and {Natoli}, T. and {Padin}, S. and 
	{Plagge}, T. and {Pryke}, C. and {Rest}, A. and {Ruel}, J. and 
	{Ruhl}, J.~E. and {Saliwanchik}, B.~R. and {Saro}, A. and {Sayre}, J.~T. and 
	{Schaffer}, K.~K. and {Shaw}, L. and {Shirokoff}, E. and {Song}, J. and 
	{Spieler}, H.~G. and {Stalder}, B. and {Staniszewski}, Z. and 
	{Stark}, A.~A. and {Story}, K. and {Stubbs}, C.~W. and {Suhada}, R. and 
	{van Engelen}, A. and {Vanderlinde}, K. and {Vieira}, J.~D. and 
	{Vikhlinin}, A. and {Williamson}, R. and {Zahn}, O. and {Zenteno}, A.
	},
    title = "{Cosmological Constraints from Sunyaev-Zel'dovich-selected Clusters with X-Ray Observations in the First 178 deg$^{2}$ of the South Pole Telescope Survey}",
  journal = {\apj},
 archivePrefix = "arXiv",
   eprint = {1112.5435},
 primaryClass = "astro-ph.CO",
     year = 2013,
    month = feb,
   volume = 763,
      eid = {147},
    pages = {147},
      doi = {10.1088/0004-637X/763/2/147},
   adsurl = {http://adsabs.harvard.edu/abs/2013ApJ...763..147B}
}

@INPROCEEDINGS{benson14,
       author = {{Benson}, B.~A. and {Ade}, P.~A.~R. and {Ahmed}, Z. and {Allen}, S.~W. and
         {Arnold}, K. and {Austermann}, J.~E. and {Bender}, A.~N. and
         {Bleem}, L.~E. and {Carlstrom}, J.~E. and {Chang}, C.~L. and
         {Cho}, H.~M. and {Cliche}, J.~F. and {Crawford}, T.~M. and
         {Cukierman}, A. and {de Haan}, T. and {Dobbs}, M.~A. and {Dutcher}, D. and
         {Everett}, W. and {Gilbert}, A. and {Halverson}, N.~W. and
         {Hanson}, D. and {Harrington}, N.~L. and {Hattori}, K. and
         {Henning}, J.~W. and {Hilton}, G.~C. and {Holder}, G.~P. and
         {Holzapfel}, W.~L. and {Irwin}, K.~D. and {Keisler}, R. and {Knox}, L. and
         {Kubik}, D. and {Kuo}, C.~L. and {Lee}, A.~T. and {Leitch}, E.~M. and
         {Li}, D. and {McDonald}, M. and {Meyer}, S.~S. and {Montgomery}, J. and
         {Myers}, M. and {Natoli}, T. and {Nguyen}, H. and {Novosad}, V. and
         {Padin}, S. and {Pan}, Z. and {Pearson}, J. and {Reichardt}, C. and
         {Ruhl}, J.~E. and {Saliwanchik}, B.~R. and {Simard}, G. and
         {Smecher}, G. and {Sayre}, J.~T. and {Shirokoff}, E. and
         {Stark}, A.~A. and {Story}, K. and {Suzuki}, A. and {Thompson}, K.~L. and
         {Tucker}, C. and {Vanderlinde}, K. and {Vieira}, J.~D. and
         {Vikhlinin}, A. and {Wang}, G. and {Yefremenko}, V. and {Yoon}, K.~W.},
        title = "{SPT-3G: A Next-Generation Cosmic Microwave Background Polarization Experiment on the South Pole Telescope}",
    booktitle = {Millimeter, Submillimeter, and Far-Infrared Detectors and Instrumentation for Astronomy VII},
         year = 2014,
       series = {\procspie},
       volume = {9153},
        month = jul,
          eid = {91531P},
        pages = {91531P},
          doi = {10.1117/12.2057305},
archivePrefix = {arXiv},
       eprint = {1407.2973},
 primaryClass = {astro-ph.IM},
       adsurl = {https://ui.adsabs.harvard.edu/abs/2014SPIE.9153E..1PB}
}

@INPROCEEDINGS{bertin12,
   author = {{Bertin}, E.},
    title = "{Displaying Digital Deep Sky Images}",
booktitle = {Astronomical Data Analysis Software and Systems XXI},
     year = 2012,
   series = {Astronomical Society of the Pacific Conference Series},
   volume = 461,
   editor = {{Ballester}, P. and {Egret}, D. and {Lorente}, N.~P.~F.},
    month = sep,
    pages = {263},
   adsurl = {http://adsabs.harvard.edu/abs/2012ASPC..461..263B}
}

@ARTICLE{bhargava25,
       author = {{Euclid Collaboration} and {Bhargava}, S. and {Benoist}, C. and {Gonzalez}, A.~H. and 
                 {Maturi}, M. and {Melin}, J. -B. and {Stanford}, S.~A. and {Munari}, E. and {Vannier}, M. and 
                 {Murray}, C. and {Maurogordato}, S. and {Biviano}, A. and {Macias-Perez}, J. and 
                 {Bartlett}, J.~G. and {Pacaud}, F. and {Widmer}, A. and {Meneghetti}, M. and {Sartoris}, B. and 
                 {Aguena}, M. and {Alguero}, G. and {Andreon}, S. and {Bardelli}, S. and {Baumont}, L. and 
                 {Bolzonella}, M. and {Cabanac}, R. and {Cappi}, A. and {Castignani}, G. and {Combet}, C. and 
                 {Comparat}, J. and {Farrens}, S. and {Ghaffari}, Z. and {Giles}, P.~A. and {Giocoli}, C. and 
                 {Girardi}, M. and {Hatch}, N.~A. and {Ingoglia}, L. and {Iovino}, A. and {Mamon}, G.~A. and 
                 {Mei}, S. and {Moscardini}, L. and {Mourre}, S. and {Odier}, J. and {Pratt}, G.~W. and 
                 {Tarr\textbackslash'io}, P. and {Toni}, G. and {Weller}, J. and {Zucca}, E. and {Aghanim}, N. and 
                 {Altieri}, B. and {Amara}, A. and {Auricchio}, N. and {Baccigalupi}, C. and {Baldi}, M. and 
                 {Balestra}, A. and {Battaglia}, P. and {Bellagamba}, F. and {Bonchi}, A. and {Branchini}, E. and 
                 {Brescia}, M. and {Brinchmann}, J. and {Camera}, S. and {Capobianco}, V. and {Carbone}, C. and 
                 {Carretero}, J. and {Casas}, S. and {Castellano}, M. and {Cavuoti}, S. and {Chambers}, K.~C. and 
                 {Cimatti}, A. and {Colodro-Conde}, C. and {Congedo}, G. and {Conselice}, C.~J. and {Conversi}, L. and 
                 {Copin}, Y. and {Courbin}, F. and {Courtois}, H.~M. and {Cropper}, M. and {Da Silva}, A. and 
                 {Degaudenzi}, H. and {De Lucia}, G. and {Di Giorgio}, A.~M. and {Dolding}, C. and {Dole}, H. and 
                 {Dubath}, F. and {Duncan}, C.~A.~J. and {Dupac}, X. and {Dusini}, S. and {Ealet}, A. and 
                 {Escoffier}, S. and {Farina}, M. and {Farinelli}, R. and {Faustini}, F. and {Ferriol}, S. and 
                 {Finelli}, F. and {Fosalba}, P. and {Fotopoulou}, S. and {Frailis}, M. and {Franceschi}, E. and 
                 {Fumana}, M. and {Galeotta}, S. and {George}, K. and {Gillis}, B. and {G\textbackslash'omez-Alvarez}, P. and 
                 {Gracia-Carpio}, J. and {Granett}, B.~R. and {Grazian}, A. and {Grupp}, F. and {Guzzo}, L. and 
                 {Gwyn}, S. and {Haugan}, S.~V.~H. and {Hoar}, J. and {Hoekstra}, H. and {Holmes}, W. and 
                 {Hormuth}, F. and {Hornstrup}, A. and {Hudelot}, P. and {Ili\textbackslash'c}, S. and {Jahnke}, K. and 
                 {Jhabvala}, M. and {Joachimi}, B. and {Keih\textbackslash''anen}, E. and {Kermiche}, S. and {Kiessling}, A. and 
                 {Kilbinger}, M. and {Kubik}, B. and {Kuijken}, K. and {K\textbackslash''ummel}, M. and {Kunz}, M. and 
                 {Kurki-Suonio}, H. and {Le Boulc'h}, Q. and {Le Brun}, A.~M.~C. and {Le Mignant}, D. and 
                 {Liebing}, P. and {Ligori}, S. and {Lilje}, P.~B. and {Lindholm}, V. and {Lloro}, I. and 
                 {Mainetti}, G. and {Maino}, D. and {Maiorano}, E. and {Mansutti}, O. and {Marcin}, S. and 
                 {Marggraf}, O. and {Martinelli}, M. and {Martinet}, N. and {Marulli}, F. and {Massey}, R. and 
                 {Medinaceli}, E. and {Melchior}, M. and {Mellier}, Y. and {Merlin}, E. and {Meylan}, G. and 
                 {Mora}, A. and {Moresco}, M. and {Nakajima}, R. and {Neissner}, C. and {Nichol}, R.~C. and 
                 {Niemi}, S. -M. and {Nightingale}, J.~W. and {Padilla}, C. and {Paltani}, S. and {Pasian}, F. and 
                 {Pedersen}, K. and {Percival}, W.~J. and {Pettorino}, V. and {Pires}, S. and {Polenta}, G. and 
                 {Poncet}, M. and {Popa}, L.~A. and {Pozzetti}, L. and {Raison}, F. and {Rebolo}, R. and 
                 {Renzi}, A. and {Rhodes}, J. and {Riccio}, G. and {Romelli}, E. and {Roncarelli}, M. and 
                 {Saglia}, R. and {Sakr}, Z. and {S\textbackslash'anchez}, A.~G. and {Sapone}, D. and {Schewtschenko}, J.~A. and 
                 {Schneider}, P. and {Schrabback}, T. and {Secroun}, A. and {Sefusatti}, E. and {Seidel}, G. and 
                 {Serrano}, S. and {Simon}, P. and {Sirignano}, C. and {Sirri}, G. and {Skottfelt}, J. and 
                 {Stanco}, L. and {Steinwagner}, J. and {Tallada-Cresp\textbackslash'i}, P. and {Taylor}, A.~N. and 
                 {Tereno}, I. and {Toft}, S. and {Toledo-Moreo}, R. and {Torradeflot}, F. and {Tutusaus}, I. and 
                 {Valenziano}, L. and {Valiviita}, J. and {Vassallo}, T. and {Verdoes Kleijn}, G. and 
                 {Veropalumbo}, A. and {Wang}, Y. and {Zacchei}, A. and {Zamorani}, G. and {Zerbi}, F.~M. and 
                 {Allevato}, V. and {Ballardini}, M. and {Bozzo}, E. and {Burigana}, C. and {Casenove}, P. and 
                 {Di Ferdinando}, D. and {Escartin Vigo}, J.~A. and {Fabbian}, G. and {Gabarra}, L. and 
                 {Mart\textbackslash'in-Fleitas}, J. and {Matthew}, S. and {Mauri}, N. and {Metcalf}, R.~B. and 
                 {Pezzotta}, A. and {P\textbackslash"ontinen}, M. and {Porciani}, C. and {Risso}, I. and 
                 {Scottez}, V. and {Sereno}, M. and {Tenti}, M. and {Viel}, M. and {Wiesmann}, M. and {Akrami}, Y. and 
                 {Andika}, I.~T. and {Anselmi}, S. and {Archidiacono}, M. and {Atrio-Barandela}, F. and 
                 {Benson}, K. and {Bergamini}, P. and {Bertacca}, D. and {Bethermin}, M. and {Blanchard}, A. and 
                 {Blot}, L. and {B\textbackslash"ohringer}, H. and {Borgani}, S. and {Brown}, M.~L. and {Bruton}, S. and 
                 {Calabro}, A. and {Camacho Quevedo}, B. and {Caro}, F. and {Carvalho}, C.~S. and {Castro}, T. and 
                 {Charles}, Y. and {Cogato}, F. and {Cooray}, A.~R. and {Costanzi}, M. and {Cucciati}, O. and 
                 {Davini}, S. and {De Paolis}, F. and {Desprez}, G. and {D{\'\i}az-S{\'a}nchez}, A. and {Diaz}, J. and 
                 {Di Domizio}, S. and {Diego}, J.~M. and {Dimauro}, P. and {Duc}, P.-A. and {Enia}, A. and 
                 {Fang}, Y. and {Ferrari}, A.~G. and {Ferreira}, P.~G. and {Finoguenov}, A. and {Fontana}, A. and 
                 {Franco}, A. and {Ganga}, K. and {Garc{\'\i}a-Bellido}, J. and {Gasparetto}, T. and {Gaztanaga}, E. and 
                 {Giacomini}, F. and {Gianotti}, F. and {Gozaliasl}, G. and {Guidi}, M. and {Gutierrez}, C.~M. and 
                 {Hall}, A. and {Hartley}, W.~G. and {Hern{\'a}ndez-Monteagudo}, C. and {Hildebrandt}, H. and 
                 {Hjorth}, J. and {Kajava}, J.~E. and {Kang}, Y. and {Kansal}, V. and {Karagiannis}, D. and 
                 {Kiiveri}, K. and {Kirkpatrick}, C. and {Kruk}, S. and {Le Graet}, J. and {Legrand}, L. and 
                 {Lembo}, M. and {Lepori}, F. and {Leroy}, G. and {Lesci}, G.~F. and {Lesgourgues}, J. and 
                 {Leuzzi}, L. and {Liaudat}, T.~I. and {Liu}, S.~J. and {Loureiro}, A. and {Maggio}, G. and 
                 {Magliocchetti}, M. and {Magnier}, E.~A. and {Mannucci}, F. and {Maoli}, R. and {Martins}, C.~J.~A.~P. and 
                 {Maurin}, L. and {Migliaccio}, M. and {Miluzio}, M. and {Monaco}, P. and {Montoro}, A. and 
                 {Moretti}, C. and {Morgante}, G. and {Naidoo}, K. and {Navarro-Alsina}, A. and {Nesseris}, S. and 
                 {Passalacqua}, F. and {Paterson}, K. and {Patrizii}, L. and {Pisani}, A. and {Potter}, D. and 
                 {Quai}, S. and {Radovich}, M. and {Rocci}, P.-F. and {Rosati}, P. and {Sacquegna}, S. and 
                 {Sahl{\'e}n}, M. and {Sanders}, D.~B. and {Sarpa}, E. and {Schaye}, J. and {Schneider}, A. and 
                 {Schultheis}, M. and {Sciotti}, D. and {Sellentin}, E. and {Smith}, L.~C. and {Tanidis}, K. and 
                 {Testera}, G. and {Teyssier}, R. and {Tosi}, S. and {Troja}, A. and {Tucci}, M. and {Valieri}, C. and 
                 {Venhola}, A. and {Vergani}, D. and {Verza}, G. and {Vielzeuf}, P. and {Walton}, N.~A. and {Soubrie}, E. and 
                 {Scott}, D.},
        title = "{Euclid Quick Data Release (Q1). First detections from the Euclid galaxy cluster workflow}",
      journal = {\aap, accepted},
         year = 2025,
        month = jul,
          pages = {arXiv:2503.19196},
          doi = {10.1051/0004-6361/202554937},
}

@ARTICLE{blain93,
   author = {{Blain}, A.~W. and {Longair}, M.~S.},
    title = "{Submillimetre Cosmology}",
  journal = {\mnras},
     year = 1993,
    month = sep,
   volume = 264,
    pages = {509},
   adsurl = {http://adsabs.harvard.edu/abs/1993MNRAS.264..509B}
}

@ARTICLE{blain02,
   author = {{Blain}, A.~W. and {Smail}, I. and {Ivison}, R.~J. and {Kneib}, J.-P. and 
	{Frayer}, D.~T.},
    title = "{Submillimeter galaxies}",
  journal = {\physrep},
   eprint = {arXiv:astro-ph/0202228},
     year = 2002,
    month = oct,
   volume = 369,
    pages = {111-176},
      doi = {10.1016/S0370-1573(02)00134-5},
   adsurl = {http://adsabs.harvard.edu/abs/2002PhR...369..111B}
}

@ARTICLE{blain04,
   author = {{Blain}, A.~W. and {Chapman}, S.~C. and {Smail}, I. and {Ivison}, R.
	},
    title = "{Clustering of Submillimeter-selected Galaxies}",
  journal = {\apj},
   eprint = {arXiv:astro-ph/0405035},
     year = 2004,
    month = aug,
   volume = 611,
    pages = {725-731},
      doi = {10.1086/422353},
   adsurl = {http://adsabs.harvard.edu/abs/2004ApJ...611..725B}
}

@ARTICLE{bleem15b,
   author = {{Bleem}, L.~E. and {Stalder}, B. and {de Haan}, T. and {Aird}, K.~A. and 
	{Allen}, S.~W. and {Applegate}, D.~E. and {Ashby}, M.~L.~N. and 
	{Bautz}, M. and {Bayliss}, M. and {Benson}, B.~A. and {Bocquet}, S. and 
	{Brodwin}, M. and {Carlstrom}, J.~E. and {Chang}, C.~L. and 
	{Chiu}, I. and {Cho}, H.~M. and {Clocchiatti}, A. and {Crawford}, T.~M. and 
	{Crites}, A.~T. and {Desai}, S. and {Dietrich}, J.~P. and {Dobbs}, M.~A. and 
	{Foley}, R.~J. and {Forman}, W.~R. and {George}, E.~M. and {Gladders}, M.~D. and 
	{Gonzalez}, A.~H. and {Halverson}, N.~W. and {Hennig}, C. and 
	{Hoekstra}, H. and {Holder}, G.~P. and {Holzapfel}, W.~L. and 
	{Hrubes}, J.~D. and {Jones}, C. and {Keisler}, R. and {Knox}, L. and 
	{Lee}, A.~T. and {Leitch}, E.~M. and {Liu}, J. and {Lueker}, M. and 
	{Luong-Van}, D. and {Mantz}, A. and {Marrone}, D.~P. and {McDonald}, M. and 
	{McMahon}, J.~J. and {Meyer}, S.~S. and {Mocanu}, L. and {Mohr}, J.~J. and 
	{Murray}, S.~S. and {Padin}, S. and {Pryke}, C. and {Reichardt}, C.~L. and 
	{Rest}, A. and {Ruel}, J. and {Ruhl}, J.~E. and {Saliwanchik}, B.~R. and 
	{Saro}, A. and {Sayre}, J.~T. and {Schaffer}, K.~K. and {Schrabback}, T. and 
	{Shirokoff}, E. and {Song}, J. and {Spieler}, H.~G. and {Stanford}, S.~A. and 
	{Staniszewski}, Z. and {Stark}, A.~A. and {Story}, K.~T. and 
	{Stubbs}, C.~W. and {Vanderlinde}, K. and {Vieira}, J.~D. and 
	{Vikhlinin}, A. and {Williamson}, R. and {Zahn}, O. and {Zenteno}, A.
	},
    title = "{Galaxy Clusters Discovered via the Sunyaev-Zel'dovich Effect in the 2500-Square-Degree SPT-SZ Survey}",
  journal = {\apjs},
archivePrefix = "arXiv",
   eprint = {1409.0850},
     year = 2015,
    month = feb,
   volume = 216,
      eid = {27},
    pages = {27},
      doi = {10.1088/0067-0049/216/2/27},
   adsurl = {http://adsabs.harvard.edu/abs/2015ApJS..216...27B}
}

@ARTICLE{bleem20,
       author = {{Bleem}, L.~E. and {Bocquet}, S. and {Stalder}, B. and {Gladders}, M.~D. and {Ade}, P.~A.~R. and {Allen}, S.~W. and {Anderson}, A.~J. and {Annis}, J. and {Ashby}, M.~L.~N. and {Austermann}, J.~E. and {Avila}, S. and {Avva}, J.~S. and {Bayliss}, M. and {Beall}, J.~A. and {Bechtol}, K. and {Bender}, A.~N. and {Benson}, B.~A. and {Bertin}, E. and {Bianchini}, F. and {Blake}, C. and {Brodwin}, M. and {Brooks}, D. and {Buckley-Geer}, E. and {Burke}, D.~L. and {Carlstrom}, J.~E. and {Rosell}, A. Carnero and {Carrasco Kind}, M. and {Carretero}, J. and {Chang}, C.~L. and {Chiang}, H.~C. and {Citron}, R. and {Moran}, C. Corbett and {Costanzi}, M. and {Crawford}, T.~M. and {Crites}, A.~T. and {da Costa}, L.~N. and {de Haan}, T. and {De Vicente}, J. and {Desai}, S. and {Diehl}, H.~T. and {Dietrich}, J.~P. and {Dobbs}, M.~A. and {Eifler}, T.~F. and {Everett}, W. and {Flaugher}, B. and {Floyd}, B. and {Frieman}, J. and {Gallicchio}, J. and {Garc{\'\i}a-Bellido}, J. and {George}, E.~M. and {Gerdes}, D.~W. and {Gilbert}, A. and {Gruen}, D. and {Gruendl}, R.~A. and {Gschwend}, J. and {Gupta}, N. and {Gutierrez}, G. and {Halverson}, N.~W. and {Harrington}, N. and {Henning}, J.~W. and {Heymans}, C. and {Holder}, G.~P. and {Hollowood}, D.~L. and {Holzapfel}, W.~L. and {Honscheid}, K. and {Hrubes}, J.~D. and {Huang}, N. and {Hubmayr}, J. and {Irwin}, K.~D. and {James}, D.~J. and {Jeltema}, T. and {Joudaki}, S. and {Khullar}, G. and {Klein}, M. and {Knox}, L. and {Kuropatkin}, N. and {Lee}, A.~T. and {Li}, D. and {Lidman}, C. and {Lowitz}, A. and {MacCrann}, N. and {Mahler}, G. and {Maia}, M.~A.~G. and {Marshall}, J.~L. and {McDonald}, M. and {McMahon}, J.~J. and {Melchior}, P. and {Menanteau}, F. and {Meyer}, S.~S. and {Miquel}, R. and {Mocanu}, L.~M. and {Mohr}, J.~J. and {Montgomery}, J. and {Nadolski}, A. and {Natoli}, T. and {Nibarger}, J.~P. and {Noble}, G. and {Novosad}, V. and {Padin}, S. and {Palmese}, A. and {Parkinson}, D. and {Patil}, S. and {Paz-Chinch{\'o}n}, F. and {Plazas}, A.~A. and {Pryke}, C. and {Ramachandra}, N.~S. and {Reichardt}, C.~L. and {Remolina Gonz{\'a}lez}, J.~D. and {Romer}, A.~K. and {Roodman}, A. and {Ruhl}, J.~E. and {Rykoff}, E.~S. and {Saliwanchik}, B.~R. and {Sanchez}, E. and {Saro}, A. and {Sayre}, J.~T. and {Schaffer}, K.~K. and {Schrabback}, T. and {Serrano}, S. and {Sharon}, K. and {Sievers}, C. and {Smecher}, G. and {Smith}, M. and {Soares-Santos}, M. and {Stark}, A.~A. and {Story}, K.~T. and {Suchyta}, E. and {Tarle}, G. and {Tucker}, C. and {Vanderlinde}, K. and {Veach}, T. and {Vieira}, J.~D. and {Wang}, G. and {Weller}, J. and {Whitehorn}, N. and {Wu}, W.~L.~K. and {Yefremenko}, V. and {Zhang}, Y.},
        title = "{The SPTpol Extended Cluster Survey}",
      journal = {\apjs},
         year = 2020,
        month = mar,
       volume = {247},
       number = {1},
          eid = {25},
        pages = {25},
          doi = {10.3847/1538-4365/ab6993},
archivePrefix = {arXiv},
       eprint = {1910.04121},
 primaryClass = {astro-ph.CO},
       adsurl = {https://ui.adsabs.harvard.edu/abs/2020ApJS..247...25B}
}

@ARTICLE{bleem22,
       author = {{Bleem}, L.~E. and {Crawford}, T.~M. and {Ansarinejad}, B. and {Benson}, B.~A. and {Bocquet}, S. and {Carlstrom}, J.~E. and {Chang}, C.~L. and {Chown}, R. and {Crites}, A.~T. and {de Haan}, T. and {Dobbs}, M.~A. and {Everett}, W.~B. and {George}, E.~M. and {Gualtieri}, R. and {Halverson}, N.~W. and {Holder}, G.~P. and {Holzapfel}, W.~L. and {Hrubes}, J.~D. and {Knox}, L. and {Lee}, A.~T. and {Luong-Van}, D. and {Marrone}, D.~P. and {McMahon}, J.~J. and {Meyer}, S.~S. and {Millea}, M. and {Mocanu}, L.~M. and {Mohr}, J.~J. and {Natoli}, T. and {Omori}, Y. and {Padin}, S. and {Pryke}, C. and {Raghunathan}, S. and {Reichardt}, C.~L. and {Ruhl}, J.~E. and {Schaffer}, K.~K. and {Shirokoff}, E. and {Staniszewski}, Z. and {Stark}, A.~A. and {Vieira}, J.~D. and {Williamson}, R.},
        title = "{CMB/kSZ and Compton-y Maps from 2500 deg$^{2}$ of SPT-SZ and Planck Survey Data}",
      journal = {\apjs},
         year = 2022,
        month = feb,
       volume = {258},
       number = {2},
          eid = {36},
        pages = {36},
          doi = {10.3847/1538-4365/ac35e9},
archivePrefix = {arXiv},
       eprint = {2102.05033},
 primaryClass = {astro-ph.CO},
       adsurl = {https://ui.adsabs.harvard.edu/abs/2022ApJS..258...36B}
}

@ARTICLE{bleem24,
       author = {{Bleem}, L.~E. and {Klein}, M. and {Abbot}, T.~M.~C. and {Ade}, P.~A.~R. and {Aguena}, M. and {Alves}, O. and {Anderson}, A.~J. and {Andrade-Oliveira}, F. and {Ansarinejad}, B. and {Archipley}, M. and {Ashby}, M.~L.~N. and {Austermann}, J.~E. and {Bacon}, D. and {Beall}, J.~A. and {Bender}, A.~N. and {Benson}, B.~A. and {Bianchini}, F. and {Bocquet}, S. and {Brooks}, D. and {Burke}, D.~L. and {Calzadilla}, M. and {Carlstrom}, J.~E. and {Carnero Rosell}, A. and {Carretero}, J. and {Chang}, C.~L. and {Chaubal}, P. and {Chiang}, H.~C. and {Chou}, T-L. and {Citron}, R. and {Corbett Moran}, C. and {Costanzi}, M. and {Constanzi}, M. and {Crawford}, T.~M. and {Crites}, A.~T. and {da Costa}, L.~N. and {de Haan}, T. and {De Vicente}, J. and {Desai}, S. and {Dobbs}, M.~A. and {Doel}, P. and {Everett}, W. and {Ferrero}, I. and {Flaugher}, B. and {Floyd}, B. and {Friedel}, D. and {Frieman}, J. and {Gallicchio}, J. and {Garc'ia-Bellido}, J. and {Gatti}, M. and {George}, E.~M. and {Giannini}, G. and {Grandis}, S. and {Gruen}, D. and {Gruendl}, R.~A. and {Gupta}, N. and {Gutierrez}, G. and {Halverson}, N.~W. and {Hinton}, S.~R. and {Hinton}, S.~R. and {Holder}, G.~P. and {Hollowood}, D.~L. and {Holzapfel}, W.~L. and {Honscheid}, K. and {Hrubes}, J.~D. and {Huang}, N. and {Hubmayr}, J. and {Irwin}, K.~D. and {Mena-Fern{\'a}ndez}, J. and {James}, D.~J. and {K{\'e}ruzor{\'e}}, F. and {Knox}, L. and {Kuehn}, K. and {Lahav}, O. and {Lee}, A.~T. and {Lee}, S. and {Li}, D. and {Lowitz}, A. and {Marshal}, J.~L. and {McDonald}, M. and {McMahon}, J.~J. and {Menanteau}, F. and {Meyer}, S.~S. and {Miquel}, R. and {Mohr}, J.~J. and {Montgomery}, J. and {Myles}, J. and {Natoli}, T. and {Nibarger}, J.~P. and {Noble}, G.~I. and {Novosad}, V. and {Ogando}, R.~L.~C. and {Padin}, S. and {Patil}, S. and {Pereira}, M.~E.~S. and {Pieres}, A. and {Plazas Malag'on}, A.~A. and {Pryke}, C. and {Reichardt}, C.~L. and {Rodr'iguez-Monroy}, M. and {Romer}, A.~K. and {Ruhl}, J.~E. and {Saliwanchik}, B.~R. and {Salvati}, L. and {Sanchez}, E. and {Saro}, A. and {Schaffer}, K.~K. and {Schrabback}, T. and {Sevilla-Noarbe}, I. and {Sievers}, C. and {Smecher}, G. and {Smith}, M. and {Somboonpanyakul}, T. and {Stalder}, B. and {Stark}, A.~A. and {Suchyta}, E. and {Swanson}, M.~E.~C. and {Tarle}, G. and {To}, C. and {Tucker}, C. and {Veach}, T. and {Vieira}, J.~D. and {Vincenzi}, M. and {Wang}, G. and {Weller}, J. and {Whitehorn}, N. and {Wiseman}, P. and {Wu}, W.~L.~K. and {Yefremenko}, V. and {Zebrowski}, J.~A. and {Zhang}, Y.},
        title = "{Galaxy Clusters Discovered via the Thermal Sunyaev-Zel'dovich Effect in the 500-square-degree SPTpol Survey}",
      journal = {The Open Journal of Astrophysics},
         year = 2024,
        month = feb,
       volume = {7},
          eid = {13},
        pages = {13},
          doi = {10.21105/astro.2311.07512},
archivePrefix = {arXiv},
       eprint = {2311.07512},
 primaryClass = {astro-ph.CO},
       adsurl = {https://ui.adsabs.harvard.edu/abs/2024OJAp....7E..13B}
}

@ARTICLE{bocquet24b,
       author = {{Bocquet}, S. and {Grandis}, S. and {Bleem}, L.~E. and {Klein}, M. and {Mohr}, J.~J. and {Schrabback}, T. and {Abbott}, T.~M.~C. and {Ade}, P.~A.~R. and {Aguena}, M. and {Alarcon}, A. and {Allam}, S. and {Allen}, S.~W. and {Alves}, O. and {Amon}, A. and {Anderson}, A.~J. and {Annis}, J. and {Ansarinejad}, B. and {Austermann}, J.~E. and {Avila}, S. and {Bacon}, D. and {Bayliss}, M. and {Beall}, J.~A. and {Bechtol}, K. and {Becker}, M.~R. and {Bender}, A.~N. and {Benson}, B.~A. and {Bernstein}, G.~M. and {Bhargava}, S. and {Bianchini}, F. and {Brodwin}, M. and {Brooks}, D. and {Bryant}, L. and {Campos}, A. and {Canning}, R.~E.~A. and {Carlstrom}, J.~E. and {Carnero Rosell}, A. and {Carrasco Kind}, M. and {Carretero}, J. and {Castander}, F.~J. and {Cawthon}, R. and {Chang}, C.~L. and {Chang}, C. and {Chaubal}, P. and {Chen}, R. and {Chiang}, H.~C. and {Choi}, A. and {Chou}, T. -L. and {Citron}, R. and {Corbett Moran}, C. and {Cordero}, J. and {Costanzi}, M. and {Crawford}, T.~M. and {Crites}, A.~T. and {da Costa}, L.~N. and {Pereira}, M.~E.~S. and {Davis}, C. and {Davis}, T.~M. and {DeRose}, J. and {Desai}, S. and {de Haan}, T. and {Diehl}, H.~T. and {Dobbs}, M.~A. and {Dodelson}, S. and {Doux}, C. and {Drlica-Wagner}, A. and {Eckert}, K. and {Elvin-Poole}, J. and {Everett}, S. and {Everett}, W. and {Ferrero}, I. and {Fert{\'e}}, A. and {Flores}, A.~M. and {Frieman}, J. and {Gallicchio}, J. and {Garc{\'\i}a-Bellido}, J. and {Gatti}, M. and {George}, E.~M. and {Giannini}, G. and {Gladders}, M.~D. and {Gruen}, D. and {Gruendl}, R.~A. and {Gupta}, N. and {Gutierrez}, G. and {Halverson}, N.~W. and {Harrison}, I. and {Hartley}, W.~G. and {Herner}, K. and {Hinton}, S.~R. and {Holder}, G.~P. and {Hollowood}, D.~L. and {Holzapfel}, W.~L. and {Honscheid}, K. and {Hrubes}, J.~D. and {Huang}, N. and {Hubmayr}, J. and {Huff}, E.~M. and {Huterer}, D. and {Irwin}, K.~D. and {James}, D.~J. and {Jarvis}, M. and {Khullar}, G. and {Kim}, K. and {Knox}, L. and {Kraft}, R. and {Krause}, E. and {Kuehn}, K. and {Kuropatkin}, N. and {K{\'e}ruzor{\'e}}, F. and {Lahav}, O. and {Lee}, A.~T. and {Leget}, P. -F. and {Li}, D. and {Lin}, H. and {Lowitz}, A. and {MacCrann}, N. and {Mahler}, G. and {Mantz}, A. and {Marshall}, J.~L. and {McCullough}, J. and {McDonald}, M. and {McMahon}, J.~J. and {Mena-Fern{\'a}ndez}, J. and {Menanteau}, F. and {Meyer}, S.~S. and {Miquel}, R. and {Montgomery}, J. and {Myles}, J. and {Natoli}, T. and {Navarro-Alsina}, A. and {Nibarger}, J.~P. and {Noble}, G.~I. and {Novosad}, V. and {Ogando}, R.~L.~C. and {Omori}, Y. and {Padin}, S. and {Pandey}, S. and {Paschos}, P. and {Patil}, S. and {Pieres}, A. and {Plazas Malag{\'o}n}, A.~A. and {Porredon}, A. and {Prat}, J. and {Pryke}, C. and {Raveri}, M. and {Reichardt}, C.~L. and {Roberson}, J. and {Rollins}, R.~P. and {Romero}, C. and {Roodman}, A. and {Ruhl}, J.~E. and {Rykoff}, E.~S. and {Saliwanchik}, B.~R. and {Salvati}, L. and {S{\'a}nchez}, C. and {Sanchez}, E. and {Sanchez Cid}, D. and {Saro}, A. and {Schaffer}, K.~K. and {Secco}, L.~F. and {Sevilla-Noarbe}, I. and {Sharon}, K. and {Sheldon}, E. and {Shin}, T. and {Sievers}, C. and {Smecher}, G. and {Smith}, M. and {Somboonpanyakul}, T. and {Sommer}, M. and {Stalder}, B. and {Stark}, A.~A. and {Stephen}, J. and {Strazzullo}, V. and {Suchyta}, E. and {Tarle}, G. and {To}, C. and {Troxel}, M.~A. and {Tucker}, C. and {Tutusaus}, I. and {Varga}, T.~N. and {Veach}, T. and {Vieira}, J.~D. and {Vikhlinin}, A. and {von der Linden}, A. and {Wang}, G. and {Weaverdyck}, N. and {Weller}, J. and {Whitehorn}, N. and {Wu}, W.~L.~K. and {Yanny}, B. and {Yefremenko}, V. and {Yin}, B. and {Young}, M. and {Zebrowski}, J.~A. and {Zhang}, Y. and {Zohren}, H. and {Zuntz}, J. and {(SPT} and {DES Collaborations)}},
        title = "{SPT clusters with DES and HST weak lensing. II. Cosmological constraints from the abundance of massive halos}",
      journal = {\prd},
         year = 2024,
        month = oct,
       volume = {110},
       number = {8},
          eid = {083510},
        pages = {083510},
          doi = {10.1103/PhysRevD.110.083510},
archivePrefix = {arXiv},
       eprint = {2401.02075},
 primaryClass = {astro-ph.CO},
       adsurl = {https://ui.adsabs.harvard.edu/abs/2024PhRvD.110h3510B}
}

@misc{bradley23,
author       = {Larry Bradley and
                Brigitta Sip{\H o}cz and
                Thomas Robitaille and
                Erik Tollerud and
                Z\`e Vin{\'{\i}}cius and
                Christoph Deil and
                Kyle Barbary and
                Tom J Wilson and
                Ivo Busko and
                Axel Donath and
                Hans Moritz G{\"u}nther and
                Mihai Cara and
                P. L. Lim and
                Sebastian Me{\ss}linger and
                Simon Conseil and
                Azalee Bostroem and
                Michael Droettboom and
                E. M. Bray and
                Lars Andersen Bratholm and
                Geert Barentsen and
                Matt Craig and
                Shivangee Rathi and
                Sergio Pascual and
                Gabriel Perren and
                Iskren Y. Georgiev and
                Miguel de Val-Borro and
                Wolfgang Kerzendorf and
                Yoonsoo P. Bach and
                Bruno Quint and
                Harrison Souchereau},
title        = {astropy/photutils: 1.8.0},
month        = may,
year         = 2023,
publisher    = {Zenodo},
version      = {1.8.0},
doi          = {10.5281/zenodo.7946442},
url          = {https://doi.org/10.5281/zenodo.7946442}
}

@ARTICLE{bulbul24,
       author = {{Bulbul}, E. and {Liu}, A. and {Kluge}, M. and {Zhang}, X. and {Sanders}, J.~S. and {Bahar}, Y.~E. and {Ghirardini}, V. and {Artis}, E. and {Seppi}, R. and {Garrel}, C. and {Ramos-Ceja}, M.~E. and {Comparat}, J. and {Balzer}, F. and {B{\"o}ckmann}, K. and {Br{\"u}ggen}, M. and {Clerc}, N. and {Dennerl}, K. and {Dolag}, K. and {Freyberg}, M. and {Grandis}, S. and {Gruen}, D. and {Kleinebreil}, F. and {Krippendorf}, S. and {Lamer}, G. and {Merloni}, A. and {Migkas}, K. and {Nandra}, K. and {Pacaud}, F. and {Predehl}, P. and {Reiprich}, T.~H. and {Schrabback}, T. and {Veronica}, A. and {Weller}, J. and {Zelmer}, S.},
        title = "{The SRG/eROSITA All-Sky Survey. The first catalog of galaxy clusters and groups in the Western Galactic Hemisphere}",
      journal = {\aap},
         year = 2024,
        month = may,
       volume = {685},
          eid = {A106},
        pages = {A106},
          doi = {10.1051/0004-6361/202348264},
archivePrefix = {arXiv},
       eprint = {2402.08452},
 primaryClass = {astro-ph.CO},
       adsurl = {https://ui.adsabs.harvard.edu/abs/2024A&A...685A.106B}
}

@ARTICLE{calabretta02,
   author = {{Calabretta}, M.~R. and {Greisen}, E.~W.},
    title = "{Representations of celestial coordinates in FITS}",
  journal = {\aap},
   eprint = {arXiv:astro-ph/0207413},
     year = 2002,
    month = dec,
   volume = 395,
    pages = {1077-1122},
      doi = {10.1051/0004-6361:20021327},
   adsurl = {http://ads.ari.uni-heidelberg.de/abs/2002A%26A...395.1077C}
}

@article{carlstrom02,
    author = {{Carlstrom}, J.~E. and {Holder}, G.~P. and {Reese}, E.~D.},
    title = "{Cosmology with the Sunyaev-Zel'dovich Effect}",
    journal = {\araa},
    year = 2002,
    volume = 40,
    pages = {643-680},
adsurl = {http://adsabs.harvard.edu/cgi-bin/nph-bib_query?bibcode=2002ARA%26A..40..643C&amp;db_key=AST}
}

@ARTICLE{carlstrom11,
   author = {{Carlstrom}, J.~E. and {Ade}, P.~A.~R. and {Aird}, K.~A. and 
	{Benson}, B.~A. and {Bleem}, L.~E. and {Busetti}, S. and {Chang}, C.~L. and 
	{Chauvin}, E. and {Cho}, H.-M. and {Crawford}, T.~M. and {Crites}, A.~T. and 
	{Dobbs}, M.~A. and {Halverson}, N.~W. and {Heimsath}, S. and 
	{Holzapfel}, W.~L. and {Hrubes}, J.~D. and {Joy}, M. and {Keisler}, R. and 
	{Lanting}, T.~M. and {Lee}, A.~T. and {Leitch}, E.~M. and {Leong}, J. and 
	{Lu}, W. and {Lueker}, M. and {Luong-van}, D. and {McMahon}, J.~J. and 
	{Mehl}, J. and {Meyer}, S.~S. and {Mohr}, J.~J. and {Montroy}, T.~E. and 
	{Padin}, S. and {Plagge}, T. and {Pryke}, C. and {Ruhl}, J.~E. and 
	{Schaffer}, K.~K. and {Schwan}, D. and {Shirokoff}, E. and {Spieler}, H.~G. and 
	{Staniszewski}, Z. and {Stark}, A.~A. and {Tucker}, C. and {Vanderlinde}, K. and 
	{Vieira}, J.~D. and {Williamson}, R.},
    title = "{The 10 Meter South Pole Telescope}",
  journal = {\pasp},
     year = 2011,
    month = may,
   volume = 123,
    pages = {568-581},
    archivePrefix = "arXiv",
   eprint = {0907.4445},
      doi = {10.1086/659879},
   adsurl = {http://adsabs.harvard.edu/abs/2011PASP..123..568C}
}

@ARTICLE{casey14,
   author = {{Casey}, C.~M. and {Narayanan}, D. and {Cooray}, A.},
    title = "{Dusty star-forming galaxies at high redshift}",
  journal = {\physrep},
archivePrefix = "arXiv",
   eprint = {1402.1456},
     year = 2014,
    month = aug,
   volume = 541,
    pages = {45-161},
      doi = {10.1016/j.physrep.2014.02.009},
   adsurl = {http://adsabs.harvard.edu/abs/2014PhR...541...45C}
}

@ARTICLE{casey16,
       author = {{Casey}, Caitlin M.},
        title = "{The Ubiquity of Coeval Starbursts in Massive Galaxy Cluster Progenitors}",
      journal = {\apj},
         year = 2016,
        month = jun,
       volume = {824},
       number = {1},
          eid = {36},
        pages = {36},
          doi = {10.3847/0004-637X/824/1/36},
archivePrefix = {arXiv},
       eprint = {1603.04437},
 primaryClass = {astro-ph.GA},
       adsurl = {https://ui.adsabs.harvard.edu/abs/2016ApJ...824...36C}
}

@ARTICLE{cavaliere76,
    author = {{Cavaliere}, A. and {Fusco-Femiano}, R.},
    title = {X-rays from Hot Plasma in Clusters of Galaxies},
    journal = {\aap},
    year = 1976,
    month = may,
    volume = 49,
    pages = {137},
    adsurl = {http://adsabs.harvard.edu/abs/1976A%26A....49..137C}
}

@ARTICLE{chichura25,
       author = {{Chichura}, P.~M. and {Rahlin}, A. and {Anderson}, A.~J. and {Ansarinejad}, B. and {Archipley}, M. and {Balkenhol}, L. and {Benabed}, K. and {Bender}, A.~N. and {Benson}, B.~A. and {Bianchini}, F. and {Bleem}, L.~E. and {Bouchet}, F.~R. and {Bryant}, L. and {Camphuis}, E. and {Carlstrom}, J.~E. and {Chang}, C.~L. and {Chaubal}, P. and {Chokshi}, A. and {Chou}, T. -L. and {Coerver}, A. and {Crawford}, T.~M. and {Daley}, C. and {Haan}, T. De and {Dibert}, K.~R. and {Dobbs}, M.~A. and {Doohan}, M. and {Doussot}, A. and {Dutcher}, D. and {Everett}, W. and {Feng}, C. and {Ferguson}, K.~R. and {Fichman}, K. and {Foster}, A. and {Galli}, S. and {Gambrel}, A.~E. and {Gardner}, R.~W. and {Ge}, F. and {Goeckner-Wald}, N. and {Gualtieri}, R. and {Guidi}, F. and {Guns}, S. and {Halverson}, N.~W. and {Hivon}, E. and {Holder}, G.~P. and {Holzapfel}, W.~L. and {Hood}, J.~C. and {Hryciuk}, A. and {Huang}, N. and {K{\'e}ruzor{\'e}}, F. and {Khalife}, A.~R. and {Kim}, J. and {Knox}, L. and {Korman}, M. and {Kornoelje}, K. and {Kuo}, C. -L. and {Levy}, K. and {Lowitz}, A.~E. and {Lu}, C. and {Maniyar}, A. and {Marrone}, D.~P. and {Martsen}, E.~S. and {Menanteau}, F. and {Millea}, M. and {Montgomery}, J. and {Nakato}, Y. and {Natoli}, T. and {Noble}, G.~I. and {Omori}, Y. and {Padin}, S. and {Pan}, Z. and {Paschos}, P. and {Phadke}, K.~A. and {Pollak}, A.~W. and {Prabhu}, K. and {Quan}, W. and {Rahimi}, M. and {Reichardt}, C.~L. and {Rouble}, M. and {Ruhl}, J.~E. and {Schiappucci}, E. and {Sobrin}, J.~A. and {Stark}, A.~A. and {Stephen}, J. and {Tandoi}, C. and {Thorne}, B. and {Trendafilova}, C. and {Umilta}, C. and {Veitch-Michaelis}, J. and {Vieira}, J.~D. and {Vitrier}, A. and {Wan}, Y. and {Whitehorn}, N. and {Wu}, W.~L.~K. and {Young}, M.~R. and {Zagorski}, K. and {Zebrowski}, J.~A.},
        title = "{Pointing Accuracy Improvements for the South Pole Telescope with Machine Learning}",
      journal = {Journal of Astronomical Instrumentation},
         year = 2025,
        month = jan,
       volume = {14},
          eid = {2550001},
        pages = {2550001},
          doi = {10.1142/S2251171725500011},
archivePrefix = {arXiv},
       eprint = {2412.15167},
 primaryClass = {astro-ph.IM},
       adsurl = {https://ui.adsabs.harvard.edu/abs/2025JAI....1450001C}
}

@ARTICLE{coerver24,
       author = {{Coerver}, A. and {Zebrowski}, J.~A. and {Takakura}, S. and {Holzapfel}, W.~L. and {Ade}, P.~A.~R. and {Anderson}, A.~J. and {Ahmed}, Z. and {Ansarinejad}, B. and {Archipley}, M. and {Balkenhol}, L. and {Barron}, D. and {Benabed}, K. and {Bender}, A.~N. and {Benson}, B.~A. and {Bianchini}, F. and {Bleem}, L.~E. and {Bouchet}, F.~R. and {Bryant}, L. and {Camphuis}, E. and {Carlstrom}, J.~E. and {Cecil}, T.~W. and {Chang}, C.~L. and {Chaubal}, P. and {Chichura}, P.~M. and {Chokshi}, A. and {Chou}, T. -L. and {Crawford}, T.~M. and {Cukierman}, A. and {Daley}, C. and {de Haan}, T. and {Dibert}, K.~R. and {Dobbs}, M.~A. and {Doussot}, A. and {Dutcher}, D. and {Everett}, W. and {Feng}, C. and {Ferguson}, K.~R. and {Fichman}, K. and {Foster}, A. and {Galli}, S. and {Gambrel}, A.~E. and {Gardner}, R.~W. and {Ge}, F. and {Goeckner-Wald}, N. and {Gualtieri}, R. and {Guidi}, F. and {Guns}, S. and {Halverson}, N.~W. and {Hivon}, E. and {Holder}, G.~P. and {Hood}, J.~C. and {Hryciuk}, A. and {Huang}, N. and {K{\'e}ruzor{\'e}}, F. and {Khalife}, A.~R. and {Knox}, L. and {Korman}, M. and {Kornoelje}, K. and {Kuo}, C. -L. and {Lee}, A.~T. and {Levy}, K. and {Lowitz}, A.~E. and {Lu}, C. and {Maniyar}, A. and {Martsen}, E.~S. and {Menanteau}, F. and {Millea}, M. and {Montgomery}, J. and {Nakato}, Y. and {Natoli}, T. and {Noble}, G.~I. and {Novosad}, V. and {Omori}, Y. and {Padin}, S. and {Pan}, Z. and {Paschos}, P. and {Phadke}, K.~A. and {Pollak}, A.~W. and {Prabhu}, K. and {Quan}, W. and {Rahimi}, M. and {Rahlin}, A. and {Reichardt}, C.~L. and {Rouble}, M. and {Ruhl}, J.~E. and {Schiappucci}, E. and {Smecher}, G. and {Sobrin}, J.~A. and {Stark}, A.~A. and {Stephen}, J. and {Suzuki}, A. and {Tandoi}, C. and {Thompson}, K.~L. and {Thorne}, B. and {Trendafilova}, C. and {Tucker}, C. and {Umilta}, C. and {Vieira}, J.~D. and {Vitrier}, A. and {Wan}, Y. and {Wang}, G. and {Whitehorn}, N. and {Wu}, W.~L.~K. and {Yefremenko}, V. and {Young}, M.~R.},
        title = "{Measurement and Modeling of Polarized Atmosphere at the South Pole with SPT-3G}",
      journal = {\apj},
         year = 2025,
        month = mar,
       volume = {982},
       number = {1},
          eid = {15},
        pages = {15},
          doi = {10.3847/1538-4357/ada35d},
archivePrefix = {arXiv},
       eprint = {2407.20579},
 primaryClass = {astro-ph.IM},
       adsurl = {https://ui.adsabs.harvard.edu/abs/2025ApJ...982...15C}
}

@ARTICLE{colless01,
       author = {{Colless}, Matthew and {Dalton}, Gavin and {Maddox}, Steve and {Sutherland}, Will and {Norberg}, Peder and {Cole}, Shaun and {Bland-Hawthorn}, Joss and {Bridges}, Terry and {Cannon}, Russell and {Collins}, Chris and {Couch}, Warrick and {Cross}, Nicholas and {Deeley}, Kathryn and {De Propris}, Roberto and {Driver}, Simon P. and {Efstathiou}, George and {Ellis}, Richard S. and {Frenk}, Carlos S. and {Glazebrook}, Karl and {Jackson}, Carole and {Lahav}, Ofer and {Lewis}, Ian and {Lumsden}, Stuart and {Madgwick}, Darren and {Peacock}, John A. and {Peterson}, Bruce A. and {Price}, Ian and {Seaborne}, Mark and {Taylor}, Keith},
        title = "{The 2dF Galaxy Redshift Survey: spectra and redshifts}",
      journal = {\mnras},
         year = 2001,
        month = dec,
       volume = {328},
       number = {4},
        pages = {1039-1063},
          doi = {10.1046/j.1365-8711.2001.04902.x},
archivePrefix = {arXiv},
       eprint = {astro-ph/0106498},
 primaryClass = {astro-ph},
       adsurl = {https://ui.adsabs.harvard.edu/abs/2001MNRAS.328.1039C}
}

@ARTICLE{condon92,
       author = {{Condon}, J.~J.},
        title = "{Radio emission from normal galaxies.}",
      journal = {\araa},
         year = 1992,
        month = jan,
       volume = {30},
        pages = {575-611},
          doi = {10.1146/annurev.aa.30.090192.003043},
       adsurl = {https://ui.adsabs.harvard.edu/abs/1992ARA&A..30..575C}
}

@ARTICLE{coppin05,
   author = {{Coppin}, K. and {Halpern}, M. and {Scott}, D. and {Borys}, C. and {Chapman}, S.},
    title = "{An 850-{$\mu$}m SCUBA map of the Groth Strip and reliable source extraction}",
  journal = {\mnras},
   eprint = {arXiv:astro-ph/0412267},
     year = 2005,
    month = mar,
   volume = 357,
    pages = {1022-1028},
      doi = {10.1111/j.1365-2966.2005.08723.x},
   adsurl = {http://adsabs.harvard.edu/abs/2005MNRAS.357.1022C}
}

@ARTICLE{cutri14,
       author = {{Cutri}, R.~M. and {et al.}},
        title = "{VizieR Online Data Catalog: AllWISE Data Release (Cutri+ 2013)}",
      journal = {VizieR Online Data Catalog},
         year = 2014,
        month = jan,
          eid = {II/328},
        pages = {II/328},
       adsurl = {https://ui.adsabs.harvard.edu/abs/2014yCat.2328....0C}
}

@ARTICLE{davies11,
       author = {{AMI Consortium} and {Davies}, Matthew L. and {Franzen}, Thomas M.~O. and {Waldram}, Elizabeth M. and {Grainge}, Keith J.~B. and {Hobson}, Michael P. and {Hurley-Walker}, Natasha and {Lasenby}, Anthony and {Olamaie}, Malak and {Pooley}, Guy G. and {Riley}, Julia M. and {Rodr{\'\i}guez-Gonz{\'a}lvez}, Carmen and {Saunders}, Richard D.~E. and {Scaife}, Anna M.~M. and {Schammel}, Michel P. and {Scott}, Paul F. and {Shimwell}, Timothy W. and {Titterington}, David J. and {Zwart}, Jonathan T.~L.},
        title = "{10C survey of radio sources at 15.7 GHz - II. First results}",
      journal = {\mnras},
         year = 2011,
        month = aug,
       volume = {415},
       number = {3},
        pages = {2708-2722},
          doi = {10.1111/j.1365-2966.2011.18925.x},
archivePrefix = {arXiv},
       eprint = {1012.3659},
 primaryClass = {astro-ph.CO},
       adsurl = {https://ui.adsabs.harvard.edu/abs/2011MNRAS.415.2708A}
}

@ARTICLE{dehaan16,
   author = {{de Haan}, T. and {Benson}, B.~A. and {Bleem}, L.~E. and {Allen}, S.~W. and 
	{Applegate}, D.~E. and {Ashby}, M.~L.~N. and {Bautz}, M. and 
	{Bayliss}, M. and {Bocquet}, S. and {Brodwin}, M. and {Carlstrom}, J.~E. and 
	{Chang}, C.~L. and {Chiu}, I. and {Cho}, H.-M. and {Clocchiatti}, A. and 
	{Crawford}, T.~M. and {Crites}, A.~T. and {Desai}, S. and {Dietrich}, J.~P. and 
	{Dobbs}, M.~A. and {Doucouliagos}, A.~N. and {Foley}, R.~J. and 
	{Forman}, W.~R. and {Garmire}, G.~P. and {George}, E.~M. and 
	{Gladders}, M.~D. and {Gonzalez}, A.~H. and {Gupta}, N. and 
	{Halverson}, N.~W. and {Hlavacek-Larrondo}, J. and {Hoekstra}, H. and 
	{Holder}, G.~P. and {Holzapfel}, W.~L. and {Hou}, Z. and {Hrubes}, J.~D. and 
	{Huang}, N. and {Jones}, C. and {Keisler}, R. and {Knox}, L. and 
	{Lee}, A.~T. and {Leitch}, E.~M. and {von der Linden}, A. and 
	{Luong-Van}, D. and {Mantz}, A. and {Marrone}, D.~P. and {McDonald}, M. and 
	{McMahon}, J.~J. and {Meyer}, S.~S. and {Mocanu}, L.~M. and 
	{Mohr}, J.~J. and {Murray}, S.~S. and {Padin}, S. and {Pryke}, C. and 
	{Rapetti}, D. and {Reichardt}, C.~L. and {Rest}, A. and {Ruel}, J. and 
	{Ruhl}, J.~E. and {Saliwanchik}, B.~R. and {Saro}, A. and {Sayre}, J.~T. and 
	{Schaffer}, K.~K. and {Schrabback}, T. and {Shirokoff}, E. and 
	{Song}, J. and {Spieler}, H.~G. and {Stalder}, B. and {Stanford}, S.~A. and 
	{Staniszewski}, Z. and {Stark}, A.~A. and {Story}, K.~T. and 
	{Stubbs}, C.~W. and {Vanderlinde}, K. and {Vieira}, J.~D. and 
	{Vikhlinin}, A. and {Williamson}, R. and {Zenteno}, A.},
    title = "{Cosmological Constraints from Galaxy Clusters in the 2500 Square-degree SPT-SZ Survey}",
  journal = {\apj},
archivePrefix = "arXiv",
   eprint = {1603.06522},
     year = 2016,
    month = nov,
   volume = 832,
      eid = {95},
    pages = {95},
      doi = {10.3847/0004-637X/832/1/95},
   adsurl = {http://adsabs.harvard.edu/abs/2016ApJ...832...95D}
}

@ARTICLE{depropris02,
       author = {{De Propris}, Roberto and {Couch}, Warrick J. and {Colless}, Matthew and {Dalton}, Gavin B. and {Collins}, Chris and {Baugh}, Carlton M. and {Bland-Hawthorn}, Joss and {Bridges}, Terry and {Cannon}, Russell and {Cole}, Shaun and {Cross}, Nicholas and {Deeley}, Kathryn and {Driver}, Simon P. and {Efstathiou}, George and {Ellis}, Richard S. and {Frenk}, Carlos S. and {Glazebrook}, Karl and {Jackson}, Carole and {Lahav}, Ofer and {Lewis}, Ian and {Lumsden}, Stuart and {Maddox}, Steve and {Madgwick}, Darren and {Moody}, Stephen and {Norberg}, Peder and {Peacock}, John A. and {Percival}, Will and {Peterson}, Bruce A. and {Sutherland}, Will and {Taylor}, Keith},
        title = "{The 2dF Galaxy Redshift Survey: a targeted study of catalogued clusters of galaxies}",
      journal = {\mnras},
         year = 2002,
        month = jan,
       volume = {329},
       number = {1},
        pages = {87-101},
          doi = {10.1046/j.1365-8711.2002.04958.x},
archivePrefix = {arXiv},
       eprint = {astro-ph/0109167},
 primaryClass = {astro-ph},
       adsurl = {https://ui.adsabs.harvard.edu/abs/2002MNRAS.329...87D}
}

@ARTICLE{dey19,
       author = {{Dey}, Arjun and {Schlegel}, David J. and {Lang}, Dustin and {Blum}, Robert and {Burleigh}, Kaylan and {Fan}, Xiaohui and {Findlay}, Joseph R. and {Finkbeiner}, Doug and {Herrera}, David and {Juneau}, St{\'e}phanie and {Landriau}, Martin and {Levi}, Michael and {McGreer}, Ian and {Meisner}, Aaron and {Myers}, Adam D. and {Moustakas}, John and {Nugent}, Peter and {Patej}, Anna and {Schlafly}, Edward F. and {Walker}, Alistair R. and {Valdes}, Francisco and {Weaver}, Benjamin A. and {Y{\`e}che}, Christophe and {Zou}, Hu and {Zhou}, Xu and {Abareshi}, Behzad and {Abbott}, T.~M.~C. and {Abolfathi}, Bela and {Aguilera}, C. and {Alam}, Shadab and {Allen}, Lori and {Alvarez}, A. and {Annis}, James and {Ansarinejad}, Behzad and {Aubert}, Marie and {Beechert}, Jacqueline and {Bell}, Eric F. and {BenZvi}, Segev Y. and {Beutler}, Florian and {Bielby}, Richard M. and {Bolton}, Adam S. and {Brice{\~n}o}, C{\'e}sar and {Buckley-Geer}, Elizabeth J. and {Butler}, Karen and {Calamida}, Annalisa and {Carlberg}, Raymond G. and {Carter}, Paul and {Casas}, Ricard and {Castander}, Francisco J. and {Choi}, Yumi and {Comparat}, Johan and {Cukanovaite}, Elena and {Delubac}, Timoth{\'e}e and {DeVries}, Kaitlin and {Dey}, Sharmila and {Dhungana}, Govinda and {Dickinson}, Mark and {Ding}, Zhejie and {Donaldson}, John B. and {Duan}, Yutong and {Duckworth}, Christopher J. and {Eftekharzadeh}, Sarah and {Eisenstein}, Daniel J. and {Etourneau}, Thomas and {Fagrelius}, Parker A. and {Farihi}, Jay and {Fitzpatrick}, Mike and {Font-Ribera}, Andreu and {Fulmer}, Leah and {G{\"a}nsicke}, Boris T. and {Gaztanaga}, Enrique and {George}, Koshy and {Gerdes}, David W. and {Gontcho}, Satya Gontcho A. and {Gorgoni}, Claudio and {Green}, Gregory and {Guy}, Julien and {Harmer}, Diane and {Hernandez}, M. and {Honscheid}, Klaus and {Huang}, Lijuan Wendy and {James}, David J. and {Jannuzi}, Buell T. and {Jiang}, Linhua and {Joyce}, Richard and {Karcher}, Armin and {Karkar}, Sonia and {Kehoe}, Robert and {Kneib}, Jean-Paul and {Kueter-Young}, Andrea and {Lan}, Ting-Wen and {Lauer}, Tod R. and {Le Guillou}, Laurent and {Le Van Suu}, Auguste and {Lee}, Jae Hyeon and {Lesser}, Michael and {Perreault Levasseur}, Laurence and {Li}, Ting S. and {Mann}, Justin L. and {Marshall}, Robert and {Mart{\'\i}nez-V{\'a}zquez}, C.~E. and {Martini}, Paul and {du Mas des Bourboux}, H{\'e}lion and {McManus}, Sean and {Meier}, Tobias Gabriel and {M{\'e}nard}, Brice and {Metcalfe}, Nigel and {Mu{\~n}oz-Guti{\'e}rrez}, Andrea and {Najita}, Joan and {Napier}, Kevin and {Narayan}, Gautham and {Newman}, Jeffrey A. and {Nie}, Jundan and {Nord}, Brian and {Norman}, Dara J. and {Olsen}, Knut A.~G. and {Paat}, Anthony and {Palanque-Delabrouille}, Nathalie and {Peng}, Xiyan and {Poppett}, Claire L. and {Poremba}, Megan R. and {Prakash}, Abhishek and {Rabinowitz}, David and {Raichoor}, Anand and {Rezaie}, Mehdi and {Robertson}, A.~N. and {Roe}, Natalie A. and {Ross}, Ashley J. and {Ross}, Nicholas P. and {Rudnick}, Gregory and {Safonova}, Sasha and {Saha}, Abhijit and {S{\'a}nchez}, F. Javier and {Savary}, Elodie and {Schweiker}, Heidi and {Scott}, Adam and {Seo}, Hee-Jong and {Shan}, Huanyuan and {Silva}, David R. and {Slepian}, Zachary and {Soto}, Christian and {Sprayberry}, David and {Staten}, Ryan and {Stillman}, Coley M. and {Stupak}, Robert J. and {Summers}, David L. and {Sien Tie}, Suk and {Tirado}, H. and {Vargas-Maga{\~n}a}, Mariana and {Vivas}, A. Katherina and {Wechsler}, Risa H. and {Williams}, Doug and {Yang}, Jinyi and {Yang}, Qian and {Yapici}, Tolga and {Zaritsky}, Dennis and {Zenteno}, A. and {Zhang}, Kai and {Zhang}, Tianmeng and {Zhou}, Rongpu and {Zhou}, Zhimin},
        title = "{Overview of the DESI Legacy Imaging Surveys}",
      journal = {\aj},
         year = 2019,
        month = may,
       volume = {157},
       number = {5},
          eid = {168},
        pages = {168},
          doi = {10.3847/1538-3881/ab089d},
archivePrefix = {arXiv},
       eprint = {1804.08657},
 primaryClass = {astro-ph.IM},
       adsurl = {https://ui.adsabs.harvard.edu/abs/2019AJ....157..168D}
}

@ARTICLE{diazsanchez17,
       author = {{D{\'\i}az-S{\'a}nchez}, A. and {Iglesias-Groth}, S. and {Rebolo}, R. and {Dannerbauer}, H.},
        title = "{Discovery of a Lensed Ultrabright Submillimeter Galaxy at z = 2.0439}",
      journal = {\apjl},
         year = 2017,
        month = jul,
       volume = {843},
       number = {2},
          eid = {L22},
        pages = {L22},
          doi = {10.3847/2041-8213/aa79ef},
archivePrefix = {arXiv},
       eprint = {1707.02454},
 primaryClass = {astro-ph.GA},
       adsurl = {https://ui.adsabs.harvard.edu/abs/2017ApJ...843L..22D}
}

@ARTICLE{diazsanchez21,
       author = {{D{\'\i}az-S{\'a}nchez}, A. and {Dannerbauer}, H. and {Sulzenauer}, N. and {Iglesias-Groth}, S. and {Rebolo}, R.},
        title = "{The Einstein Ring GAL-CLUS-022058s: a Lensed Ultrabright Submillimeter Galaxy at z = 1.4796}",
      journal = {\apj},
         year = 2021,
        month = sep,
       volume = {919},
       number = {1},
          eid = {48},
        pages = {48},
          doi = {10.3847/1538-4357/ac0f75},
archivePrefix = {arXiv},
       eprint = {2106.14281},
 primaryClass = {astro-ph.GA},
       adsurl = {https://ui.adsabs.harvard.edu/abs/2021ApJ...919...48D}
}

@ARTICLE{duchesne24,
       author = {{Duchesne}, S.~W. and {Grundy}, J.~A. and {Heald}, George H. and {Lenc}, Emil and {Leung}, James K. and {McConnell}, David and {Murphy}, Tara and {Pritchard}, Joshua and {Rose}, Kovi and {Thomson}, Alec J.~M. and {Wang}, Yuanming and {Wang}, Ziteng and {Whiting}, Matthew T.},
        title = "{The Rapid ASKAP Continuum Survey V: Cataloguing the sky at 1 367.5 MHz and the second data release of RACS-mid}",
      journal = {\pasa},
         year = 2024,
        month = jan,
       volume = {41},
          eid = {e003},
        pages = {e003},
          doi = {10.1017/pasa.2023.60},
archivePrefix = {arXiv},
       eprint = {2311.12369},
 primaryClass = {astro-ph.GA},
       adsurl = {https://ui.adsabs.harvard.edu/abs/2024PASA...41....3D}
}

@ARTICLE{duchesne25,
       author = {{Duchesne}, S. and {Ross}, K. and {Thomson}, A.~J.~M. and {Lenc}, E. and {Murphy}, T. and {Galvin}, T.~J. and {Hotan}, A.~W. and {Moss}, V.~A. and {Whiting}, M.~T.},
        title = "{The Rapid ASKAP Continuum Survey (RACS) VI: The RACS-high 1655.5 MHz images and catalogue.}",
      journal = {\pasa},
         year = 2025,
        month = jan,
       volume = {42},
          eid = {38},
        pages = {38},
          doi = {10.1017/pasa.2025.2},
archivePrefix = {arXiv},
       eprint = {2501.04978},
 primaryClass = {astro-ph.GA},
       adsurl = {https://ui.adsabs.harvard.edu/abs/2025PASA...42...38D}
}

@ARTICLE{dusserre25,
       author = {{Euclid Collaboration} and {Dusserre}, T. and {Dole}, H. and {Sarron}, F. and {Castignani}, G. and {Ramos-Chernenko}, N. and {Aghanim}, N. and {Garic}, A. and {Mellouki}, I. -E. and {Dagoneau}, N. and {Chapuis}, O. and {Frye}, B.~L. and {Polletta}, M. and {Dannerbauer}, H. and {Langer}, M. and {Maurin}, L. and {Soubrie}, E. and {Biviano}, A. and {Mei}, S. and {Mai}, N. and {Altieri}, B. and {Amara}, A. and {Andreon}, S. and {Auricchio}, N. and {Baccigalupi}, C. and {Baldi}, M. and {Balestra}, A. and {Bardelli}, S. and {Battaglia}, P. and {Bonchi}, A. and {Bonino}, D. and {Branchini}, E. and {Brescia}, M. and {Brinchmann}, J. and {Camera}, S. and {Ca{\~n}as-Herrera}, G. and {Capobianco}, V. and {Carbone}, C. and {Carretero}, J. and {Casas}, S. and {Castellano}, M. and {Cavuoti}, S. and {Chambers}, K.~C. and {Cimatti}, A. and {Colodro-Conde}, C. and {Congedo}, G. and {Conselice}, C.~J. and {Conversi}, L. and {Copin}, Y. and {Costille}, A. and {Courbin}, F. and {Courtois}, H.~M. and {Cropper}, M. and {Da Silva}, A. and {Degaudenzi}, H. and {De Lucia}, G. and {Di Giorgio}, A.~M. and {Dolding}, C. and {Dubath}, F. and {Duncan}, C.~A.~J. and {Dupac}, X. and {Dusini}, S. and {Ealet}, A. and {Escoffier}, S. and {Farina}, M. and {Farinelli}, R. and {Faustini}, F. and {Ferriol}, S. and {Finelli}, F. and {Fosalba}, P. and {Fotopoulou}, S. and {Frailis}, M. and {Franceschi}, E. and {Fumana}, M. and {Galeotta}, S. and {George}, K. and {Gillis}, B. and {Giocoli}, C. and {G\textbackslash'omez-Alvarez}, P. and {Gracia-Carpio}, J. and {Granett}, B.~R. and {Grazian}, A. and {Grupp}, F. and {Guzzo}, L. and {Gwyn}, S. and {Haugan}, S.~V.~H. and {Hoar}, J. and {Holmes}, W. and {Hormuth}, F. and {Hornstrup}, A. and {Hudelot}, P. and {Jahnke}, K. and {Jhabvala}, M. and {Joachimi}, B. and {Keih\textbackslash''anen}, E. and {Kermiche}, S. and {Kiessling}, A. and {Kubik}, B. and {Kuijken}, K. and {K\textbackslash''ummel}, M. and {Kunz}, M. and {Kurki-Suonio}, H. and {Le Boulc'h}, Q. and {Le Brun}, A.~M.~C. and {Le Mignant}, D. and {Ligori}, S. and {Lilje}, P.~B. and {Lindholm}, V. and {Lloro}, I. and {Mainetti}, G. and {Maino}, D. and {Maiorano}, E. and {Mansutti}, O. and {Marcin}, S. and {Marggraf}, O. and {Martinelli}, M. and {Martinet}, N. and {Marulli}, F. and {Massey}, R. and {Maurogordato}, S. and {Medinaceli}, E. and {Melchior}, M. and {Mellier}, Y. and {Meneghetti}, M. and {Merlin}, E. and {Meylan}, G. and {Mora}, A. and {Moresco}, M. and {Moscardini}, L. and {Nakajima}, R. and {Neissner}, C. and {Nichol}, R.~C. and {Niemi}, S. -M. and {Nightingale}, J.~W. and {Padilla}, C. and {Paltani}, S. and {Pasian}, F. and {Pedersen}, K. and {Percival}, W.~J. and {Pettorino}, V. and {Pires}, S. and {Polenta}, G. and {Poncet}, M. and {Popa}, L.~A. and {Pozzetti}, L. and {Raison}, F. and {Rebolo}, R. and {Renzi}, A. and {Rhodes}, J. and {Riccio}, G. and {Romelli}, E. and {Roncarelli}, M. and {Saglia}, R. and {Sakr}, Z. and {S\textbackslash'anchez}, A.~G. and {Sapone}, D. and {Sartoris}, B. and {Schewtschenko}, J.~A. and {Schirmer}, M. and {Schneider}, P. and {Schrabback}, T. and {Secroun}, A. and {Seidel}, G. and {Serrano}, S. and {Simon}, P. and {Sirignano}, C. and {Sirri}, G. and {Stanco}, L. and {Steinwagner}, J. and {Tallada-Cresp\textbackslash'i}, P. and {Taylor}, A.~N. and {Teplitz}, H.~I. and {Tereno}, I. and {Toft}, S. and {Toledo-Moreo}, R. and {Torradeflot}, F. and {Tutusaus}, I. and {Valenziano}, L. and {Valiviita}, J. and {Vassallo}, T. and {Verdoes Kleijn}, G. and {Veropalumbo}, A. and {Wang}, Y. and {Weller}, J. and {Zacchei}, A. and {Zamorani}, G. and {Zerbi}, F.~M. and {Zinchenko}, I.~A. and {Zucca}, E. and {Allevato}, V. and {Ballardini}, M. and {Bolzonella}, M. and {Bozzo}, E. and {Burigana}, C. and {Cabanac}, R. and {Cappi}, A. and {Di Ferdinando}, D. and {Escartin Vigo}, J.~A. and {Fabbian}, G. and {Gabarra}, L.},
        title = "{Euclid Quick Data Release (Q1). The Euclid view on Planck galaxy protocluster candidates: towards a probe of the highest sites of star formation at cosmic noon}",
      journal = {\aap, submitted},
         year = 2025,
        month = mar,
          eid = {arXiv:2503.21304},
        pages = {arXiv:2503.21304},
          doi = {10.48550/arXiv.2503.21304},
archivePrefix = {arXiv},
       eprint = {2503.21304},
 primaryClass = {astro-ph.CO},
       adsurl = {https://ui.adsabs.harvard.edu/abs/2025arXiv250321304E}
}

@ARTICLE{dutcher21,
       author = {{Dutcher}, D. and {Balkenhol}, L. and {Ade}, P.~A.~R. and {Ahmed}, Z. and {Anderes}, E. and {Anderson}, A.~J. and {Archipley}, M. and {Avva}, J.~S. and {Aylor}, K. and {Barry}, P.~S. and {Basu Thakur}, R. and {Benabed}, K. and {Bender}, A.~N. and {Benson}, B.~A. and {Bianchini}, F. and {Bleem}, L.~E. and {Bouchet}, F.~R. and {Bryant}, L. and {Byrum}, K. and {Carlstrom}, J.~E. and {Carter}, F.~W. and {Cecil}, T.~W. and {Chang}, C.~L. and {Chaubal}, P. and {Chen}, G. and {Cho}, H. -M. and {Chou}, T. -L. and {Cliche}, J. -F. and {Crawford}, T.~M. and {Cukierman}, A. and {Daley}, C. and {de Haan}, T. and {Denison}, E.~V. and {Dibert}, K. and {Ding}, J. and {Dobbs}, M.~A. and {Everett}, W. and {Feng}, C. and {Ferguson}, K.~R. and {Foster}, A. and {Fu}, J. and {Galli}, S. and {Gambrel}, A.~E. and {Gardner}, R.~W. and {Goeckner-Wald}, N. and {Gualtieri}, R. and {Guns}, S. and {Gupta}, N. and {Guyser}, R. and {Halverson}, N.~W. and {Harke-Hosemann}, A.~H. and {Harrington}, N.~L. and {Henning}, J.~W. and {Hilton}, G.~C. and {Hivon}, E. and {Holder}, G.~P. and {Holzapfel}, W.~L. and {Hood}, J.~C. and {Howe}, D. and {Huang}, N. and {Irwin}, K.~D. and {Jeong}, O.~B. and {Jonas}, M. and {Jones}, A. and {Khaire}, T.~S. and {Knox}, L. and {Kofman}, A.~M. and {Korman}, M. and {Kubik}, D.~L. and {Kuhlmann}, S. and {Kuo}, C. -L. and {Lee}, A.~T. and {Leitch}, E.~M. and {Lowitz}, A.~E. and {Lu}, C. and {Meyer}, S.~S. and {Michalik}, D. and {Millea}, M. and {Montgomery}, J. and {Nadolski}, A. and {Natoli}, T. and {Nguyen}, H. and {Noble}, G.~I. and {Novosad}, V. and {Omori}, Y. and {Padin}, S. and {Pan}, Z. and {Paschos}, P. and {Pearson}, J. and {Posada}, C.~M. and {Prabhu}, K. and {Quan}, W. and {Raghunathan}, S. and {Rahlin}, A. and {Reichardt}, C.~L. and {Riebel}, D. and {Riedel}, B. and {Rouble}, M. and {Ruhl}, J.~E. and {Sayre}, J.~T. and {Schiappucci}, E. and {Shirokoff}, E. and {Smecher}, G. and {Sobrin}, J.~A. and {Stark}, A.~A. and {Stephen}, J. and {Story}, K.~T. and {Suzuki}, A. and {Thompson}, K.~L. and {Thorne}, B. and {Tucker}, C. and {Umilta}, C. and {Vale}, L.~R. and {Vanderlinde}, K. and {Vieira}, J.~D. and {Wang}, G. and {Whitehorn}, N. and {Wu}, W.~L.~K. and {Yefremenko}, V. and {Yoon}, K.~W. and {Young}, M.~R. and {SPT-3G Collaboration}},
        title = "{Measurements of the E -mode polarization and temperature-E -mode correlation of the CMB from SPT-3G 2018 data}",
      journal = {\prd},
         year = 2021,
        month = jul,
       volume = {104},
       number = {2},
          eid = {022003},
        pages = {022003},
          doi = {10.1103/PhysRevD.104.022003},
archivePrefix = {arXiv},
       eprint = {2101.01684},
 primaryClass = {astro-ph.CO},
       adsurl = {https://ui.adsabs.harvard.edu/abs/2021PhRvD.104b2003D}
}

@ARTICLE{dutka17,
       author = {{Dutka}, Michael S. and {Carpenter}, Bryce D. and {Ojha}, Roopesh and {Finke}, Justin D. and {D'Ammando}, Filippo and {Kadler}, Matthias and {Edwards}, Philip G. and {Stevens}, Jamie and {Torresi}, Eleonora and {Grandi}, Paola and {Nesci}, Roberto and {Krau{\ss}}, Felicia and {M{\"u}ller}, Cornelia and {Wilms}, Joern and {Gehrels}, Neil},
        title = "{Multiband Observations of the Quasar PKS 2326-502 during Active and Quiescent Gamma-Ray States in 2010-2012}",
      journal = {\apj},
         year = 2017,
        month = feb,
       volume = {835},
       number = {2},
          eid = {182},
        pages = {182},
          doi = {10.3847/1538-4357/835/2/182},
archivePrefix = {arXiv},
       eprint = {1612.08061},
 primaryClass = {astro-ph.GA},
       adsurl = {https://ui.adsabs.harvard.edu/abs/2017ApJ...835..182D}
}

@ARTICLE{erler18,
   author = {{Erler}, J. and {Basu}, K. and {Chluba}, J. and {Bertoldi}, F.
	},
    title = "{Planck's view on the spectrum of the Sunyaev-Zeldovich effect}",
  journal = {\mnras},
archivePrefix = "arXiv",
   eprint = {1709.01187},
     year = 2018,
    month = may,
   volume = 476,
    pages = {3360-3381},
      doi = {10.1093/mnras/sty327},
   adsurl = {https://ui.adsabs.harvard.edu/abs/2018MNRAS.476.3360E}
}

@ARTICLE{everett20,
       author = {{Everett}, W.~B. and {Zhang}, L. and {Crawford}, T.~M. and
         {Vieira}, J.~D. and {Aravena}, M. and {Archipley}, M.~A. and
         {Austermann}, J.~E. and {Benson}, B.~A. and {Bleem}, L.~E. and
         {Carlstrom}, J.~E. and {Chang}, C.~L. and {Chapman}, S. and
         {Crites}, A.~T. and {de Haan}, T. and {Dobbs}, M.~A. and
         {George}, E.~M. and {Halverson}, N.~W. and {Harrington}, N. and
         {Holder}, G.~P. and {Holzapfel}, W.~L. and {Hrubes}, J.~D. and
         {Knox}, L. and {Lee}, A.~T. and {Luong-Van}, D. and {Mangian}, A.~C. and
         {Marrone}, D.~P. and {McMahon}, J.~J. and {Meyer}, S.~S. and
         {Mocanu}, L.~M. and {Mohr}, J.~J. and {Natoli}, T. and {Padin}, S. and
         {Pryke}, C. and {Reichardt}, C.~L. and {Reuter}, C.~A. and
         {Ruhl}, J.~E. and {Sayre}, J.~T. and {Schaffer}, K.~K. and
         {Shirokoff}, E. and {Spilker}, J.~S. and {Stalder}, B. and
         {Staniszewski}, Z. and {Stark}, A.~A. and {Story}, K.~T. and
         {Switzer}, E.~R. and {Vanderlinde}, K. and {Wei{\ss}}, A. and
         {Williamson}, R.},
        title = "{Millimeter-wave Point Sources from the 2500 Square Degree SPT-SZ Survey: Catalog and Population Statistics}",
      journal = {\apj},
         year = 2020,
        month = sep,
       volume = {900},
       number = {1},
          eid = {55},
        pages = {55},
          doi = {10.3847/1538-4357/ab9df7},
archivePrefix = {arXiv},
       eprint = {2003.03431},
 primaryClass = {astro-ph.IM},
       adsurl = {https://ui.adsabs.harvard.edu/abs/2020ApJ...900...55E}
}

@ARTICLE{flaugher15,
   author = {{Flaugher}, B. and {Diehl}, H.~T. and {Honscheid}, K. and {Abbott}, T.~M.~C. and 
	{Alvarez}, O. and {Angstadt}, R. and {Annis}, J.~T. and {Antonik}, M. and 
	{Ballester}, O. and {Beaufore}, L. and {Bernstein}, G.~M. and 
	{Bernstein}, R.~A. and {Bigelow}, B. and {Bonati}, M. and {Boprie}, D. and 
	{Brooks}, D. and {Buckley-Geer}, E.~J. and {Campa}, J. and {Cardiel-Sas}, L. and 
	{Castander}, F.~J. and {Castilla}, J. and {Cease}, H. and {Cela-Ruiz}, J.~M. and 
	{Chappa}, S. and {Chi}, E. and {Cooper}, C. and {da Costa}, L.~N. and 
	{Dede}, E. and {Derylo}, G. and {DePoy}, D.~L. and {de Vicente}, J. and 
	{Doel}, P. and {Drlica-Wagner}, A. and {Eiting}, J. and {Elliott}, A.~E. and 
	{Emes}, J. and {Estrada}, J. and {Fausti Neto}, A. and {Finley}, D.~A. and 
	{Flores}, R. and {Frieman}, J. and {Gerdes}, D. and {Gladders}, M.~D. and 
	{Gregory}, B. and {Gutierrez}, G.~R. and {Hao}, J. and {Holland}, S.~E. and 
	{Holm}, S. and {Huffman}, D. and {Jackson}, C. and {James}, D.~J. and 
	{Jonas}, M. and {Karcher}, A. and {Karliner}, I. and {Kent}, S. and 
	{Kessler}, R. and {Kozlovsky}, M. and {Kron}, R.~G. and {Kubik}, D. and 
	{Kuehn}, K. and {Kuhlmann}, S. and {Kuk}, K. and {Lahav}, O. and 
	{Lathrop}, A. and {Lee}, J. and {Levi}, M.~E. and {Lewis}, P. and 
	{Li}, T.~S. and {Mandrichenko}, I. and {Marshall}, J.~L. and 
	{Martinez}, G. and {Merritt}, K.~W. and {Miquel}, R. and {Mu{\~n}oz}, F. and 
	{Neilsen}, E.~H. and {Nichol}, R.~C. and {Nord}, B. and {Ogando}, R. and 
	{Olsen}, J. and {Palaio}, N. and {Patton}, K. and {Peoples}, J. and 
	{Plazas}, A.~A. and {Rauch}, J. and {Reil}, K. and {Rheault}, J.-P. and 
	{Roe}, N.~A. and {Rogers}, H. and {Roodman}, A. and {Sanchez}, E. and 
	{Scarpine}, V. and {Schindler}, R.~H. and {Schmidt}, R. and 
	{Schmitt}, R. and {Schubnell}, M. and {Schultz}, K. and {Schurter}, P. and 
	{Scott}, L. and {Serrano}, S. and {Shaw}, T.~M. and {Smith}, R.~C. and 
	{Soares-Santos}, M. and {Stefanik}, A. and {Stuermer}, W. and 
	{Suchyta}, E. and {Sypniewski}, A. and {Tarle}, G. and {Thaler}, J. and 
	{Tighe}, R. and {Tran}, C. and {Tucker}, D. and {Walker}, A.~R. and 
	{Wang}, G. and {Watson}, M. and {Weaverdyck}, C. and {Wester}, W. and 
	{Woods}, R. and {Yanny}, B. and {DES Collaboration}},
    title = "{The Dark Energy Camera}",
  journal = {\aj},
archivePrefix = "arXiv",
   eprint = {1504.02900},
 primaryClass = "astro-ph.IM",
     year = 2015,
    month = nov,
   volume = 150,
      eid = {150},
    pages = {150},
      doi = {10.1088/0004-6256/150/5/150},
   adsurl = {http://adsabs.harvard.edu/abs/2015AJ....150..150F}
}

@article{ge24,
  title         = {Cosmology from CMB lensing and delensed $EE$ power spectra using 2019--2020 SPT-3G polarization data},
  author        = {Ge, F. and Millea, M. and Camphuis, E. and Daley, C. and Huang, N. and Omori, Y. and Quan, W. and Anderes, E. and Anderson, A. J. and Ansarinejad, B. and Archipley, M. and Balkenhol, L. and Benabed, K. and Bender, A. N. and Benson, B. A. and Bianchini, F. and Bleem, L. E. and Bouchet, F. R. and Bryant, L. and Carlstrom, J. E. and Chang, C. L. and Chaubal, P. and Chen, G. and Chichura, P. M. and Chokshi, A. and Chou, T.-L. and Coerver, A. and Crawford, T. M. and de Haan, T. and Dibert, K. R. and Dobbs, M. A. and Doohan, M. and Doussot, A. and Dutcher, D. and Everett, W. and Feng, C. and Ferguson, K. R. and Fichman, K. and Foster, A. and Galli, S. and Gambrel, A. E. and Gardner, R. W. and Goeckner-Wald, N. and Gualtieri, R. and Guidi, F. and Guns, S. and Halverson, N. W. and Hivon, E. and Holder, G. P. and Holzapfel, W. L. and Hood, J. C. and Howe, D. and Hryciuk, A. and K\'eruzor\'e, F. and Khalife, A. R. and Knox, L. and Korman, M. and Kornoelje, K. and Kuo, C.-L. and Lee, A. T. and Levy, K. and Lowitz, A. E. and Lu, C. and Maniyar, A. and Martsen, E. S. and Menanteau, F. and Montgomery, J. and Nakato, Y. and Natoli, T. and Noble, G. I. and Pan, Z. and Paschos, P. and Phadke, K. A. and Pollak, A. W. and Prabhu, K. and Rahimi, M. and Rahlin, A. and Reichardt, C. L. and Riebel, D. and Rouble, M. and Ruhl, J. E. and Schiappucci, E. and Sobrin, J. A. and Stark, A. A. and Stephen, J. and Tandoi, C. and Thorne, B. and Trendafilova, C. and Umilta, C. and Vieira, J. D. and Vitrier, A. and Wan, Y. and Whitehorn, N. and Wu, W. L. K. and Young, M. R. and Zebrowski, J. A.},
  collaboration = {SPT-3G Collaboration},
  journal       = {Phys. Rev. D},
  volume        = {111},
  issue         = {8},
  pages         = {083534},
  numpages      = {40},
  year          = {2025},
  month         = {Apr},
  publisher     = {American Physical Society},
  doi           = {10.1103/PhysRevD.111.083534},
  url           = {https://link.aps.org/doi/10.1103/PhysRevD.111.083534}
}

@ARTICLE{giulietti24,
       author = {{Giulietti}, Marika and {Gandolfi}, Giovanni and {Massardi}, Marcella and {Behiri}, Meriem and {Lapi}, Andrea},
        title = "{Observing Dusty Star-Forming Galaxies at the Cosmic Noon through Gravitational Lensing: Perspectives from New-Generation Telescopes}",
      journal = {Galaxies},
         year = 2024,
        month = mar,
       volume = {12},
       number = {2},
          eid = {9},
        pages = {9},
          doi = {10.3390/galaxies12020009},
archivePrefix = {arXiv},
       eprint = {2403.02123},
 primaryClass = {astro-ph.GA},
       adsurl = {https://ui.adsabs.harvard.edu/abs/2024Galax..12....9G}
}

@ARTICLE{gladders00,
    author = {{Gladders}, M.\ D. and {Yee}, H.\ K.\ C.},
    title = "A New Method For Galaxy Cluster Detection. I. The Algorithm",
    journal = {\aj},
    year = 2000,
    month = oct,
    volume = 120,
    pages = {2148--2162},
    adsurl = {http://adsabs.harvard.edu/cgi-bin/nph-bib_query?bibcode=2000AJ....120.2148G&db_key=AST}
}

@article{guns21,
	doi = {10.3847/1538-4357/ac06a3},
	url = {https://doi.org/10.3847/1538-4357/ac06a3},
	year = 2021,
	month = {aug},
	publisher = {American Astronomical Society},
	volume = {916},
	number = {2},
	pages = {98},
	author = {S. Guns and A. Foster and C. Daley and A. Rahlin and N. Whitehorn and P. A. R. Ade and Z. Ahmed and E. Anderes and A. J. Anderson and M. Archipley and J. S. Avva and K. Aylor and L. Balkenhol and P. S. Barry and R. Basu Thakur and K. Benabed and A. N. Bender and B. A. Benson and F. Bianchini and L. E. Bleem and F. R. Bouchet and L. Bryant and K. Byrum and J. E. Carlstrom and F. W. Carter and T. W. Cecil and C. L. Chang and P. Chaubal and G. Chen and H.-M. Cho and T.-L. Chou and J.-F. Cliche and T. M. Crawford and A. Cukierman and T. de Haan and E. V. Denison and K. Dibert and J. Ding and M. A. Dobbs and D. Dutcher and W. Everett and C. Feng and K. R. Ferguson and J. Fu and S. Galli and A. E. Gambrel and R. W. Gardner and N. Goeckner-Wald and R. Gualtieri and N. Gupta and R. Guyser and N. W. Halverson and A. H. Harke-Hosemann and N. L. Harrington and J. W. Henning and G. C. Hilton and E. Hivon and G. P. Holder and W. L. Holzapfel and J. C. Hood and D. Howe and N. Huang and K. D. Irwin and O. B. Jeong and M. Jonas and A. Jones and T. S. Khaire and L. Knox and A. M. Kofman and M. Korman and D. L. Kubik and S. Kuhlmann and C.-L. Kuo and A. T. Lee and E. M. Leitch and A. E. Lowitz and C. Lu and D. P. Marrone and S. S. Meyer and D. Michalik and M. Millea and J. Montgomery and A. Nadolski and T. Natoli and H. Nguyen and G. I. Noble and V. Novosad and Y. Omori and S. Padin and Z. Pan and P. Paschos and J. Pearson and K. A. Phadke and C. M. Posada and K. Prabhu and W. Quan and C. L. Reichardt and D. Riebel and B. Riedel and M. Rouble and J. E. Ruhl and J. T. Sayre and E. Schiappucci and E. Shirokoff and G. Smecher and J. A. Sobrin and A. A. Stark and J. Stephen and K. T. Story and A. Suzuki and K. L. Thompson and B. Thorne and C. Tucker and C. Umilta and L. R. Vale and J. D. Vieira and G. Wang and W. L. K. Wu and V. Yefremenko and K. W. Yoon and M. R. Young and L. Zhang},
	title = {Detection of Galactic and Extragalactic Millimeter-wavelength Transient Sources with {SPT}-3G},
	journal = {The Astrophysical Journal}
}

@ARTICLE{hale21,
       author = {{Hale}, Catherine L. and {McConnell}, D. and {Thomson}, A.~J.~M. and {Lenc}, E. and {Heald}, G.~H. and {Hotan}, A.~W. and {Leung}, J.~K. and {Moss}, V.~A. and {Murphy}, T. and {Pritchard}, J. and {Sadler}, E.~M. and {Stewart}, A.~J. and {Whiting}, M.~T.},
        title = "{The Rapid ASKAP Continuum Survey Paper II: First Stokes I Source Catalogue Data Release}",
      journal = {\pasa},
         year = 2021,
        month = nov,
       volume = {38},
          eid = {e058},
        pages = {e058},
          doi = {10.1017/pasa.2021.47},
archivePrefix = {arXiv},
       eprint = {2109.00956},
 primaryClass = {astro-ph.GA},
       adsurl = {https://ui.adsabs.harvard.edu/abs/2021PASA...38...58H}
}

@ARTICLE{hayward21,
       author = {{Hayward}, Christopher C. and {Sparre}, Martin and {Chapman}, Scott C. and {Hernquist}, Lars and {Nelson}, Dylan and {Pakmor}, R{\"u}diger and {Pillepich}, Annalisa and {Springel}, Volker and {Torrey}, Paul and {Vogelsberger}, Mark and {Weinberger}, Rainer},
        title = "{Submillimetre galaxies in cosmological hydrodynamical simulations - an opportunity for constraining feedback models}",
      journal = {\mnras},
         year = 2021,
        month = apr,
       volume = {502},
       number = {2},
        pages = {2922-2933},
          doi = {10.1093/mnras/stab246},
archivePrefix = {arXiv},
       eprint = {2007.01885},
 primaryClass = {astro-ph.GA},
       adsurl = {https://ui.adsabs.harvard.edu/abs/2021MNRAS.502.2922H}
}

@ARTICLE{heitmann19,
       author = {{Heitmann}, Katrin and {Finkel}, Hal and {Pope}, Adrian and {Morozov}, Vitali and {Frontiere}, Nicholas and {Habib}, Salman and {Rangel}, Esteban and {Uram}, Thomas and {Korytov}, Danila and {Child}, Hillary and {Flender}, Samuel and {Insley}, Joe and {Rizzi}, Silvio},
        title = "{The Outer Rim Simulation: A Path to Many-core Supercomputers}",
      journal = {\apjs},
         year = 2019,
        month = nov,
       volume = {245},
       number = {1},
          eid = {16},
        pages = {16},
          doi = {10.3847/1538-4365/ab4da1},
archivePrefix = {arXiv},
       eprint = {1904.11970},
 primaryClass = {astro-ph.CO},
       adsurl = {https://ui.adsabs.harvard.edu/abs/2019ApJS..245...16H}
}

@ARTICLE{hilton21,
       author = {{Hilton}, M. and {Sif{\'o}n}, C. and {Naess}, S. and {Madhavacheril}, M. and {Oguri}, M. and {Rozo}, E. and {Rykoff}, E. and {Abbott}, T.~M.~C. and {Adhikari}, S. and {Aguena}, M. and {Aiola}, S. and {Allam}, S. and {Amodeo}, S. and {Amon}, A. and {Annis}, J. and {Ansarinejad}, B. and {Aros-Bunster}, C. and {Austermann}, J.~E. and {Avila}, S. and {Bacon}, D. and {Battaglia}, N. and {Beall}, J.~A. and {Becker}, D.~T. and {Bernstein}, G.~M. and {Bertin}, E. and {Bhandarkar}, T. and {Bhargava}, S. and {Bond}, J.~R. and {Brooks}, D. and {Burke}, D.~L. and {Calabrese}, E. and {Carrasco Kind}, M. and {Carretero}, J. and {Choi}, S.~K. and {Choi}, A. and {Conselice}, C. and {da Costa}, L.~N. and {Costanzi}, M. and {Crichton}, D. and {Crowley}, K.~T. and {D{\"u}nner}, R. and {Denison}, E.~V. and {Devlin}, M.~J. and {Dicker}, S.~R. and {Diehl}, H.~T. and {Dietrich}, J.~P. and {Doel}, P. and {Duff}, S.~M. and {Duivenvoorden}, A.~J. and {Dunkley}, J. and {Everett}, S. and {Ferraro}, S. and {Ferrero}, I. and {Fert{\'e}}, A. and {Flaugher}, B. and {Frieman}, J. and {Gallardo}, P.~A. and {Garc{\'\i}a-Bellido}, J. and {Gaztanaga}, E. and {Gerdes}, D.~W. and {Giles}, P. and {Golec}, J.~E. and {Gralla}, M.~B. and {Grandis}, S. and {Gruen}, D. and {Gruendl}, R.~A. and {Gschwend}, J. and {Gutierrez}, G. and {Han}, D. and {Hartley}, W.~G. and {Hasselfield}, M. and {Hill}, J.~C. and {Hilton}, G.~C. and {Hincks}, A.~D. and {Hinton}, S.~R. and {Ho}, S. -P.~P. and {Honscheid}, K. and {Hoyle}, B. and {Hubmayr}, J. and {Huffenberger}, K.~M. and {Hughes}, J.~P. and {Jaelani}, A.~T. and {Jain}, B. and {James}, D.~J. and {Jeltema}, T. and {Kent}, S. and {Knowles}, K. and {Koopman}, B.~J. and {Kuehn}, K. and {Lahav}, O. and {Lima}, M. and {Lin}, Y. -T. and {Lokken}, M. and {Loubser}, S.~I. and {MacCrann}, N. and {Maia}, M.~A.~G. and {Marriage}, T.~A. and {Martin}, J. and {McMahon}, J. and {Melchior}, P. and {Menanteau}, F. and {Miquel}, R. and {Miyatake}, H. and {Moodley}, K. and {Morgan}, R. and {Mroczkowski}, T. and {Nati}, F. and {Newburgh}, L.~B. and {Niemack}, M.~D. and {Nishizawa}, A.~J. and {Ogando}, R.~L.~C. and {Orlowski-Scherer}, J. and {Page}, L.~A. and {Palmese}, A. and {Partridge}, B. and {Paz-Chinch{\'o}n}, F. and {Phakathi}, P. and {Plazas}, A.~A. and {Robertson}, N.~C. and {Romer}, A.~K. and {Carnero Rosell}, A. and {Salatino}, M. and {Sanchez}, E. and {Schaan}, E. and {Schillaci}, A. and {Sehgal}, N. and {Serrano}, S. and {Shin}, T. and {Simon}, S.~M. and {Smith}, M. and {Soares-Santos}, M. and {Spergel}, D.~N. and {Staggs}, S.~T. and {Storer}, E.~R. and {Suchyta}, E. and {Swanson}, M.~E.~C. and {Tarle}, G. and {Thomas}, D. and {To}, C. and {Trac}, H. and {Ullom}, J.~N. and {Vale}, L.~R. and {Van Lanen}, J. and {Vavagiakis}, E.~M. and {De Vicente}, J. and {Wilkinson}, R.~D. and {Wollack}, E.~J. and {Xu}, Z. and {Zhang}, Y.},
        title = "{The Atacama Cosmology Telescope: A Catalog of >4000 Sunyaev-Zel{\textquoteright}dovich Galaxy Clusters}",
      journal = {\apjs},
         year = 2021,
        month = mar,
       volume = {253},
       number = {1},
          eid = {3},
        pages = {3},
          doi = {10.3847/1538-4365/abd023},
archivePrefix = {arXiv},
       eprint = {2009.11043},
 primaryClass = {astro-ph.CO},
       adsurl = {https://ui.adsabs.harvard.edu/abs/2021ApJS..253....3H}
}

@ARTICLE{hodge20,
       author = {{Hodge}, J.~A. and {da Cunha}, E.},
        title = "{High-redshift star formation in the Atacama large millimetre/submillimetre array era}",
      journal = {Royal Society Open Science},
         year = 2020,
        month = dec,
       volume = {7},
       number = {12},
          eid = {200556},
        pages = {200556},
          doi = {10.1098/rsos.200556},
archivePrefix = {arXiv},
       eprint = {2004.00934},
 primaryClass = {astro-ph.GA},
       adsurl = {https://ui.adsabs.harvard.edu/abs/2020RSOS....700556H}
}

@ARTICLE{hogbom74,
    author = {{H{\" o}gbom}, J.~A.},
    title = "{Aperture Synthesis with a Non-Regular Distribution of Interferometer Baselines}",
    journal = {\aaps},
    year = 1974,
    month = jun,
    volume = 15,
    pages = {417-426},
    url = {http://adsabs.harvard.edu/cgi-bin/nph-bib_query?bibcode=1974A%26AS...15..417H&db_key=AST}
}

@article{hughes98,
author = "Hughes, J.~P. and Birkinshaw, M.",
title = "Another X-Ray-Discovered Poor Cluster of
Galaxies Associated with CL 0016+16",
year = 1998,
journal = {\apj},
volume = 497,
pages = 645}

@ARTICLE{ivison07,
   author = {{Ivison}, R.~J. and {Greve}, T.~R. and {Dunlop}, J.~S. and {Peacock}, J.~A. and 
	{Egami}, E. and {Smail}, I. and {Ibar}, E. and {van Kampen}, E. and 
	{Aretxaga}, I. and {Babbedge}, T. and {Biggs}, A.~D. and {Blain}, A.~W. and 
	{Chapman}, S.~C. and {Clements}, D.~L. and {Coppin}, K. and 
	{Farrah}, D. and {Halpern}, M. and {Hughes}, D.~H. and {Jarvis}, M.~J. and 
	{Jenness}, T. and {Jones}, J.~R. and {Mortier}, A.~M.~J. and 
	{Oliver}, S. and {Papovich}, C. and {P{\'e}rez-Gonz{\'a}lez}, P.~G. and 
	{Pope}, A. and {Rawlings}, S. and {Rieke}, G.~H. and {Rowan-Robinson}, M. and 
	{Savage}, R.~S. and {Scott}, D. and {Seigar}, M. and {Serjeant}, S. and 
	{Simpson}, C. and {Stevens}, J.~A. and {Vaccari}, M. and {Wagg}, J. and 
	{Willott}, C.~J.},
    title = "{The SCUBA HAlf Degree Extragalactic Survey - III. Identification of radio and mid-infrared counterparts to submillimetre galaxies}",
  journal = {\mnras},
   eprint = {arXiv:astro-ph/0702544},
     year = 2007,
    month = sep,
   volume = 380,
    pages = {199-228},
      doi = {10.1111/j.1365-2966.2007.12044.x},
   adsurl = {http://adsabs.harvard.edu/abs/2007MNRAS.380..199I}
}

@INPROCEEDINGS{jonas16,
       author = {{Jonas}, J. and {MeerKAT Team}},
        title = "{The MeerKAT Radio Telescope}",
    booktitle = {MeerKAT Science: On the Pathway to the SKA},
         year = 2016,
        month = jan,
          eid = {1},
        pages = {1},
          doi = {10.22323/1.277.0001},
       adsurl = {https://ui.adsabs.harvard.edu/abs/2016mks..confE...1J}
}

@ARTICLE{klein18,
       author = {{Klein}, M. and {Mohr}, J.~J. and {Desai}, S. and {Israel}, H. and {Allam}, S. and {Benoit-L{\'e}vy}, A. and {Brooks}, D. and {Buckley-Geer}, E. and {Carnero Rosell}, A. and {Carrasco Kind}, M. and {Cunha}, C.~E. and {da Costa}, L.~N. and {Dietrich}, J.~P. and {Eifler}, T.~F. and {Evrard}, A.~E. and {Frieman}, J. and {Gruen}, D. and {Gruendl}, R.~A. and {Gutierrez}, G. and {Honscheid}, K. and {James}, D.~J. and {Kuehn}, K. and {Lima}, M. and {Maia}, M.~A.~G. and {March}, M. and {Melchior}, P. and {Menanteau}, F. and {Miquel}, R. and {Plazas}, A.~A. and {Reil}, K. and {Romer}, A.~K. and {Sanchez}, E. and {Santiago}, B. and {Scarpine}, V. and {Schubnell}, M. and {Sevilla-Noarbe}, I. and {Smith}, M. and {Soares-Santos}, M. and {Sobreira}, F. and {Suchyta}, E. and {Swanson}, M.~E.~C. and {Tarle}, G. and {DES Collaboration}},
        title = "{A multicomponent matched filter cluster confirmation tool for eROSITA: initial application to the RASS and DES-SV data sets}",
      journal = {\mnras},
         year = 2018,
        month = mar,
       volume = {474},
       number = {3},
        pages = {3324-3343},
          doi = {10.1093/mnras/stx2929},
archivePrefix = {arXiv},
       eprint = {1706.06577},
 primaryClass = {astro-ph.CO},
       adsurl = {https://ui.adsabs.harvard.edu/abs/2018MNRAS.474.3324K}
}

@ARTICLE{klein19,
       author = {{Klein}, M. and {Grandis}, S. and {Mohr}, J.~J. and {Paulus}, M. and
         {Abbott}, T.~M.~C. and {Annis}, J. and {Avila}, S. and {Bertin}, E. and
         {Brooks}, D. and {Buckley-Geer}, E. and {Rosell}, A. Carnero and
         {Kind}, M. Carrasco and {Carretero}, J. and {Castander}, F.~J. and
         {Cunha}, C.~E. and {D'Andrea}, C.~B. and {da Costa}, L.~N. and
         {De Vicente}, J. and {Desai}, S. and {Diehl}, H.~T. and
         {Dietrich}, J.~P. and {Doel}, P. and {Evrard}, A.~E. and
         {Flaugher}, B. and {Fosalba}, P. and {Frieman}, J. and
         {Garc{\'\i}a-Bellido}, J. and {Gaztanaga}, E. and {Giles}, P.~A. and
         {Gruen}, D. and {Gruendl}, R.~A. and {Gschwend}, J. and
         {Gutierrez}, G. and {Hartley}, W.~G. and {Hollowood}, D.~L. and
         {Honscheid}, K. and {Hoyle}, B. and {James}, D.~J. and {Jeltema}, T. and
         {Kuehn}, K. and {Kuropatkin}, N. and {Lima}, M. and {Maia}, M.~A.~G. and
         {March}, M. and {Marshall}, J.~L. and {Menanteau}, F. and {Miquel}, R. and
         {Ogando}, R.~L.~C. and {Plazas}, A.~A. and {Romer}, A.~K. and
         {Roodman}, A. and {Sanchez}, E. and {Scarpine}, V. and {Schindler}, R. and
         {Serrano}, S. and {Sevilla-Noarbe}, I. and {Smith}, M. and
         {Smith}, R.~C. and {Soares-Santos}, M. and {Sobreira}, F. and
         {Suchyta}, E. and {Swanson}, M.~E.~C. and {Tarle}, G. and {Thomas}, D. and
         {Vikram}, V. and {DES Collaboration}},
        title = "{A new RASS galaxy cluster catalogue with low contamination extending to z ̃ 1 in the DES overlap region}",
      journal = {\mnras},
         year = "2019",
        month = "Sep",
       volume = {488},
       number = {1},
        pages = {739-769},
          doi = {10.1093/mnras/stz1463},
archivePrefix = {arXiv},
       eprint = {1812.09956},
 primaryClass = {astro-ph.CO},
       adsurl = {https://ui.adsabs.harvard.edu/abs/2019MNRAS.488..739K}
}

@ARTICLE{klein23,
       author = {{Klein}, Matthias and {Hern{\'a}ndez-Lang}, Daniel and {Mohr}, Joseph J. and {Bocquet}, Sebastian and {Singh}, Aditya},
        title = "{RASS-MCMF: a full-sky X-ray selected galaxy cluster catalogue}",
      journal = {\mnras},
         year = 2023,
        month = dec,
       volume = {526},
       number = {3},
        pages = {3757-3778},
          doi = {10.1093/mnras/stad2729},
archivePrefix = {arXiv},
       eprint = {2305.20066},
 primaryClass = {astro-ph.CO},
       adsurl = {https://ui.adsabs.harvard.edu/abs/2023MNRAS.526.3757K}
}

@ARTICLE{klein24a,
       author = {{Klein}, M. and {Mohr}, J.~J. and {Bocquet}, S. and {Aguena}, M. and {Allen}, S.~W. and {Alves}, O. and {Ansarinejad}, B. and {Ashby}, M.~L.~N. and {Bacon}, D. and {Bayliss}, M. and {Benson}, B.~A. and {Bleem}, L.~E. and {Brodwin}, M. and {Brooks}, D. and {Bulbul}, E. and {Burke}, D.~L. and {Canning}, R.~E.~A. and {Carlstrom}, J.~E. and {Rosell}, A. Carnero and {Carretero}, J. and {Chang}, C.~L. and {Conselice}, C. and {Costanzi}, M. and {Crites}, A.~T. and {da Costa}, L.~N. and {Pereira}, M.~E.~S. and {Davis}, T.~M. and {De Vicente}, J. and {Desai}, S. and {de Haan}, T. and {Dobbs}, M.~A. and {Doel}, P. and {Ferrero}, I. and {Flores}, A.~M. and {Frieman}, J. and {George}, E.~M. and {Giannini}, G. and {Gladders}, M.~D. and {Gonzalez}, A.~H. and {Grandis}, S. and {Gruen}, D. and {Gruendl}, R.~A. and {Gutierrez}, G. and {Halverson}, N.~W. and {Hinton}, S.~R. and {Holder}, G.~P. and {Hollowood}, D.~L. and {Holzapfel}, W.~L. and {Honscheid}, K. and {Hrubes}, J.~D. and {Huang}, N. and {James}, D.~J. and {Khullar}, G. and {Kim}, K. and {Knox}, L. and {Kraft}, R. and {K{\'e}ruzor{\'e}}, F. and {Lee}, A.~T. and {Luong-Van}, D. and {Mahler}, G. and {Mantz}, A. and {Marrone}, D.~P. and {Marshall}, J.~L. and {McDonald}, M. and {McMahon}, J.~J. and {Mena-Fern{\'a}ndez}, J. and {Menanteau}, F. and {Meyer}, S.~S. and {Miquel}, R. and {Myles}, J. and {Padin}, S. and {Pieres}, A. and {Plazas Malag{\'o}n}, A.~A. and {Pryke}, C. and {Reichardt}, C.~L. and {Reil}, K. and {Roberson}, J. and {Romer}, A.~K. and {Romero}, C. and {Ruhl}, J.~E. and {Saliwanchik}, B.~R. and {Salvati}, L. and {Sanchez}, E. and {Saro}, A. and {Schaffer}, K.~K. and {Schrabback}, T. and {Schubnell}, M. and {Sevilla-Noarbe}, I. and {Sharon}, K. and {Shirokoff}, E. and {Smith}, M. and {Somboonpanyakul}, T. and {Stalder}, B. and {Stanford}, S.~A. and {Stark}, A.~A. and {Strazzullo}, V. and {Suchyta}, E. and {Swanson}, M.~E.~C. and {Tarle}, G. and {To}, C. and {Vanderlinde}, K. and {Vieira}, J.~D. and {von der Linden}, A. and {Weaverdyck}, N. and {Williamson}, R. and {Wiseman}, P. and {Young}, M.},
        title = "{SPT-SZ MCMF: an extension of the SPT-SZ catalogue over the DES region}",
      journal = {\mnras},
         year = 2024,
        month = jul,
       volume = {531},
       number = {4},
        pages = {3973-3990},
          doi = {10.1093/mnras/stae1359},
archivePrefix = {arXiv},
       eprint = {2309.09908},
 primaryClass = {astro-ph.CO},
       adsurl = {https://ui.adsabs.harvard.edu/abs/2024MNRAS.531.3973K}
}

@ARTICLE{klein24b,
       author = {{Klein}, M. and {Mohr}, J.~J. and {Davies}, C.~T.},
        title = "{The ACT-DR5 MCMF galaxy cluster catalog}",
      journal = {\aap},
         year = 2024,
        month = oct,
       volume = {690},
          eid = {A322},
        pages = {A322},
          doi = {10.1051/0004-6361/202451203},
archivePrefix = {arXiv},
       eprint = {2406.14754},
 primaryClass = {astro-ph.CO},
       adsurl = {https://ui.adsabs.harvard.edu/abs/2024A&A...690A.322K}
}

@ARTICLE{klein25,
       author = {{Klein}, M. and {George}, K. and {Mohr}, J.~J. and {Altieri}, B. and {Amendola}, L. and {Andreon}, S. and {Auricchio}, N. and {Baccigalupi}, C. and {Baldi}, M. and {Balestra}, A. and {Bardelli}, S. and {Biviano}, A. and {Branchini}, E. and {Brescia}, M. and {Camera}, S. and {Ca{\~n}as-Herrera}, G. and {Capobianco}, V. and {Carbone}, C. and {Carretero}, J. and {Casas}, S. and {Castellano}, M. and {Castignani}, G. and {Cavuoti}, S. and {Chambers}, K.~C. and {Cimatti}, A. and {Colodro-Conde}, C. and {Congedo}, G. and {Conversi}, L. and {Copin}, Y. and {Courbin}, F. and {Courtois}, H.~M. and {Cropper}, M. and {Da Silva}, A. and {Degaudenzi}, H. and {De Lucia}, G. and {Dolding}, C. and {Dole}, H. and {Dubath}, F. and {Ducret}, F. and {Dupac}, X. and {Dusini}, S. and {Farina}, M. and {Farinelli}, R. and {Faustini}, F. and {Ferriol}, S. and {Finelli}, F. and {Frailis}, M. and {Franceschi}, E. and {Fumana}, M. and {Galeotta}, S. and {Gillis}, B. and {Giocoli}, C. and {Gracia-Carpio}, J. and {Grazian}, A. and {Grupp}, F. and {Haugan}, S.~V.~H. and {Holmes}, W. and {Hook}, I.~M. and {Hormuth}, F. and {Hornstrup}, A. and {Jahnke}, K. and {Jhabvala}, M. and {Keih{\"a}nen}, E. and {Kermiche}, S. and {Kubik}, B. and {K{\"u}mmel}, M. and {Kunz}, M. and {Kurki-Suonio}, H. and {Le Brun}, A.~M.~C. and {Le Mignant}, D. and {Ligori}, S. and {Lilje}, P.~B. and {Lindholm}, V. and {Lloro}, I. and {Mainetti}, G. and {Maino}, D. and {Maiorano}, E. and {Mansutti}, O. and {Marggraf}, O. and {Martinelli}, M. and {Martinet}, N. and {Marulli}, F. and {Massey}, R. and {Maurogordato}, S. and {Medinaceli}, E. and {Mei}, S. and {Mellier}, Y. and {Meneghetti}, M. and {Merlin}, E. and {Meylan}, G. and {Moscardini}, L. and {Nakajima}, R. and {Neissner}, C. and {Niemi}, S. -M. and {Padilla}, C. and {Paltani}, S. and {Pasian}, F. and {Pedersen}, K. and {Percival}, W.~J. and {Pettorino}, V. and {Pires}, S. and {Polenta}, G. and {Poncet}, M. and {Popa}, L.~A. and {Pozzetti}, L. and {Raison}, F. and {Rebolo}, R. and {Renzi}, A. and {Rhodes}, J. and {Riccio}, G. and {Romelli}, E. and {Roncarelli}, M. and {Rosset}, C. and {Rottgering}, H.~J.~A. and {Saglia}, R. and {Sakr}, Z. and {Sapone}, D. and {Schirmer}, M. and {Schneider}, P. and {Schrabback}, T. and {Secroun}, A. and {Sefusatti}, E. and {Seidel}, G. and {Serrano}, S. and {Sirignano}, C. and {Sirri}, G. and {Stanco}, L. and {Steinwagner}, J. and {Tallada-Cresp{\'\i}}, P. and {Taylor}, A.~N. and {Tereno}, I. and {Toft}, S. and {Toledo-Moreo}, R. and {Torradeflot}, F. and {Tutusaus}, I. and {Valenziano}, L. and {Valiviita}, J. and {Vassallo}, T. and {Veropalumbo}, A. and {Wang}, Y. and {Weller}, J. and {Zerbi}, F.~M. and {Zucca}, E. and {Burigana}, C. and {Scottez}, V. and {Sereno}, M. and {Viel}, M.},
        title = "{Euclid: Quick Data Release (Q1) -- Watching ICM-selected galaxy clusters with Euclid eyes -- prospects of Euclid data in the context of large SZ and X-ray based surveys}",
      journal = {arXiv e-prints},
         year = 2025,
        month = jun,
          eid = {arXiv:2506.19566},
        pages = {arXiv:2506.19566},
          doi = {10.48550/arXiv.2506.19566},
archivePrefix = {arXiv},
       eprint = {2506.19566},
 primaryClass = {astro-ph.CO},
       adsurl = {https://ui.adsabs.harvard.edu/abs/2025arXiv250619566K}
}

@article{kluge24,
   title={The SRG/eROSITA All-Sky Survey: Optical identification and properties of galaxy clusters and groups in the western galactic hemisphere},
   volume={688},
   ISSN={1432-0746},
   url={http://dx.doi.org/10.1051/0004-6361/202349031},
   DOI={10.1051/0004-6361/202349031},
   journal={Astronomy \& Astrophysics},
   publisher={EDP Sciences},
   author={Kluge, M. and Comparat, J. and Liu, A. and Balzer, F. and Bulbul, E. and Ider Chitham, J. and Ghirardini, V. and Garrel, C. and Bahar, Y. E. and Artis, E. and Bender, R. and Clerc, N. and Dwelly, T. and Fabricius, M. H. and Grandis, S. and Hernández-Lang, D. and Hill, G. J. and Joshi, J. and Lamer, G. and Merloni, A. and Nandra, K. and Pacaud, F. and Predehl, P. and Ramos-Ceja, M. E. and Reiprich, T. H. and Salvato, M. and Sanders, J. S. and Schrabback, T. and Seppi, R. and Zelmer, S. and Zenteno, A. and Zhang, X.},
   year={2024},
   month=aug, pages={A210} }

@ARTICLE{kornoelje25,
       author = {{Kornoelje}, K. and {Bleem}, L.~E. and {Rykoff}, E.~S. and {Abbott}, T.~M.~C. and {Ade}, P.~A.~R. and {Aguena}, M. and {Alves}, O. and {Anderson}, A.~J. and {Andrade-Oliveira}, F. and {Ansarinejad}, B. and {Archipley}, M. and {Ashby}, M.~L.~N. and {Austermann}, J.~E. and {Bacon}, D. and {Balkenhol}, L. and {Beall}, J.~A. and {Benabed}, K. and {Bender}, A.~N. and {Benson}, B.~A. and {Bianchini}, F. and {Bocquet}, S. and {Bouchet}, F.~R. and {Brooks}, D. and {Burke}, D.~L. and {Calzadilla}, M. and {Camphuis}, E. and {Carlstrom}, J.~E. and {Carnero Rosell}, A. and {Carretero}, J. and {Chang}, C.~L. and {Chaubal}, P. and {Chiang}, H.~C. and {Chichura}, P.~M. and {Chokshi}, A. and {Chou}, T. -L. and {Citron}, R. and {Coerver}, A. and {Corbett Moran}, C. and {Costanzi}, M. and {Crawford}, T.~M. and {Crites}, A.~T. and {da Costa}, L.~N. and {Daley}, C. and {de Haan}, T. and {De Vicente}, J. and {Desai}, S. and {Dibert}, K.~R. and {Dobbs}, M.~A. and {Doel}, P. and {Doohan}, M. and {Doussot}, A. and {Dutcher}, D. and {Everett}, W. and {Everett}, S. and {Feng}, C. and {Ferguson}, K.~R. and {Ferrero}, I. and {Fichman}, K. and {Flaugher}, B. and {Floyd}, B. and {Foster}, A. and {Friedel}, D. and {Frieman}, J. and {Galli}, S. and {Gallicchio}, J. and {Gambrel}, A.~E. and {Garc'ia-Bellido}, J. and {Gardner}, R.~W. and {Gassis}, R. and {Gatti}, M. and {Ge}, F. and {George}, E.~M. and {Giannini}, G. and {Goeckner-Wald}, N. and {Grandis}, S. and {Gruen}, D. and {Gruendl}, R.~A. and {Gualtieri}, R. and {Guidi}, F. and {Mahler}, Guillaume and {Guns}, S. and {Gupta}, N. and {Gutierrez}, G. and {Halverson}, N.~W. and {Hinton}, S.~R. and {Hivon}, E. and {Holder}, G.~P. and {Hollowood}, D.~L. and {Holzapfel}, W.~L. and {Honscheid}, K. and {Hood}, J.~C. and {Hrubes}, J.~D. and {Hryciuk}, A. and {Huang}, N. and {Hubmayr}, J. and {Irwin}, K.~D. and {Mena-Fern\textbackslash'andez}, J. and {James}, D.~J. and {K\textbackslash'eruzor\textbackslash'e}, F. and {Khalife}, A.~R. and {Klein}, M. and {Knox}, L. and {Korman}, M. and {Kuehn}, K. and {Kuo}, C. -L. and {Lahav}, O. and {Lee}, A.~T. and {Lee}, S. and {Levy}, K. and {Li}, D. and {Lima}, M. and {Lowitz}, A.~E. and {Lowitz}, A. and {Lu}, C. and {Maniyar}, A. and {Marshal}, J.~L. and {Marshall}, J.~L. and {Martsen}, E.~S. and {Bayliss}, Matthew B. and {McDonald}, M. and {McMahon}, J.~J. and {Menanteau}, F. and {Millea}, M. and {Miquel}, R. and {Mohr}, J.~J. and {Montgomery}, J. and {Myles}, J. and {Nakato}, Y. and {Natoli}, T. and {Nibarger}, J.~P. and {Noble}, G.~I. and {Novosad}, V. and {Ogando}, R.~L.~C. and {Omori}, Y. and {Ouellette}, A. and {Padin}, S. and {Pan}, Z. and {Patil}, S. and {Pereira}, M.~E.~S. and {Phadke}, K.~A. and {Pieres}, A. and {Plazas Malag'on}, A.~A. and {Pollak}, A.~W. and {Prabhu}, K. and {Pryke}, C. and {Quan}, W. and {Raghunathan}, S. and {Rahimi}, M. and {Rahlin}, A. and {Reichardt}, C.~L. and {Rodr'iguez-Monroy}, M. and {Romer}, A.~K. and {Rouble}, M. and {Ruhl}, J.~E. and {Saliwanchik}, B.~R. and {Salvati}, L. and {Samuroff}, S. and {Sanchez}, E. and {Saro}, A. and {Schaffer}, K.~K. and {Schiappucci}, E. and {Schrabback}, T. and {Sevilla-Noarbe}, I. and {Sievers}, C. and {Smecher}, G. and {Smith}, M. and {Sobrin}, J.~A. and {Somboonpanyakul}, T. and {Stalder}, B. and {Stark}, A.~A. and {Suchyta}, E. and {Swanson}, M.~E.~C. and {Tandoi}, C. and {Tarle}, G. and {Thorne}, B. and {To}, C. and {Trendafilova}, C. and {Tucker}, C. and {Umilta}, C. and {Veach}, T. and {Vieira}, J.~D. and {Vincenzi}, M. and {Vitrier}, A. and {Wan}, Y. and {Wang}, G. and {Weaverdyck}, N. and {Weller}, J. and {Whitehorn}, N. and {Wiseman}, P. and {Wu}, W.~L.~K. and {Yefremenko}, V. and {Young}, M.~R. and {Zebrowski}, J.~A. and {Zhang}, Y.},
        title = "{The SPT-Deep Cluster Catalog: Sunyaev-Zel'dovich Selected Clusters from Combined SPT-3G and SPTpol Measurements over 100 Square Degrees}",
      journal = {\apj, submitted},
         year = 2025,
        month = mar,
          eid = {arXiv:2503.17271},
        pages = {arXiv:2503.17271},
          doi = {10.48550/arXiv.2503.17271},
archivePrefix = {arXiv},
       eprint = {2503.17271},
 primaryClass = {astro-ph.CO},
       adsurl = {https://ui.adsabs.harvard.edu/abs/2025arXiv250317271K}
}

@ARTICLE{kravtsov12,
   author = {{Kravtsov}, A.~V. and {Borgani}, S.},
    title = "{Formation of Galaxy Clusters}",
  journal = {\araa},
   eprint = {arXiv:astro-ph/1205.5556},
     year = 2012,
    month = sep,
   volume = 50,
    pages = {353-409},
      doi = {10.1146/annurev-astro-081811-125502},
   adsurl = {http://adsabs.harvard.edu/abs/2012ARA\%26A..50..353K},
}

@ARTICLE{koprowski17,
       author = {{Koprowski}, M.~P. and {Dunlop}, J.~S. and {Micha{\l}owski}, M.~J. and {Coppin}, K.~E.~K. and {Geach}, J.~E. and {McLure}, R.~J. and {Scott}, D. and {van der Werf}, P.~P.},
        title = "{The evolving far-IR galaxy luminosity function and dust-obscured star formation rate density out to z≃5.}",
      journal = {\mnras},
         year = 2017,
        month = nov,
       volume = {471},
       number = {4},
        pages = {4155-4169},
          doi = {10.1093/mnras/stx1843},
archivePrefix = {arXiv},
       eprint = {1706.00426},
 primaryClass = {astro-ph.GA},
       adsurl = {https://ui.adsabs.harvard.edu/abs/2017MNRAS.471.4155K}
}

@ARTICLE{lagache20,
       author = {{Lagache}, G. and {B{\'e}thermin}, M. and {Montier}, L. and {Serra}, P. and {Tucci}, M.},
        title = "{Impact of polarised extragalactic sources on the measurement of CMB B-mode anisotropies}",
      journal = {\aap},
         year = 2020,
        month = oct,
       volume = {642},
          eid = {A232},
        pages = {A232},
          doi = {10.1051/0004-6361/201937147},
archivePrefix = {arXiv},
       eprint = {1911.09466},
 primaryClass = {astro-ph.CO},
       adsurl = {https://ui.adsabs.harvard.edu/abs/2020A&A...642A.232L}
}

@ARTICLE{madau14,
   author = {{Madau}, P. and {Dickinson}, M.},
    title = "{Cosmic Star-Formation History}",
  journal = {\araa},
archivePrefix = "arXiv",
   eprint = {1403.0007},
     year = 2014,
    month = aug,
   volume = 52,
    pages = {415-486},
      doi = {10.1146/annurev-astro-081811-125615},
   adsurl = {http://adsabs.harvard.edu/abs/2014ARA%26A..52..415M}
}

@ARTICLE{mahony11,
       author = {{Mahony}, Elizabeth K. and {Sadler}, Elaine M. and {Croom}, Scott M. and {Ekers}, Ronald D. and {Bannister}, Keith W. and {Chhetri}, Rajan and {Hancock}, Paul J. and {Johnston}, Helen M. and {Massardi}, Marcella and {Murphy}, Tara},
        title = "{Optical properties of high-frequency radio sources from the Australia Telescope 20 GHz (AT20G) Survey}",
      journal = {\mnras},
         year = 2011,
        month = nov,
       volume = {417},
       number = {4},
        pages = {2651-2675},
          doi = {10.1111/j.1365-2966.2011.19427.x},
archivePrefix = {arXiv},
       eprint = {1107.2454},
 primaryClass = {astro-ph.GA},
       adsurl = {https://ui.adsabs.harvard.edu/abs/2011MNRAS.417.2651M}
}

@ARTICLE{maturi18,
       author = {{Maturi}, Matteo and {Bellagamba}, Fabio and {Radovich}, Mario and {Roncarelli}, Mauro and {Sereno}, Mauro and {Moscardini}, Lauro and {Bardelli}, Sandro and {Puddu}, Emanuella},
        title = "{AMICO galaxy clusters in KiDS-DR3: sample properties and selection function}",
      journal = {\mnras},
         year = 2019,
        month = may,
       volume = {485},
       number = {1},
        pages = {498-512},
          doi = {10.1093/mnras/stz294},
archivePrefix = {arXiv},
       eprint = {1810.02811},
 primaryClass = {astro-ph.CO},
       adsurl = {https://ui.adsabs.harvard.edu/abs/2019MNRAS.485..498M}
}

@ARTICLE{melin06,
   author = {{Melin}, J.-B. and {Bartlett}, J.~G. and {Delabrouille}, J.},
    title = "{Catalog extraction in SZ cluster surveys: a matched filter approach}",
  journal = {\aap},
   eprint = {arXiv:astro-ph/0602424},
     year = 2006,
    month = nov,
   volume = 459,
    pages = {341-352},
      doi = {10.1051/0004-6361:20065034},
   adsurl = {http://adsabs.harvard.edu/abs/2006A%26A...459..341M}
}

@ARTICLE{mellier25,
       author = {{Euclid Collaboration} and {Mellier}, Y. and {Abdurro'uf} and {Acevedo Barroso}, J.~A. and {Ach{\'u}carro}, A. and {Adamek}, J. and {Adam}, R. and {Addison}, G.~E. and {Aghanim}, N. and {Aguena}, M. and {Ajani}, V. and {Akrami}, Y. and {Al-Bahlawan}, A. and {Alavi}, A. and {Albuquerque}, I.~S. and {Alestas}, G. and {Alguero}, G. and {Allaoui}, A. and {Allen}, S.~W. and {Allevato}, V. and {Alonso-Tetilla}, A.~V. and {Altieri}, B. and {Alvarez-Candal}, A. and {Alvi}, S. and {Amara}, A. and {Amendola}, L. and {Amiaux}, J. and {Andika}, I.~T. and {Andreon}, S. and {Andrews}, A. and {Angora}, G. and {Angulo}, R.~E. and {Annibali}, F. and {Anselmi}, A. and {Anselmi}, S. and {Arcari}, S. and {Archidiacono}, M. and {Aric{\`o}}, G. and {Arnaud}, M. and {Arnouts}, S. and {Asgari}, M. and {Asorey}, J. and {Atayde}, L. and {Atek}, H. and {Atrio-Barandela}, F. and {Aubert}, M. and {Aubourg}, E. and {Auphan}, T. and {Auricchio}, N. and {Aussel}, B. and {Aussel}, H. and {Avelino}, P.~P. and {Avgoustidis}, A. and {Avila}, S. and {Awan}, S. and {Azzollini}, R. and {Baccigalupi}, C. and {Bachelet}, E. and {Bacon}, D. and {Baes}, M. and {Bagley}, M.~B. and {Bahr-Kalus}, B. and {Balaguera-Antolinez}, A. and {Balbinot}, E. and {Balcells}, M. and {Baldi}, M. and {Baldry}, I. and {Balestra}, A. and {Ballardini}, M. and {Ballester}, O. and {Balogh}, M. and {Ba{\~n}ados}, E. and {Barbier}, R. and {Bardelli}, S. and {Baron}, M. and {Barreiro}, T. and {Barrena}, R. and {Barriere}, J.-C. and {Barros}, B.~J. and {Barthelemy}, A. and {Bartolo}, N. and {Basset}, A. and {Battaglia}, P. and {Battisti}, A.~J. and {Baugh}, C.~M. and {Baumont}, L. and {Bazzanini}, L. and {Beaulieu}, J.-P. and {Beckmann}, V. and {Belikov}, A.~N. and {Bel}, J. and {Bellagamba}, F. and {Bella}, M. and {Bellini}, E. and {Benabed}, K. and {Bender}, R. and {Benevento}, G. and {Bennett}, C.~L. and {Benson}, K. and {Bergamini}, P. and {Bermejo-Climent}, J.~R. and {Bernardeau}, F. and {Bertacca}, D. and {Berthe}, M. and {Berthier}, J. and {Bethermin}, M. and {Beutler}, F. and {Bevillon}, C. and {Bhargava}, S. and {Bhatawdekar}, R. and {Bianchi}, D. and {Bisigello}, L. and {Biviano}, A. and {Blake}, R.~P. and {Blanchard}, A. and {Blazek}, J. and {Blot}, L. and {Bosco}, A. and {Bodendorf}, C. and {Boenke}, T. and {B{\"o}hringer}, H. and {Boldrini}, P. and {Bolzonella}, M. and {Bonchi}, A. and {Bonici}, M. and {Bonino}, D. and {Bonino}, L. and {Bonvin}, C. and {Bon}, W. and {Booth}, J.~T. and {Borgani}, S. and {Borlaff}, A.~S. and {Borsato}, E. and {Bose}, B. and {Botticella}, M.~T. and {Boucaud}, A. and {Bouche}, F. and {Boucher}, J.~S. and {Boutigny}, D. and {Bouvard}, T. and {Bouwens}, R. and {Bouy}, H. and {Bowler}, R.~A.~A. and {Bozza}, V. and {Bozzo}, E. and {Branchini}, E. and {Brando}, G. and {Brau-Nogue}, S. and {Brekke}, P. and {Bremer}, M.~N. and {Brescia}, M. and {Breton}, M.-A. and {Brinchmann}, J. and {Brinckmann}, T. and {Brockley-Blatt}, C. and {Brodwin}, M. and {Brouard}, L. and {Brown}, M.~L. and {Bruton}, S. and {Bucko}, J. and {Buddelmeijer}, H. and {Buenadicha}, G. and {Buitrago}, F. and {Burger}, P. and {Burigana}, C. and {Busillo}, V. and {Busonero}, D. and {Cabanac}, R. and {Cabayol-Garcia}, L. and {Cagliari}, M.~S. and {Caillat}, A. and {Caillat}, L. and {Calabrese}, M. and {Calabro}, A. and {Calderone}, G. and {Calura}, F. and {Camacho Quevedo}, B. and {Camera}, S. and {Campos}, L. and {Ca{\~n}as-Herrera}, G. and {Candini}, G.~P. and {Cantiello}, M. and {Capobianco}, V. and {Cappellaro}, E. and {Cappelluti}, N. and {Cappi}, A. and {Caputi}, K.~I. and {Cara}, C. and {Carbone}, C. and {Cardone}, V.~F. and {Carella}, E. and {Carlberg}, R.~G. and {Carle}, M. and {Carminati}, L. and {Caro}, F. and {Carrasco}, J.~M. and {Carretero}, J. and {Carrilho}, P. and {Carron Duque}, J. and {Carry}, B.},
        title = "{Euclid: I. Overview of the Euclid mission}",
      journal = {\aap},
         year = 2025,
        month = may,
       volume = {697},
          eid = {A1},
        pages = {A1},
          doi = {10.1051/0004-6361/202450810},
archivePrefix = {arXiv},
       eprint = {2405.13491},
 primaryClass = {astro-ph.CO},
       adsurl = {https://ui.adsabs.harvard.edu/abs/2025A&A...697A...1E}
}

@ARTICLE{miller18,
   author = {{Miller}, T.~B. and {Chapman}, S.~C. and {Aravena}, M. and {Ashby}, M.~L.~N. and 
	{Hayward}, C.~C. and {Vieira}, J.~D. and {Wei{\ss}}, A. and 
	{Babul}, A. and {B{\'e}thermin}, M. and {Bradford}, C.~M. and 
	{Brodwin}, M. and {Carlstrom}, J.~E. and {Chen}, C.-C. and {Cunningham}, D.~J.~M. and 
	{De Breuck}, C. and {Gonzalez}, A.~H. and {Greve}, T.~R. and 
	{Harnett}, J. and {Hezaveh}, Y. and {Lacaille}, K. and {Litke}, K.~C. and 
	{Ma}, J. and {Malkan}, M. and {Marrone}, D.~P. and {Morningstar}, W. and 
	{Murphy}, E.~J. and {Narayanan}, D. and {Pass}, E. and {Perry}, R. and 
	{Phadke}, K.~A. and {Rennehan}, D. and {Rotermund}, K.~M. and 
	{Simpson}, J. and {Spilker}, J.~S. and {Sreevani}, J. and {Stark}, A.~A. and 
	{Strandet}, M.~L. and {Strom}, A.~L.},
    title = "{A massive core for a cluster of galaxies at a redshift of 4.3}",
  journal = {\nat},
archivePrefix = "arXiv",
   eprint = {1804.09231},
     year = 2018,
    month = apr,
   volume = 556,
    pages = {469-472},
      doi = {10.1038/s41586-018-0025-2},
   adsurl = {http://adsabs.harvard.edu/abs/2018Natur.556..469M}
}

@ARTICLE{mitra24,
       author = {{Mitra}, Dipanjan and {Negrello}, Mattia and {De Zotti}, Gianfranco and {Cai}, Zhen-Yi},
        title = "{Euclid view of the dusty star-forming galaxies at {\ensuremath{\gtrsim}} detected in wide area submillimetre surveys}",
      journal = {\mnras},
         year = 2024,
        month = may,
       volume = {530},
       number = {2},
        pages = {2292-2314},
          doi = {10.1093/mnras/stae976},
archivePrefix = {arXiv},
       eprint = {2404.11551},
 primaryClass = {astro-ph.GA},
       adsurl = {https://ui.adsabs.harvard.edu/abs/2024MNRAS.530.2292M}
}

@ARTICLE{mocanu13,
   author = {{Mocanu}, L.~M. and {Crawford}, T.~M. and {Vieira}, J.~D. and 
	{Aird}, K.~A. and {Aravena}, M. and {Austermann}, J.~E. and 
	{Benson}, B.~A. and {B{\'e}thermin}, M. and {Bleem}, L.~E. and 
	{Bothwell}, M. and {Carlstrom}, J.~E. and {Chang}, C.~L. and 
	{Chapman}, S. and {Cho}, H.-M. and {Crites}, A.~T. and {de Haan}, T. and 
	{Dobbs}, M.~A. and {Everett}, W.~B. and {George}, E.~M. and 
	{Halverson}, N.~W. and {Harrington}, N. and {Hezaveh}, Y. and 
	{Holder}, G.~P. and {Holzapfel}, W.~L. and {Hoover}, S. and 
	{Hrubes}, J.~D. and {Keisler}, R. and {Knox}, L. and {Lee}, A.~T. and 
	{Leitch}, E.~M. and {Lueker}, M. and {Luong-Van}, D. and {Marrone}, D.~P. and 
	{McMahon}, J.~J. and {Mehl}, J. and {Meyer}, S.~S. and {Mohr}, J.~J. and 
	{Montroy}, T.~E. and {Natoli}, T. and {Padin}, S. and {Plagge}, T. and 
	{Pryke}, C. and {Rest}, A. and {Reichardt}, C.~L. and {Ruhl}, J.~E. and 
	{Sayre}, J.~T. and {Schaffer}, K.~K. and {Shirokoff}, E. and 
	{Spieler}, H.~G. and {Spilker}, J.~S. and {Stalder}, B. and 
	{Staniszewski}, Z. and {Stark}, A.~A. and {Story}, K.~T. and 
	{Switzer}, E.~R. and {Vanderlinde}, K. and {Williamson}, R.},
    title = "{Extragalactic Millimeter-wave Point-source Catalog, Number Counts and Statistics from 771 deg$^{2}$ of the SPT-SZ Survey}",
  journal = {\apj},
archivePrefix = "arXiv",
   eprint = {1306.3470},
 primaryClass = "astro-ph.CO",
     year = 2013,
    month = dec,
   volume = 779,
      eid = {61},
    pages = {61},
      doi = {10.1088/0004-637X/779/1/61},
   adsurl = {http://adsabs.harvard.edu/abs/2013ApJ...779...61M}
}

@ARTICLE{mocanu19,
       author = {{Mocanu}, L.~M. and {Crawford}, T.~M. and {Aylor}, K. and
         {Benson}, B.~A. and {Bleem}, L.~E. and {Carlstrom}, J.~E. and
         {Chang}, C.~L. and {Cho}, H. -M. and {Chown}, R. and {Crites}, A.~T. and
         {de Haan}, T. and {Dobbs}, M.~A. and {Everett}, W.~B. and
         {George}, E.~M. and {Halverson}, N.~W. and {Harrington}, N.~L. and
         {Henning}, J.~W. and {Holder}, G.~P. and {Holzapfel}, W.~L. and
         {Hou}, Z. and {Hrubes}, J.~D. and {Knox}, L. and {Lee}, A.~T. and
         {Luong-Van}, D. and {Marrone}, D.~P. and {McMahon}, J.~J. and
         {Meyer}, S.~S. and {Millea}, M. and {Mohr}, J.~J. and {Natoli}, T. and
         {Omori}, Y. and {Padin}, S. and {Pryke}, C. and {Reichardt}, C.~L. and
         {Ruhl}, J.~E. and {Sayre}, J.~T. and {Schaffer}, K.~K. and
         {Shirokoff}, E. and {Staniszewski}, Z. and {Stark}, A.~A. and
         {Story}, K.~T. and {Vanderlinde}, K. and {Vieira}, J.~D. and
         {Williamson}, R. and {Wu}, W.~L.~K.},
        title = "{Consistency of cosmic microwave background temperature measurements in three frequency bands in the 2500-square-degree SPT-SZ survey}",
      journal = {\jcap},
         year = "2019",
        month = "Jul",
       volume = {2019},
       number = {7},
          eid = {038},
        pages = {038},
          doi = {10.1088/1475-7516/2019/07/038},
archivePrefix = {arXiv},
       eprint = {1904.12995},
 primaryClass = {astro-ph.CO},
       adsurl = {https://ui.adsabs.harvard.edu/abs/2019JCAP...07..038M}
}

@ARTICLE{moneti22,
       author = {{Euclid Collaboration} and {Moneti}, A. and {McCracken}, H.~J. and {Shuntov}, M. and {Kauffmann}, O.~B. and {Capak}, P. and {Davidzon}, I. and {Ilbert}, O. and {Scarlata}, C. and {Toft}, S. and {Weaver}, J. and {Chary}, R. and {Cuby}, J. and {Faisst}, A.~L. and {Masters}, D.~C. and {McPartland}, C. and {Mobasher}, B. and {Sanders}, D.~B. and {Scaramella}, R. and {Stern}, D. and {Szapudi}, I. and {Teplitz}, H. and {Zalesky}, L. and {Amara}, A. and {Auricchio}, N. and {Bodendorf}, C. and {Bonino}, D. and {Branchini}, E. and {Brau-Nogue}, S. and {Brescia}, M. and {Brinchmann}, J. and {Capobianco}, V. and {Carbone}, C. and {Carretero}, J. and {Castander}, F.~J. and {Castellano}, M. and {Cavuoti}, S. and {Cimatti}, A. and {Cledassou}, R. and {Congedo}, G. and {Conselice}, C.~J. and {Conversi}, L. and {Copin}, Y. and {Corcione}, L. and {Costille}, A. and {Cropper}, M. and {Da Silva}, A. and {Degaudenzi}, H. and {Douspis}, M. and {Dubath}, F. and {Duncan}, C.~A.~J. and {Dupac}, X. and {Dusini}, S. and {Farrens}, S. and {Ferriol}, S. and {Fosalba}, P. and {Frailis}, M. and {Franceschi}, E. and {Fumana}, M. and {Garilli}, B. and {Gillis}, B. and {Giocoli}, C. and {Granett}, B.~R. and {Grazian}, A. and {Grupp}, F. and {Haugan}, S.~V.~H. and {Hoekstra}, H. and {Holmes}, W. and {Hormuth}, F. and {Hudelot}, P. and {Jahnke}, K. and {Kermiche}, S. and {Kiessling}, A. and {Kilbinger}, M. and {Kitching}, T. and {Kohley}, R. and {K{\"u}mmel}, M. and {Kunz}, M. and {Kurki-Suonio}, H. and {Ligori}, S. and {Lilje}, P.~B. and {Lloro}, I. and {Maiorano}, E. and {Mansutti}, O. and {Marggraf}, O. and {Markovic}, K. and {Marulli}, F. and {Massey}, R. and {Maurogordato}, S. and {Meneghetti}, M. and {Merlin}, E. and {Meylan}, G. and {Moresco}, M. and {Moscardini}, L. and {Munari}, E. and {Niemi}, S.~M. and {Padilla}, C. and {Paltani}, S. and {Pasian}, F. and {Pedersen}, K. and {Pires}, S. and {Poncet}, M. and {Popa}, L. and {Pozzetti}, L. and {Raison}, F. and {Rebolo}, R. and {Rhodes}, J. and {Rix}, H. and {Roncarelli}, M. and {Rossetti}, E. and {Saglia}, R. and {Schneider}, P. and {Secroun}, A. and {Seidel}, G. and {Serrano}, S. and {Sirignano}, C. and {Sirri}, G. and {Stanco}, L. and {Tallada-Cresp{\'\i}}, P. and {Taylor}, A.~N. and {Tereno}, I. and {Toledo-Moreo}, R. and {Torradeflot}, F. and {Wang}, Y. and {Welikala}, N. and {Weller}, J. and {Zamorani}, G. and {Zoubian}, J. and {Andreon}, S. and {Bardelli}, S. and {Camera}, S. and {Graci{\'a}-Carpio}, J. and {Medinaceli}, E. and {Mei}, S. and {Polenta}, G. and {Romelli}, E. and {Sureau}, F. and {Tenti}, M. and {Vassallo}, T. and {Zacchei}, A. and {Zucca}, E. and {Baccigalupi}, C. and {Balaguera-Antol{\'\i}nez}, A. and {Bernardeau}, F. and {Biviano}, A. and {Bolzonella}, M. and {Bozzo}, E. and {Burigana}, C. and {Cabanac}, R. and {Cappi}, A. and {Carvalho}, C.~S. and {Casas}, S. and {Castignani}, G. and {Colodro-Conde}, C. and {Coupon}, J. and {Courtois}, H.~M. and {Di Ferdinando}, D. and {Farina}, M. and {Finelli}, F. and {Flose-Reimberg}, P. and {Fotopoulou}, S. and {Galeotta}, S. and {Ganga}, K. and {Garcia-Bellido}, J. and {Gaztanaga}, E. and {Gozaliasl}, G. and {Hook}, I. and {Joachimi}, B. and {Kansal}, V. and {Keihanen}, E. and {Kirkpatrick}, C.~C. and {Lindholm}, V. and {Mainetti}, G. and {Maino}, D. and {Maoli}, R. and {Martinelli}, M. and {Martinet}, N. and {Maturi}, M. and {Metcalf}, R.~B. and {Morgante}, G. and {Morisset}, N. and {Nucita}, A. and {Patrizii}, L. and {Potter}, D. and {Renzi}, A. and {Riccio}, G. and {S{\'a}nchez}, A.~G. and {Sapone}, D. and {Schirmer}, M. and {Schultheis}, M. and {Scottez}, V. and {Sefusatti}, E. and {Teyssier}, R. and {Tubio}, O. and {Tutusaus}, I. and {Valiviita}, J. and {Viel}, M. and {Hildebrandt}, H.},
        title = "{Euclid preparation. XVII. Cosmic Dawn Survey: Spitzer Space Telescope observations of the Euclid deep fields and calibration fields}",
      journal = {\aap},
         year = 2022,
        month = feb,
       volume = {658},
          eid = {A126},
        pages = {A126},
          doi = {10.1051/0004-6361/202142361},
archivePrefix = {arXiv},
       eprint = {2110.13928},
 primaryClass = {astro-ph.GA},
       adsurl = {https://ui.adsabs.harvard.edu/abs/2022A&A...658A.126E}
}

@ARTICLE{moshir90,
        author = {{Moshir}, M. and {et al.}},
         title = "{IRAS Faint Source Catalogue, version 2.0.}",
       journal = {IRAS Faint Source Catalogue},
          year = 1990,
         month = jan,
         pages = {0},
        adsurl = {https://ui.adsabs.harvard.edu/abs/1990IRASF.C......0M}
 }

@ARTICLE{murphy10,
   author = {{Murphy}, T. and {Sadler}, E.~M. and {Ekers}, R.~D. and {Massardi}, M. and 
	{Hancock}, P.~J. and {Mahony}, E. and {Ricci}, R. and {Burke-Spolaor}, S. and 
	{Calabretta}, M. and {Chhetri}, R. and {de Zotti}, G. and {Edwards}, P.~G. and 
	{Ekers}, J.~A. and {Jackson}, C.~A. and {Kesteven}, M.~J. and 
	{Lindley}, E. and {Newton-McGee}, K. and {Phillips}, C. and 
	{Roberts}, P. and {Sault}, R.~J. and {Staveley-Smith}, L. and 
	{Subrahmanyan}, R. and {Walker}, M.~A. and {Wilson}, W.~E.},
    title = "{The Australia Telescope 20 GHz Survey: the source catalogue}",
  journal = {\mnras},
archivePrefix = "arXiv",
   eprint = {0911.0002},
     year = 2010,
    month = mar,
   volume = 402,
    pages = {2403-2423},
      doi = {10.1111/j.1365-2966.2009.15961.x},
   adsurl = {http://adsabs.harvard.edu/abs/2010MNRAS.402.2403M}
}

@ARTICLE{negrello07,
   author = {{Negrello}, M. and {Perrotta}, F. and {Gonz{\'a}lez-Nuevo}, J. and 
	{Silva}, L. and {de Zotti}, G. and {Granato}, G.~L. and {Baccigalupi}, C. and 
	{Danese}, L.},
    title = "{Astrophysical and cosmological information from large-scale submillimetre surveys of extragalactic sources}",
  journal = {\mnras},
   eprint = {arXiv:astro-ph/0703210},
     year = 2007,
    month = jun,
   volume = 377,
    pages = {1557-1568},
      doi = {10.1111/j.1365-2966.2007.11708.x},
   adsurl = {http://adsabs.harvard.edu/abs/2007MNRAS.377.1557N}
}

@misc{osg06,
  doi = {10.21231/906P-4D78},
  url = {https://osg-htc.org/services/open_science_pool.html},
  author = {{OSG}},
  title = {OSPool},
  publisher = {OSG},
  year = {2006}
}

@inproceedings{osg07,
  title  = {The open science grid},
  author = {
    Pordes, Ruth 
    and Petravick, Don 
    and Kramer, Bill 
    and Olson, Doug 
    and Livny, Miron 
    and Roy, Alain 
    and Avery, Paul 
    and Blackburn, Kent 
    and Wenaus, Torre 
    and W{\"u}rthwein, Frank 
    and Foster, Ian
    and Gardner, Rob
    and Wilde, Mike
    and Blatecky, Alan
    and McGee, John
    and Quick, Rob
  },
  doi       = {10.1088/1742-6596/78/1/012057},
  booktitle = {J. Phys. Conf. Ser.},
  volume    = {78},
  series    = {78},
  pages     = {012057},
  year      = {2007},
}

@inproceedings{osg09,
  title        = {The pilot way to grid resources using glideinWMS},
  author       = {
    Sfiligoi, Igor    
    and Bradley, Daniel C 
    and Holzman, Burt     
    and Mhashilkar, Parag 
    and Padhi, Sanjay     
    and Wurthwein, Frank
  },
  doi          = {10.1109/CSIE.2009.950},
  booktitle    = {2009 WRI World Congress on Computer Science and Information Engineering},
  volume       = {2},
  series       = {2},
  pages        = {428--432},
  year         = {2009},
}

@misc{osg15,
  doi = {10.21231/0KVZ-VE57},
  url = {https://osdf.osg-htc.org/},
  author = {{OSG}},
  title = {Open Science Data Federation},
  publisher = {OSG},
  year = {2015}
}

@ARTICLE{overzier16,
       author = {{Overzier}, Roderik A.},
        title = "{The realm of the galaxy protoclusters. A review}",
      journal = {\aapr},
         year = 2016,
        month = nov,
       volume = {24},
       number = {1},
          eid = {14},
        pages = {14},
          doi = {10.1007/s00159-016-0100-3},
archivePrefix = {arXiv},
       eprint = {1610.05201},
 primaryClass = {astro-ph.GA},
       adsurl = {https://ui.adsabs.harvard.edu/abs/2016A&ARv..24...14O}
}

@ARTICLE{padovani17,
       author = {{Padovani}, P. and {Alexander}, D.~M. and {Assef}, R.~J. and {De Marco}, B. and {Giommi}, P. and {Hickox}, R.~C. and {Richards}, G.~T. and {Smol{\v{c}}i{\'c}}, V. and {Hatziminaoglou}, E. and {Mainieri}, V. and {Salvato}, M.},
        title = "{Active galactic nuclei: what's in a name?}",
      journal = {\aapr},
         year = 2017,
        month = aug,
       volume = {25},
       number = {1},
          eid = {2},
        pages = {2},
          doi = {10.1007/s00159-017-0102-9},
archivePrefix = {arXiv},
       eprint = {1707.07134},
 primaryClass = {astro-ph.GA},
       adsurl = {https://ui.adsabs.harvard.edu/abs/2017A&ARv..25....2P}
}

@ARTICLE{pan18,
       author = {{Pan}, Z. and {Ade}, P.~A.~R. and {Ahmed}, Z. and {Anderson}, A.~J. and
         {Austermann}, J.~E. and {Avva}, J.~S. and {Thakur}, R. Basu and
         {Bender}, A.~N. and {Benson}, B.~A. and {Carlstrom}, J.~E. and
         {Carter}, F.~W. and {Cecil}, T. and {Chang}, C.~L. and {Cliche}, J.~F. and
         {Cukierman}, A. and {Denison}, E.~V. and {de Haan}, T. and {Ding}, J. and
         {Dobbs}, M.~A. and {Dutcher}, D. and {Everett}, W. and {Foster}, A. and
         {Gannon}, R.~N. and {Gilbert}, A. and {Groh}, J.~C. and
         {Halverson}, N.~W. and {Harke-Hosemann}, A.~H. and {Harrington}, N.~L. and
         {Henning}, J.~W. and {Hilton}, G.~C. and {Holzapfel}, W.~L. and
         {Huang}, N. and {Irwin}, K.~D. and {Jeong}, O.~B. and {Jonas}, M. and
         {Khaire}, T. and {Kofman}, A.~M. and {Korman}, M. and {Kubik}, D. and
         {Kuhlmann}, S. and {Kuo}, C.~L. and {Lee}, A.~T. and {Lowitz}, A.~E. and
         {Meyer}, S.~S. and {Michalik}, D. and {Montgomery}, J. and
         {Nadolski}, A. and {Natoli}, T. and {Nguyen}, H. and {Noble}, G.~I. and
         {Novosad}, V. and {Padin}, S. and {Pearson}, J. and {Posada}, C.~M. and
         {Rahlin}, A. and {Ruhl}, J.~E. and {Saunders}, L.~J. and
         {Sayre}, J.~T. and {Shirley}, I. and {Shirokoff}, E. and {Smecher}, G. and
         {Sobrin}, J.~A. and {Stark}, A.~A. and {Story}, K.~T. and {Suzuki}, A. and
         {Tang}, Q.~Y. and {Thompson}, K.~L. and {Tucker}, C. and {Vale}, L.~R. and
         {Vanderlinde}, K. and {Vieira}, J.~D. and {Wang}, G. and
         {Whitehorn}, N. and {Yefremenko}, V. and {Yoon}, K.~W. and
         {Young}, M.~R.},
        title = "{Optical Characterization of the SPT-3G Camera}",
      journal = {Journal of Low Temperature Physics},
         year = "2018",
        month = "Nov",
       volume = {193},
       number = {3-4},
        pages = {305-313},
          doi = {10.1007/s10909-018-1935-y},
archivePrefix = {arXiv},
       eprint = {1805.03219},
 primaryClass = {astro-ph.IM},
       adsurl = {https://ui.adsabs.harvard.edu/abs/2018JLTP..193..305P}
}

@ARTICLE{pan23,
       author = {{Pan}, Z. and {Bianchini}, F. and {Wu}, W.~L.~K. and {Ade}, P.~A.~R. and {Ahmed}, Z. and {Anderes}, E. and {Anderson}, A.~J. and {Ansarinejad}, B. and {Archipley}, M. and {Aylor}, K. and {Balkenhol}, L. and {Barry}, P.~S. and {Basu Thakur}, R. and {Benabed}, K. and {Bender}, A.~N. and {Benson}, B.~A. and {Bleem}, L.~E. and {Bouchet}, F.~R. and {Bryant}, L. and {Byrum}, K. and {Camphuis}, E. and {Carlstrom}, J.~E. and {Carter}, F.~W. and {Cecil}, T.~W. and {Chang}, C.~L. and {Chaubal}, P. and {Chen}, G. and {Chichura}, P.~M. and {Cho}, H. -M. and {Chou}, T. -L. and {Cliche}, J. -F. and {Coerver}, A. and {Crawford}, T.~M. and {Cukierman}, A. and {Daley}, C. and {de Haan}, T. and {Denison}, E.~V. and {Dibert}, K.~R. and {Ding}, J. and {Dobbs}, M.~A. and {Doussot}, A. and {Dutcher}, D. and {Everett}, W. and {Feng}, C. and {Ferguson}, K.~R. and {Fichman}, K. and {Foster}, A. and {Fu}, J. and {Galli}, S. and {Gambrel}, A.~E. and {Gardner}, R.~W. and {Ge}, F. and {Goeckner-Wald}, N. and {Gualtieri}, R. and {Guidi}, F. and {Guns}, S. and {Gupta}, N. and {Halverson}, N.~W. and {Harke-Hosemann}, A.~H. and {Harrington}, N.~L. and {Henning}, J.~W. and {Hilton}, G.~C. and {Hivon}, E. and {Holder}, G.~P. and {Holzapfel}, W.~L. and {Hood}, J.~C. and {Howe}, D. and {Huang}, N. and {Irwin}, K.~D. and {Jeong}, O. and {Jonas}, M. and {Jones}, A. and {K{\'e}ruzor{\'e}}, F. and {Khaire}, T.~S. and {Knox}, L. and {Kofman}, A.~M. and {Korman}, M. and {Kubik}, D.~L. and {Kuhlmann}, S. and {Kuo}, C. -L. and {Lee}, A.~T. and {Leitch}, E.~M. and {Levy}, K. and {Lowitz}, A.~E. and {Lu}, C. and {Maniyar}, A. and {Menanteau}, F. and {Meyer}, S.~S. and {Michalik}, D. and {Millea}, M. and {Montgomery}, J. and {Nadolski}, A. and {Nakato}, Y. and {Natoli}, T. and {Nguyen}, H. and {Noble}, G.~I. and {Novosad}, V. and {Omori}, Y. and {Padin}, S. and {Paschos}, P. and {Pearson}, J. and {Posada}, C.~M. and {Prabhu}, K. and {Quan}, W. and {Raghunathan}, S. and {Rahimi}, M. and {Rahlin}, A. and {Reichardt}, C.~L. and {Riebel}, D. and {Riedel}, B. and {Ruhl}, J.~E. and {Sayre}, J.~T. and {Schiappucci}, E. and {Shirokoff}, E. and {Smecher}, G. and {Sobrin}, J.~A. and {Stark}, A.~A. and {Stephen}, J. and {Story}, K.~T. and {Suzuki}, A. and {Takakura}, S. and {Tandoi}, C. and {Thompson}, K.~L. and {Thorne}, B. and {Trendafilova}, C. and {Tucker}, C. and {Umilta}, C. and {Vale}, L.~R. and {Vanderlinde}, K. and {Vieira}, J.~D. and {Wang}, G. and {Whitehorn}, N. and {Yefremenko}, V. and {Yoon}, K.~W. and {Young}, M.~R. and {Zebrowski}, J.~A.},
        title = "{Measurement of gravitational lensing of the cosmic microwave background using SPT-3G 2018 data}",
      journal = {\prd},
         year = 2023,
        month = dec,
       volume = {108},
       number = {12},
          eid = {122005},
        pages = {122005},
          doi = {10.1103/PhysRevD.108.122005},
archivePrefix = {arXiv},
       eprint = {2308.11608},
 primaryClass = {astro-ph.CO},
       adsurl = {https://ui.adsabs.harvard.edu/abs/2023PhRvD.108l2005P}
}

@misc{pettorino24,
  author = {Valeria Pettorino and R. Laureijs},
  title = {Q1 fields definition},
  year = {2024},
  month = {September},
  howpublished = {Memo},
  institution = {European Space Agency (ESA)},
  note = {ESA UNCLASSIFIED – For ESA Official Use Only, Ref. EUCL-EST-ME-8-018, Version 1, Revision 1, Date: 23/09/2024, Approved by Euclid Science Team},
  url = {https://www.cosmos.esa.int/web/euclid},
  address = {European Space Agency}
}

@ARTICLE{planck15-10,
       author = {{Planck Collaboration} and {Adam}, R. and {Ade}, P.~A.~R. and
         {Aghanim}, N. and {Alves}, M.~I.~R. and {Arnaud}, M. and {Ashdown}, M. and
         {Aumont}, J. and {Baccigalupi}, C. and {Banday}, A.~J. and
         {Barreiro}, R.~B. and {Bartlett}, J.~G. and {Bartolo}, N. and
         {Battaner}, E. and {Benabed}, K. and {Beno{\^\i}t}, A. and
         {Benoit-L{\'e}vy}, A. and {Bernard}, J. -P. and {Bersanelli}, M. and
         {Bielewicz}, P. and {Bock}, J.~J. and {Bonaldi}, A. and {Bonavera}, L. and
         {Bond}, J.~R. and {Borrill}, J. and {Bouchet}, F.~R. and
         {Boulanger}, F. and {Bucher}, M. and {Burigana}, C. and
         {Butler}, R.~C. and {Calabrese}, E. and {Cardoso}, J. -F. and
         {Catalano}, A. and {Challinor}, A. and {Chamballu}, A. and
         {Chary}, R. -R. and {Chiang}, H.~C. and {Christensen}, P.~R. and
         {Clements}, D.~L. and {Colombi}, S. and {Colombo}, L.~P.~L. and
         {Combet}, C. and {Couchot}, F. and {Coulais}, A. and {Crill}, B.~P. and
         {Curto}, A. and {Cuttaia}, F. and {Danese}, L. and {Davies}, R.~D. and
         {Davis}, R.~J. and {de Bernardis}, P. and {de Rosa}, A. and
         {de Zotti}, G. and {Delabrouille}, J. and {D{\'e}sert}, F. -X. and
         {Dickinson}, C. and {Diego}, J.~M. and {Dole}, H. and {Donzelli}, S. and
         {Dor{\'e}}, O. and {Douspis}, M. and {Ducout}, A. and {Dupac}, X. and
         {Efstathiou}, G. and {Elsner}, F. and {En{\ss}lin}, T.~A. and
         {Eriksen}, H.~K. and {Falgarone}, E. and {Fergusson}, J. and
         {Finelli}, F. and {Forni}, O. and {Frailis}, M. and {Fraisse}, A.~A. and
         {Franceschi}, E. and {Frejsel}, A. and {Galeotta}, S. and {Galli}, S. and
         {Ganga}, K. and {Ghosh}, T. and {Giard}, M. and
         {Giraud-H{\'e}raud}, Y. and {Gjerl{\o}w}, E. and
         {Gonz{\'a}lez-Nuevo}, J. and {G{\'o}rski}, K.~M. and {Gratton}, S. and
         {Gregorio}, A. and {Gruppuso}, A. and {Gudmundsson}, J.~E. and
         {Hansen}, F.~K. and {Hanson}, D. and {Harrison}, D.~L. and {Helou}, G. and
         {Henrot-Versill{\'e}}, S. and {Hern{\'a}ndez-Monteagudo}, C. and
         {Herranz}, D. and {Hildebrandt}, S.~R. and {Hivon}, E. and
         {Hobson}, M. and {Holmes}, W.~A. and {Hornstrup}, A. and {Hovest}, W. and
         {Huffenberger}, K.~M. and {Hurier}, G. and {Jaffe}, A.~H. and
         {Jaffe}, T.~R. and {Jones}, W.~C. and {Juvela}, M. and
         {Keih{\"a}nen}, E. and {Keskitalo}, R. and {Kisner}, T.~S. and
         {Kneissl}, R. and {Knoche}, J. and {Kunz}, M. and {Kurki-Suonio}, H. and
         {Lagache}, G. and {L{\"a}hteenm{\"a}ki}, A. and {Lamarre}, J. -M. and
         {Lasenby}, A. and {Lattanzi}, M. and {Lawrence}, C.~R. and
         {Le Jeune}, M. and {Leahy}, J.~P. and {Leonardi}, R. and
         {Lesgourgues}, J. and {Levrier}, F. and {Liguori}, M. and
         {Lilje}, P.~B. and {Linden-V{\o}rnle}, M. and {L{\'o}pez-Caniego}, M. and
         {Lubin}, P.~M. and {Mac{\'\i}as-P{\'e}rez}, J.~F. and {Maggio}, G. and
         {Maino}, D. and {Mandolesi}, N. and {Mangilli}, A. and {Maris}, M. and
         {Marshall}, D.~J. and {Martin}, P.~G. and
         {Mart{\'\i}nez-Gonz{\'a}lez}, E. and {Masi}, S. and {Matarrese}, S. and
         {McGehee}, P. and {Meinhold}, P.~R. and {Melchiorri}, A. and
         {Mendes}, L. and {Mennella}, A. and {Migliaccio}, M. and {Mitra}, S. and
         {Miville-Desch{\^e}nes}, M. -A. and {Moneti}, A. and {Montier}, L. and
         {Morgante}, G. and {Mortlock}, D. and {Moss}, A. and {Munshi}, D. and
         {Murphy}, J.~A. and {Naselsky}, P. and {Nati}, F. and {Natoli}, P. and
         {Netterfield}, C.~B. and {N{\o}rgaard-Nielsen}, H.~U. and
         {Noviello}, F. and {Novikov}, D. and {Novikov}, I. and {Orlando}, E. and
         {Oxborrow}, C.~A. and {Paci}, F. and {Pagano}, L. and {Pajot}, F. and
         {Paladini}, R. and {Paoletti}, D. and {Partridge}, B. and {Pasian}, F. and
         {Patanchon}, G. and {Pearson}, T.~J. and {Perdereau}, O. and
         {Perotto}, L. and {Perrotta}, F. and {Pettorino}, V. and
         {Piacentini}, F. and {Piat}, M. and {Pierpaoli}, E. and
         {Pietrobon}, D. and {Plaszczynski}, S. and {Pointecouteau}, E. and
         {Polenta}, G. and {Pratt}, G.~W. and {Pr{\'e}zeau}, G. and
         {Prunet}, S. and {Puget}, J. -L. and {Rachen}, J.~P. and
         {Reach}, W.~T. and {Rebolo}, R. and {Reinecke}, M. and
         {Remazeilles}, M. and {Renault}, C. and {Renzi}, A. and
         {Ristorcelli}, I. and {Rocha}, G. and {Rosset}, C. and {Rossetti}, M. and
         {Roudier}, G. and {Rubi{\~n}o-Mart{\'\i}n}, J.~A. and {Rusholme}, B. and
         {Sandri}, M. and {Santos}, D. and {Savelainen}, M. and {Savini}, G. and
         {Scott}, D. and {Seiffert}, M.~D. and {Shellard}, E.~P.~S. and
         {Spencer}, L.~D. and {Stolyarov}, V. and {Stompor}, R. and
         {Strong}, A.~W. and {Sudiwala}, R. and {Sunyaev}, R. and {Sutton}, D. and
         {Suur-Uski}, A. -S. and {Sygnet}, J. -F. and {Tauber}, J.~A. and
         {Terenzi}, L. and {Toffolatti}, L. and {Tomasi}, M. and {Tristram}, M. and
         {Tucci}, M. and {Tuovinen}, J. and {Umana}, G. and {Valenziano}, L. and
         {Valiviita}, J. and {Van Tent}, F. and {Vielva}, P. and {Villa}, F. and
         {Wade}, L.~A. and {Wandelt}, B.~D. and {Wehus}, I.~K. and
         {Wilkinson}, A. and {Yvon}, D. and {Zacchei}, A. and {Zonca}, A.},
        title = "{Planck 2015 results. X. Diffuse component separation: Foreground maps}",
      journal = {\aap},
         year = "2016",
        month = "Sep",
       volume = {594},
          eid = {A10},
        pages = {A10},
          doi = {10.1051/0004-6361/201525967},
archivePrefix = {arXiv},
       eprint = {1502.01588},
 primaryClass = {astro-ph.CO},
       adsurl = {https://ui.adsabs.harvard.edu/abs/2016A&A...594A..10P}
}

@article{planck15-27,
       author = {{Planck Collaboration} and {Ade}, P.~A.~R. and {Aghanim}, N. and
         {Arnaud}, M. and {Ashdown}, M. and {Aumont}, J. and {Baccigalupi}, C. and
         {Banday}, A.~J. and {Barreiro}, R.~B. and {Barrena}, R. and
         {Bartlett}, J.~G. and {Bartolo}, N. and {Battaner}, E. and
         {Battye}, R. and {Benabed}, K. and {Beno{\^\i}t}, A. and
         {Benoit-L{\'e}vy}, A. and {Bernard}, J. -P. and {Bersanelli}, M. and
         {Bielewicz}, P. and {Bikmaev}, I. and {B{\"o}hringer}, H. and
         {Bonaldi}, A. and {Bonavera}, L. and {Bond}, J.~R. and {Borrill}, J. and
         {Bouchet}, F.~R. and {Bucher}, M. and {Burenin}, R. and {Burigana}, C. and
         {Butler}, R.~C. and {Calabrese}, E. and {Cardoso}, J. -F. and
         {Carvalho}, P. and {Catalano}, A. and {Challinor}, A. and
         {Chamballu}, A. and {Chary}, R. -R. and {Chiang}, H.~C. and {Chon}, G. and
         {Christensen}, P.~R. and {Clements}, D.~L. and {Colombi}, S. and
         {Colombo}, L.~P.~L. and {Combet}, C. and {Comis}, B. and {Couchot}, F. and
         {Coulais}, A. and {Crill}, B.~P. and {Curto}, A. and {Cuttaia}, F. and
         {Dahle}, H. and {Danese}, L. and {Davies}, R.~D. and {Davis}, R.~J. and
         {de Bernardis}, P. and {de Rosa}, A. and {de Zotti}, G. and
         {Delabrouille}, J. and {D{\'e}sert}, F. -X. and {Dickinson}, C. and
         {Diego}, J.~M. and {Dolag}, K. and {Dole}, H. and {Donzelli}, S. and
         {Dor{\'e}}, O. and {Douspis}, M. and {Ducout}, A. and {Dupac}, X. and
         {Efstathiou}, G. and {Eisenhardt}, P.~R.~M. and {Elsner}, F. and
         {En{\ss}lin}, T.~A. and {Eriksen}, H.~K. and {Falgarone}, E. and
         {Fergusson}, J. and {Feroz}, F. and {Ferragamo}, A. and {Finelli}, F. and
         {Forni}, O. and {Frailis}, M. and {Fraisse}, A.~A. and
         {Franceschi}, E. and {Frejsel}, A. and {Galeotta}, S. and {Galli}, S. and
         {Ganga}, K. and {G{\'e}nova-Santos}, R.~T. and {Giard}, M. and
         {Giraud-H{\'e}raud}, Y. and {Gjerl{\o}w}, E. and
         {Gonz{\'a}lez-Nuevo}, J. and {G{\'o}rski}, K.~M. and
         {Grainge}, K.~J.~B. and {Gratton}, S. and {Gregorio}, A. and
         {Gruppuso}, A. and {Gudmundsson}, J.~E. and {Hansen}, F.~K. and
         {Hanson}, D. and {Harrison}, D.~L. and {Hempel}, A. and
         {Henrot-Versill{\'e}}, S. and {Hern{\'a}ndez-Monteagudo}, C. and
         {Herranz}, D. and {Hildebrandt}, S.~R. and {Hivon}, E. and
         {Hobson}, M. and {Holmes}, W.~A. and {Hornstrup}, A. and {Hovest}, W. and
         {Huffenberger}, K.~M. and {Hurier}, G. and {Jaffe}, A.~H. and
         {Jaffe}, T.~R. and {Jin}, T. and {Jones}, W.~C. and {Juvela}, M. and
         {Keih{\"a}nen}, E. and {Keskitalo}, R. and {Khamitov}, I. and
         {Kisner}, T.~S. and {Kneissl}, R. and {Knoche}, J. and {Kunz}, M. and
         {Kurki-Suonio}, H. and {Lagache}, G. and {Lamarre}, J. -M. and
         {Lasenby}, A. and {Lattanzi}, M. and {Lawrence}, C.~R. and
         {Leonardi}, R. and {Lesgourgues}, J. and {Levrier}, F. and
         {Liguori}, M. and {Lilje}, P.~B. and {Linden-V{\o}rnle}, M. and
         {L{\'o}pez-Caniego}, M. and {Lubin}, P.~M. and
         {Mac{\'\i}as-P{\'e}rez}, J.~F. and {Maggio}, G. and {Maino}, D. and
         {Mak}, D.~S.~Y. and {Mandolesi}, N. and {Mangilli}, A. and
         {Martin}, P.~G. and {Mart{\'\i}nez-Gonz{\'a}lez}, E. and {Masi}, S. and
         {Matarrese}, S. and {Mazzotta}, P. and {McGehee}, P. and {Mei}, S. and
         {Melchiorri}, A. and {Melin}, J. -B. and {Mendes}, L. and
         {Mennella}, A. and {Migliaccio}, M. and {Mitra}, S. and
         {Miville-Desch{\^e}nes}, M. -A. and {Moneti}, A. and {Montier}, L. and
         {Morgante}, G. and {Mortlock}, D. and {Moss}, A. and {Munshi}, D. and
         {Murphy}, J.~A. and {Naselsky}, P. and {Nastasi}, A. and {Nati}, F. and
         {Natoli}, P. and {Netterfield}, C.~B. and {N{\o}rgaard-Nielsen}, H.~U. and
         {Noviello}, F. and {Novikov}, D. and {Novikov}, I. and {Olamaie}, M. and
         {Oxborrow}, C.~A. and {Paci}, F. and {Pagano}, L. and {Pajot}, F. and
         {Paoletti}, D. and {Pasian}, F. and {Patanchon}, G. and
         {Pearson}, T.~J. and {Perdereau}, O. and {Perotto}, L. and
         {Perrott}, Y.~C. and {Perrotta}, F. and {Pettorino}, V. and
         {Piacentini}, F. and {Piat}, M. and {Pierpaoli}, E. and
         {Pietrobon}, D. and {Plaszczynski}, S. and {Pointecouteau}, E. and
         {Polenta}, G. and {Pratt}, G.~W. and {Pr{\'e}zeau}, G. and
         {Prunet}, S. and {Puget}, J. -L. and {Rachen}, J.~P. and
         {Reach}, W.~T. and {Rebolo}, R. and {Reinecke}, M. and
         {Remazeilles}, M. and {Renault}, C. and {Renzi}, A. and
         {Ristorcelli}, I. and {Rocha}, G. and {Rosset}, C. and {Rossetti}, M. and
         {Roudier}, G. and {Rozo}, E. and {Rubi{\~n}o-Mart{\'\i}n}, J.~A. and
         {Rumsey}, C. and {Rusholme}, B. and {Rykoff}, E.~S. and {Sandri}, M. and
         {Santos}, D. and {Saunders}, R.~D.~E. and {Savelainen}, M. and
         {Savini}, G. and {Schammel}, M.~P. and {Scott}, D. and
         {Seiffert}, M.~D. and {Shellard}, E.~P.~S. and {Shimwell}, T.~W. and
         {Spencer}, L.~D. and {Stanford}, S.~A. and {Stern}, D. and
         {Stolyarov}, V. and {Stompor}, R. and {Streblyanska}, A. and
         {Sudiwala}, R. and {Sunyaev}, R. and {Sutton}, D. and
         {Suur-Uski}, A. -S. and {Sygnet}, J. -F. and {Tauber}, J.~A. and
         {Terenzi}, L. and {Toffolatti}, L. and {Tomasi}, M. and {Tramonte}, D. and
         {Tristram}, M. and {Tucci}, M. and {Tuovinen}, J. and {Umana}, G. and
         {Valenziano}, L. and {Valiviita}, J. and {Van Tent}, B. and
         {Vielva}, P. and {Villa}, F. and {Wade}, L.~A. and {Wandelt}, B.~D. and
         {Wehus}, I.~K. and {White}, S.~D.~M. and {Wright}, E.~L. and
         {Yvon}, D. and {Zacchei}, A. and {Zonca}, A.},
        title = "{Planck 2015 results. XXVII. The second Planck catalogue of Sunyaev-Zeldovich sources}",
      journal = {\aap},
         year = "2016",
        month = "Sep",
       volume = {594},
          eid = {A27},
        pages = {A27},
          doi = {10.1051/0004-6361/201525823},
archivePrefix = {arXiv},
       eprint = {1502.01598},
 primaryClass = {astro-ph.CO},
       adsurl = {https://ui.adsabs.harvard.edu/\#abs/2016A&A...594A..27P}
}

@misc{prabhu24,
  doi = {10.48550/ARXIV.2403.17925},
  url = {https://arxiv.org/abs/2403.17925},
  author = {{Prabhu}, K. and {Raghunathan}, S. and {Millea}, M. and {Lynch}, G. and
  {Ade}, P.~A.~R. and {Anderes}, E. and {Anderson}, A.~J. and {Ansarinejad}, B. and
  {Archipley}, M. and {Balkenhol}, L. and {Benabed}, K. and {Bender}, A.~N. and
  {Benson}, B.~A. and {Bianchini}, F. and {Bleem}, L.~E. and {Bouchet},  F.~R. and
  {Bryant}, L. and {Camphuis}, E. and {Carlstrom}, J.~E. and {Cecil}, T.~W. and
  {Chang}, C.~L. and {Chaubal}, P. and {Chichura}, P.~M. and {Chou}, T.-L. and
  {Coerver}, A. and {Crawford}, T.~M. and {Cukierman}, A. and {Daley}, C. and
  {de Haan}, T. and {Dibert}, K.~R. and {Dobbs}, M.~A. and {Doussot}, A. and
  {Dutcher}, D. and {Everett}, W. and {Feng}, C. and {Ferguson}, K.~R. and
  {Fichman}, K. and {Foster}, A. and {Galli}, S. and {Gambrel}, A.~E. and
  {Gardner},  R.~W. and {Ge}, F. and {Goeckner-Wald}, N. and {Gualtieri}, R. and
  {Guidi}, F. and {Guns}, S. and {Halverson}, N.~W. and {Hivon},  E. and
  {Holder}, G.~P. and {Holzapfel},  W.~L. and {Hood},  J.~C. and {Hryciuk}, A. and 
  {Huang}, N. and {Kéruzoré}, F. and {Knox}, L. and {Korman}, M. and {Kornoelje}, K. and
  {Kuo},  C.-L. and {Lee}, A.~T. and {Levy},  K. and {Lowitz},  A.~E. and {Lu},  C. and
  {Maniyar}, A. and Menanteau, F. and {Montgomery}, J. and {Nakato}, Y. and
  {Natoli}, T. and {Noble},  G.~I. and {Novosad}, V. and {Omori}, Y. and {Padin}, S. and
  {Pan}, Z. and {Paschos}, P. and {Phadke}, K.~A. and {Quan},  W. and {Rahimi}, M. and
  {Rahlin}, A. and {Reichardt},  C.~L. and {Rouble}, M. and {Ruhl}, J.~E. and
  {Schiappucci}, E. and {Smecher}, G. and {Sobrin}, J.~A. and {Stark}, A.~A. and
  {Stephen}, J. and {Suzuki}, A. and {Tandoi}, C. and {Thompson}, K.~L. and
  {Thorne}, B. and {Trendafilova}, C. and {Tucker}, C. and {Umilta}, C. and
  {Vitrier}, A. and {Vieira}, J.~D. and {Wan}, Y. and {Wang}, G. and {Whitehorn}, N. and
  {Wu}, W.~L.~K. and {Yefremenko}, V. and {Young}, M.~R. and {Zebrowski},  J.~A.},
  title = {Testing the $\mathbf{\Lambda}$CDM Cosmological Model with Forthcoming Measurements of the Cosmic Microwave Background with SPT-3G},
  publisher = {arXiv},
  year = {2024},
  copyright = {Creative Commons Attribution 4.0 International}
}

@ARTICLE{raghunathan24,
       author = {{Raghunathan}, S. and {Ade}, P.~A.~R. and {Anderson}, A.~J. and {Ansarinejad}, B. and {Archipley}, M. and {Austermann}, J.~E. and {Balkenhol}, L. and {Beall}, J.~A. and {Benabed}, K. and {Bender}, A.~N. and {Benson}, B.~A. and {Bianchini}, F. and {Bleem}, L.~E. and {Bock}, J. and {Bouchet}, F.~R. and {Bryant}, L. and {Camphuis}, E. and {Carlstrom}, J.~E. and {Cecil}, T.~W. and {Chang}, C.~L. and {Chaubal}, P. and {Chiang}, H.~C. and {Chichura}, P.~M. and {Chou}, T. -L. and {Citron}, R. and {Coerver}, A. and {Crawford}, T.~M. and {Crites}, A.~T. and {Cukierman}, A. and {Daley}, C. and {Dibert}, K.~R. and {Dobbs}, M.~A. and {Doussot}, A. and {Dutcher}, D. and {Everett}, W. and {Feng}, C. and {Ferguson}, K.~R. and {Fichman}, K. and {Foster}, A. and {Galli}, S. and {Gallicchio}, J. and {Gambrel}, A.~E. and {Gardner}, R.~W. and {Ge}, F. and {George}, E.~M. and {Goeckner-Wald}, N. and {Gualtieri}, R. and {Guidi}, F. and {Guns}, S. and {Gupta}, N. and {de Haan}, T. and {Halverson}, N.~W. and {Hivon}, E. and {Holder}, G.~P. and {Holzapfel}, W.~L. and {Hood}, J.~C. and {Hrubes}, J.~D. and {Hryciuk}, A. and {Huang}, N. and {Hubmayr}, J. and {Irwin}, K.~D. and {K{\'e}ruzor{\'e}}, F. and {Khalife}, A.~R. and {Knox}, L. and {Korman}, M. and {Kornoelje}, K. and {Kuo}, C. -L. and {Lee}, A.~T. and {Levy}, K. and {Li}, D. and {Lowitz}, A.~E. and {Lu}, C. and {Maniyar}, A. and {Martsen}, E.~S. and {McMahon}, J.~J. and {Menanteau}, F. and {Millea}, M. and {Montgomery}, J. and {Corbett Moran}, C. and {Nakato}, Y. and {Natoli}, T. and {Nibarger}, J.~P. and {Noble}, G.~I. and {Novosad}, V. and {Omori}, Y. and {Padin}, S. and {Pan}, Z. and {Paschos}, P. and {Patil}, S. and {Phadke}, K.~A. and {Prabhu}, K. and {Pryke}, C. and {Quan}, W. and {Rahimi}, M. and {Rahlin}, A. and {Reichardt}, C.~L. and {Rouble}, M. and {Ruhl}, J.~E. and {Saliwanchik}, B.~R. and {Schaffer}, K.~K. and {Schiappucci}, E. and {Sievers}, C. and {Smecher}, G. and {Sobrin}, J.~A. and {Stark}, A.~A. and {Stephen}, J. and {Suzuki}, A. and {Tandoi}, C. and {Thompson}, K.~L. and {Thorne}, B. and {Trendafilova}, C. and {Tucker}, C. and {Umilta}, C. and {Veach}, T. and {Vieira}, J.~D. and {Viero}, M.~P. and {Wan}, Y. and {Wang}, G. and {Whitehorn}, N. and {Wu}, W.~L.~K. and {Yefremenko}, V. and {Young}, M.~R. and {Zebrowski}, J.~A. and {Zemcov}, M. and {SPT-3G} and {SPTpol Collaboration}},
        title = "{First Constraints on the Epoch of Reionization Using the Non-Gaussianity of the Kinematic Sunyaev-Zel'dovich Effect from the South Pole Telescope and Herschel-SPIRE Observations}",
      journal = {\prl},
         year = 2024,
        month = sep,
       volume = {133},
       number = {12},
          eid = {121004},
        pages = {121004},
          doi = {10.1103/PhysRevLett.133.121004},
archivePrefix = {arXiv},
       eprint = {2403.02337},
 primaryClass = {astro-ph.CO},
       adsurl = {https://ui.adsabs.harvard.edu/abs/2024PhRvL.133l1004R}
}

@ARTICLE{reichardt13,
   author = {{Reichardt}, C.~L. and {Stalder}, B. and {Bleem}, L.~E. and 
	{Montroy}, T.~E. and {Aird}, K.~A. and {Andersson}, K. and {Armstrong}, R. and 
	{Ashby}, M.~L.~N. and {Bautz}, M. and {Bayliss}, M. and {Bazin}, G. and 
	{Benson}, B.~A. and {Brodwin}, M. and {Carlstrom}, J.~E. and 
	{Chang}, C.~L. and {Cho}, H.~M. and {Clocchiatti}, A. and {Crawford}, T.~M. and 
	{Crites}, A.~T. and {de Haan}, T. and {Desai}, S. and {Dobbs}, M.~A. and 
	{Dudley}, J.~P. and {Foley}, R.~J. and {Forman}, W.~R. and {George}, E.~M. and 
	{Gladders}, M.~D. and {Gonzalez}, A.~H. and {Halverson}, N.~W. and 
	{Harrington}, N.~L. and {High}, F.~W. and {Holder}, G.~P. and 
	{Holzapfel}, W.~L. and {Hoover}, S. and {Hrubes}, J.~D. and 
	{Jones}, C. and {Joy}, M. and {Keisler}, R. and {Knox}, L. and 
	{Lee}, A.~T. and {Leitch}, E.~M. and {Liu}, J. and {Lueker}, M. and 
	{Luong-Van}, D. and {Mantz}, A. and {Marrone}, D.~P. and {McDonald}, M. and 
	{McMahon}, J.~J. and {Mehl}, J. and {Meyer}, S.~S. and {Mocanu}, L. and 
	{Mohr}, J.~J. and {Murray}, S.~S. and {Natoli}, T. and {Padin}, S. and 
	{Plagge}, T. and {Pryke}, C. and {Rest}, A. and {Ruel}, J. and 
	{Ruhl}, J.~E. and {Saliwanchik}, B.~R. and {Saro}, A. and {Sayre}, J.~T. and 
	{Schaffer}, K.~K. and {Shaw}, L. and {Shirokoff}, E. and {Song}, J. and 
	{Spieler}, H.~G. and {Staniszewski}, Z. and {Stark}, A.~A. and 
	{Story}, K. and {Stubbs}, C.~W. and {{\v S}uhada}, R. and {van Engelen}, A. and 
	{Vanderlinde}, K. and {Vieira}, J.~D. and {Vikhlinin}, A. and 
	{Williamson}, R. and {Zahn}, O. and {Zenteno}, A.},
    title = "{Galaxy Clusters Discovered via the Sunyaev-Zel'dovich Effect in the First 720 Square Degrees of the South Pole Telescope Survey}",
  journal = {\apj},
archivePrefix = "arXiv",
   eprint = {1203.5775},
 primaryClass = "astro-ph.CO",
     year = 2013,
    month = feb,
   volume = 763,
      eid = {127},
    pages = {127},
      doi = {10.1088/0004-637X/763/2/127},
   adsurl = {http://adsabs.harvard.edu/abs/2013ApJ...763..127R}
}

@ARTICLE{reichardt21,
       author = {{Reichardt}, C.~L. and {Patil}, S. and {Ade}, P.~A.~R. and {Anderson}, A.~J. and {Austermann}, J.~E. and {Avva}, J.~S. and {Baxter}, E. and {Beall}, J.~A. and {Bender}, A.~N. and {Benson}, B.~A. and {Bianchini}, F. and {Bleem}, L.~E. and {Carlstrom}, J.~E. and {Chang}, C.~L. and {Chaubal}, P. and {Chiang}, H.~C. and {Chou}, T.~L. and {Citron}, R. and {Moran}, C. Corbett and {Crawford}, T.~M. and {Crites}, A.~T. and {de Haan}, T. and {Dobbs}, M.~A. and {Everett}, W. and {Gallicchio}, J. and {George}, E.~M. and {Gilbert}, A. and {Gupta}, N. and {Halverson}, N.~W. and {Harrington}, N. and {Henning}, J.~W. and {Hilton}, G.~C. and {Holder}, G.~P. and {Holzapfel}, W.~L. and {Hrubes}, J.~D. and {Huang}, N. and {Hubmayr}, J. and {Irwin}, K.~D. and {Knox}, L. and {Lee}, A.~T. and {Li}, D. and {Lowitz}, A. and {Luong-Van}, D. and {McMahon}, J.~J. and {Mehl}, J. and {Meyer}, S.~S. and {Millea}, M. and {Mocanu}, L.~M. and {Mohr}, J.~J. and {Montgomery}, J. and {Nadolski}, A. and {Natoli}, T. and {Nibarger}, J.~P. and {Noble}, G. and {Novosad}, V. and {Omori}, Y. and {Padin}, S. and {Pryke}, C. and {Ruhl}, J.~E. and {Saliwanchik}, B.~R. and {Sayre}, J.~T. and {Schaffer}, K.~K. and {Shirokoff}, E. and {Sievers}, C. and {Smecher}, G. and {Spieler}, H.~G. and {Staniszewski}, Z. and {Stark}, A.~A. and {Tucker}, C. and {Vanderlinde}, K. and {Veach}, T. and {Vieira}, J.~D. and {Wang}, G. and {Whitehorn}, N. and {Williamson}, R. and {Wu}, W.~L.~K. and {Yefremenko}, V.},
        title = "{An Improved Measurement of the Secondary Cosmic Microwave Background Anisotropies from the SPT-SZ + SPTpol Surveys}",
      journal = {\apj},
         year = 2021,
        month = feb,
       volume = {908},
       number = {2},
          eid = {199},
        pages = {199},
          doi = {10.3847/1538-4357/abd407},
archivePrefix = {arXiv},
       eprint = {2002.06197},
 primaryClass = {astro-ph.CO},
       adsurl = {https://ui.adsabs.harvard.edu/abs/2021ApJ...908..199R}
}

@misc{spt3gsoftware,  
  author       = {{CMB-S4/SPT-3G Collaboration}},  
  title        = {spt3g\_software: Analysis software for the South Pole Telescope},  
  year         = {2024},  
  publisher    = {GitHub},  
  journal      = {GitHub repository},  
  url          = {https://github.com/CMB-S4/spt3g_software},
  version      = {0.4}
}
\bibliographystyle{aa}

\begin{appendix}
\onecolumn
\section{Catalog samples and detailed column descriptions} \label{sec:appendixb}
Tables \ref{tab:catalog_flux} and \ref{tab:catalog_flags} include 12 catalog entries from the emissive source catalog, where the sample data are sources matched to \textit{Euclid} Q1 strong lenses \citep{walmsley25} and \textit{Planck} protoclusters \citep{dusserre25}. Table \ref{tab:cluster_catalog} includes ten random cluster catalog entries. Table \ref{tab:columnlabels} shows the exact name and format of the columns in the emissive and cluster catalog online files, as well as detailed explanations of each quantity.

    \begin{table*}[h]
    \caption{SPT-3G EDF-S emissive source catalog preview, featuring matches to \textit{Planck} protoclusters and \textit{Euclid} Q1 strong lenses.}
        \centering
        \begin{tabular}{lccccccccc}
            \hline
            \multicolumn{4}{c}{\textbf{Name and coordinates}} & 
            \multicolumn{2}{c}{\textbf{95 GHz data}} &
            \multicolumn{2}{c}{\textbf{150 GHz data}} &
            \multicolumn{2}{c}{\textbf{220 GHz data}} \\
                Index & SPT IAU name & R.A. (deg) & Dec. (deg) & S/N & $S$ (mJy) & S/N & $S$ (mJy) & S/N & $S$ (mJy) \\
                21 & SPT-S J040310-5030.2 & 60.800 & $-$50.504 & 73.62 & $24.79$ & 40.3 & $16.35$ & 8.83 & $11.53$ \\ 
                36 & SPT-S J034743-4947.5 & 56.938 & $-$49.792 & 37.75 & $12.81$ & 28.09 & $11.09$ & 8.13 & $10.57$ \\ 
                131 & SPT-S J040259-4711.6 & 60.749 & $-$47.194 & 11.26 & $3.76$ & 8.63 & $3.33$ & 2.35 & $3.04$ \\ 
                150 & SPT-S J041604-4926.0 & 64.020 & $-$49.435 & 3.52 & $1.19$ & 7.62 & $3.01$ & 7.7 & $10.07$ \\ 
                199 & SPT-S J035538-5114.2 & 58.911 & $-$51.237 & 7.17 & $2.58$ & 6.15 & $2.56$ & 2.76 & $3.77$ \\ 
                244 & SPT-S J040308-4715.7 & 60.784 & $-$47.263 & 1.57 & $0.53$ & 5.6 & $2.16$ & 4.53 & $5.84$ \\ 
                323 & SPT-S J041614-4911.8 & 64.060 & $-$49.197 & 2.16 & $0.73$ & 5.06 & $2.0$ & 2.63 & $3.43$ \\ 
                359 & SPT-S J040539-4914.9 & 61.416 & $-$49.250 & 2.04 & $0.69$ & 4.72 & $1.86$ & 5.75 & $7.51$ \\ 
                372 & SPT-S J035223-4839.2 & 58.098 & $-$48.655 & 0.19 & $0.07$ & 4.62 & $1.82$ & 5.2 & $6.79$ \\ 
                380 & SPT-S J042029-4726.6 & 65.125 & $-$47.445 & 2.03 & $0.68$ & 4.58 & $1.78$ & 5.16 & $6.66$ \\ 
                388 & SPT-S J035326-5035.4 & 58.362 & $-$50.590 & $-$0.32 & $-0.11$ & 4.51 & $1.8$ & 5.2 & $6.79$ \\ 
                463 & SPT-S J035845-4918.0 & 59.688 & $-$49.301 & 2.66 & $0.9$ & 3.64 & $1.44$ & 5.37 & $7.01$ \\ 
            \hline
        \end{tabular}
    \tablefoot{The catalog file is sorted in descending order of 150 GHz signal-to-noise ratio.}
    \label{tab:catalog_flux}
    \end{table*}
\vspace{-0.6cm}
    \begin{table*}[h]
    \caption{SPT-3G EDF-S emissive source catalog preview continued.}
        \centering
        \begin{tabular}{l cc ccccc r}
            \hline
            \multicolumn{1}{c}{} & 
            \multicolumn{2}{c}{\textbf{Spectral indices}} &
            \multicolumn{5}{c}{\textbf{Flags}} &
            \multicolumn{1}{c}{\textbf{Notes}} \\
                Index & $\alpha^{95}_{150}$ & $\alpha^{150}_{220}$ &
                Thumbnail & Extended & Inside EDF-S & DSFG & Counterparts & Angular separation \\
                21 & $-0.92 \pm 0.06$ & $-0.87 \pm 0.29$ & 0 & 0 & 1 & 0 & 1,3 & 3$^{\prime\prime}$ C grade SL \\ 
                36 & $-0.32 \pm 0.1$ & $-0.12 \pm 0.32$ & 0 & 0 & 1 & 0 & 1,3 & 2$^{\prime\prime}$ B grade SL \\ 
                131 & $-0.27 \pm 0.32$ & $-0.23 \pm 1.1$ & 0 & 0 & 1 & 0 & 1 & 281$^{\prime\prime}$ PC \\ 
                150 & $2.06 \pm 0.7$ & $3.01 \pm 0.46$ & 0 & 0 & 1 & 1 & -- & 68$^{\prime\prime}$ PC \\ 
                199 & $-0.02 \pm 0.48$ & $0.97 \pm 0.99$ & 0 & 0 & 1 & 0 & 1 & 25$^{\prime\prime}$ C grade SL \\ 
                244 & $3.14 \pm 1.47$ & $2.48 \pm 0.71$ & 0 & 0 & 1 & 1 & -- & 117$^{\prime\prime}$ PC \\ 
                323 & $2.23 \pm 1.12$ & $1.35 \pm 1.07$ & 0 & 0 & 1 & 0 & 1 & 25$^{\prime\prime}$ C grade SL \\ 
                359 & $2.2 \pm 1.19$ & $3.47 \pm 0.68$ & 0 & 0 & 1 & 1 & -- & 22$^{\prime\prime}$ C grade SL \\ 
                372 & $7.4 \pm 11.55$ & $3.28 \pm 0.72$ & 0 & 0 & 1 & 1 & -- & 68$^{\prime\prime}$ PC \\ 
                380 & $2.13 \pm 1.2$ & $3.29 \pm 0.73$ & 0 & 1 & 1 & 1 & -- & 40$^{\prime\prime}$ C grade SL \\ 
                388 & $-$ & $3.31 \pm 0.73$ & 0 & 0 & 1 & 0 & -- & 15$^{\prime\prime}$ C grade SL \\ 
                463 & $1.04 \pm 1.03$ & $3.95 \pm 0.83$ & 0 & 0 & 1 & 1 & -- & 113$^{\prime\prime}$ PC \\ 
            \hline
        \end{tabular}
    \tablefoot{The external counterparts correspond to the following wavelengths: 1 = radio (ASKAP low, mid, or high), 2 = infrared (WISE), and 3 = millimeter (SPT-SZ). The notes column indicates angular separations from \textit{Planck} protoclusters (PC) and \textit{Euclid} Q1 strong lenses (SL), including the grade of strong lens.}
    \label{tab:catalog_flags}
    \end{table*}
\vspace{-0.25cm}
    \begin{table*}[h]
    \caption{SPT-3G EDF-S cluster catalog preview featuring ten cluster candidates chosen at random.}
        \centering
        \begin{tabular}{lccccccc}
            \hline
            \multicolumn{3}{c}{\textbf{Name and coordinates}} & 
            \multicolumn{2}{c}{\textbf{Best}} &
            \multicolumn{3}{c}{\textbf{Cluster attributes}} \\
                SPT IAU name & R.A. (deg) & Dec. (deg) & $\xi$ & $\theta_{\mathrm{core}}$ (arcmin) & $z$ & $M_{500c}\times10^{-14}$(\msun) & $\lambda$ \\
                
                SPT-CL J0339-4800 & 54.786 & $-$48.011 & 10.8 & 1 & $0.49 \pm 0.008$ & $3.21^{+0.39}_{-0.46}$ & $103 \pm 10.6$ \\
                SPT-CL J0350-4620 & 57.717 & $-$46.334 & 25.7 & 0.75 & $0.79 \pm 0.02$ & $4.91^{+0.54}_{-0.64}$ & $194 \pm 17.6$ \\
                SPT-CL J0357-4530 & 59.387 & $-$45.51 & 4.52 & 0.25 & $0.88 \pm 0.04$ & $1.54^{+0.27}_{-0.33}$ & $80.2 \pm 10.7$ \\
                SPT-CL J0401-4911 & 60.415 & $-$49.200 & 5.18 & 0.5 & $0.94 \pm 0.06$ & $1.65^{+0.29}_{-0.34}$ & $46.7 \pm 10.4$ \\
                SPT-CL J0403-4828 & 60.882 & $-$48.479 & 5.94 & 1.75 & $0.34 \pm 0.008$ & $2.25^{+0.36}_{-0.41}$ & $92.1 \pm 10.6$ \\
                SPT-CL J0411-4819 & 62.812 & $-$48.322 & 45.2 & 1.25 & $0.42 \pm 0.007$ & $7.68^{+0.8}_{-0.97}$ & $313 \pm 19.5$ \\
                SPT-CL J0418-5120 & 64.734 & $-$51.345 & 10.1 & 0.25 & $1.1 \pm 0.09$ & $2.50^{+0.29}_{-0.35}$ & $26.9 \pm 10.9$ \\
                SPT-CL J0421-4740 & 65.461 & $-$47.671 & 4.41 & 0.25 & $1.5 \pm 0.07$ & $1.23^{+0.2}_{-0.27}$ & $47.8 \pm 7.97$ \\
                SPT-CL J0426-4749 & 66.55 & $-$47.83 & 5.67 & 0.75 & $1.3 \pm 0.07$ & $1.56^{+0.26}_{-0.3}$ & $59 \pm 8.71$ \\
                SPT-CL J0431-4928 & 67.912 & $-$49.475 & 4.61 & 1 & $0.45 \pm 0.01$ & $1.81^{+0.32}_{-0.39}$ & $54.2 \pm 8.43$ \\
            \hline
        \end{tabular}
    \tablefoot{The catalog file is sorted by ascending R.A. value.}
    \label{tab:cluster_catalog}
    \end{table*}

    \begin{table*}[h]
    \caption{Exact column labels and their descriptions for the catalogs.}
                \centering
                \begin{tabularx}{\textwidth}{ l  X }
            \hline
            \textbf{Emissive catalog column label} & \textbf{Emissive source catalog description} \\
            \texttt{\fontfamily{pcr}\selectfont index} & Integer index where the catalog is sorted in descending order of 150 GHz signal-to-noise. \\
            \texttt{\fontfamily{pcr}\selectfont iau\_name} & IAU name follows the convention of a catalog prefix (``SPT-S''), R.A., and Dec. in \mbox{sexagesimal} format. \\
            \texttt{\fontfamily{pcr}\selectfont ra} & Right ascension (J2000) of the source in the highest detection band in degrees. \\
            \texttt{\fontfamily{pcr}\selectfont dec} & Declination (J2000) of the source in the highest detection band in degrees. \\
            \texttt{\fontfamily{pcr}\selectfont sn95} & 95 GHz signal-to-noise ratio of the detection. Values below 5$\sigma$ are pulled from the signal-to-noise map when the source was detected above 5$\sigma$ in another band. \\
            \texttt{\fontfamily{pcr}\selectfont sn150} & 150 GHz signal-to-noise ratio of the detection. \\
            \texttt{\fontfamily{pcr}\selectfont sn220} & 220 GHz signal-to-noise ratio of the detection. \\
            \texttt{\fontfamily{pcr}\selectfont s95} & Flux at 95 GHz in mJy converted from CMB temperature. Fluxes are from forced-photometry in bands where the source was not detected $>5\sigma$. \\
            \texttt{\fontfamily{pcr}\selectfont s150} & Flux at 150 GHz in mJy. \\
            \texttt{\fontfamily{pcr}\selectfont s220} & Flux at 220 GHz in mJy. \\
            \texttt{\fontfamily{pcr}\selectfont alpha95} & Spectral index between 95 and 150 GHz ($\alpha^{95}_{150}$, Eq. \ref{eqn:alpha}). \\
            \texttt{\fontfamily{pcr}\selectfont erroralpha95} & The error on $\alpha^{95}_{150}$ propagated from the flux errors. \\
            \texttt{\fontfamily{pcr}\selectfont alpha220} & Spectral index between 150 and 220 GHz ($\alpha^{150}_{220}$, Eq. \ref{eqn:alpha}). \\
            \texttt{\fontfamily{pcr}\selectfont erroralpha220} & The error on $\alpha^{150}_{220}$ propagated from the flux errors. \\
            \texttt{\fontfamily{pcr}\selectfont thumb} & ``Coadd[frame name]'' if the source is one of the specially treated bright source thumbnails (Sect. \ref{sec:brightsrcs}), 0 (for \texttt{False}) otherwise.\tablefootmark{a}\\
            \texttt{\fontfamily{pcr}\selectfont extended} & 1 for \texttt{True} if flagged according to the methods described in Sect. \ref{sec:flags}, 0 for \texttt{False} if point-like. \\
            \texttt{\fontfamily{pcr}\selectfont edfs\_flag} & 1 for \texttt{True} if inside the EDF-S, 0 for \texttt{False} if outside the EDF-S. \\
            \texttt{\fontfamily{pcr}\selectfont dsfg\_flag} & 1 for \texttt{True} if flagged as a DSFG candidate (see Sect. \ref{sec:flags}), 0 for \texttt{False} otherwise. \\
            \texttt{\fontfamily{pcr}\selectfont counterparts} & External catalog counterparts according to Sect. \ref{sec:crossmatch} where 1 = radio (ASKAP low, mid, or high), 2 = infrared (WISE), and 3 = millimeter (SPT-SZ). \\[1cm]
            \hline
            \textbf{Cluster catalog column label} & \textbf{Cluster catalog description} \\
            \texttt{\fontfamily{pcr}\selectfont iau\_name} & Name of the SPT cluster candidate. The IAU name follows the convention of a catalog prefix (``SPT-CL''), R.A., and Dec. in sexagesimal format. \\
            \texttt{\fontfamily{pcr}\selectfont ra} & Right ascension (J2000) of the tSZ detection in degrees. \\
            \texttt{\fontfamily{pcr}\selectfont dec} & Declination (J2000) of the tSZ detection in degrees. \\
            \texttt{\fontfamily{pcr}\selectfont xi} & SPT cluster detection significance (see Sect. \ref{sec:clusterfinding}). \\
            \texttt{\fontfamily{pcr}\selectfont theta\_core} & $\beta$-profile core radii used in the matched filter corresponding to \texttt{xi}. \\
            \texttt{\fontfamily{pcr}\selectfont redshift} & Redshift of associated optical/infrared galaxy overdensity. If no associated redshift is available, the column value is $-$1. \\
            \texttt{\fontfamily{pcr}\selectfont redshift\_unc} & Redshift uncertainty if available, $-1$ otherwise. \\
            \texttt{\fontfamily{pcr}\selectfont specz} & Equals 1 if \texttt{redshift} is spectroscopic. \\
            \texttt{\fontfamily{pcr}\selectfont m500} & Mass $M_\textrm{500c}$ in units of $ 10^{14}$\msun. \\
            \texttt{\fontfamily{pcr}\selectfont m500\_uerr} & 1 sigma upper uncertainty on mass. \\
            \texttt{\fontfamily{pcr}\selectfont m500\_lerr} & 1 sigma lower uncertainty on mass. \\
            \texttt{\fontfamily{pcr}\selectfont lambda} & Richness of associated optical/infrared galaxy overdensity. \\
            \texttt{\fontfamily{pcr}\selectfont lambda\_unc} & Richness uncertainty. \\
            \texttt{\fontfamily{pcr}\selectfont fcont} & Integrated optical/infrared contamination for richness $>$ \texttt{lambda} ($f_\textrm{cont}$, see Sect. \ref{sec:redshifts}). \\
            \texttt{\fontfamily{pcr}\selectfont edfs\_flag} & 1 for \texttt{True} if inside the EDF-S, 0 for \texttt{False} if outside the EDF-S. \\
            \texttt{\fontfamily{pcr}\selectfont los} & Equals 1 if there is a secondary structure along the LOS with $f_\textrm{cont}<$0.2. \\
            \texttt{\fontfamily{pcr}\selectfont redshift2} & Redshift secondary structure. \\
            \texttt{\fontfamily{pcr}\selectfont redshift2\_unc} & Redshift uncertainty secondary structure. \\
            \texttt{\fontfamily{pcr}\selectfont lambda2} & Richness secondary structure. \\
            \texttt{\fontfamily{pcr}\selectfont lambda2\_unc} & Richness uncertainty secondary structure. \\
            \texttt{\fontfamily{pcr}\selectfont fcont2} & Integrated optical/infrared contamination for richness $>$ \texttt{lambda2} for secondary structure. \\
            \texttt{\fontfamily{pcr}\selectfont unconfirmed\_redshift} & Redshift for LOS overdensity below confirmation threshold (only provided when candidate is not confirmed). \\
            \texttt{\fontfamily{pcr}\selectfont unconfirmed\_redshift\_unc} & Redshift uncertainty for LOS overdensity below confirmation threshold. \\
            \texttt{\fontfamily{pcr}\selectfont unconfirmed\_lambda} & Richness of most significant LOS galaxy overdensity when candidate is not confirmed. \\
            \texttt{\fontfamily{pcr}\selectfont unconfirmed\_lambda\_unc} & Richness uncertainty for LOS overdensity below confirmation threshold. \\
            \texttt{\fontfamily{pcr}\selectfont unconfirmed\_fcont} & Integrated optical/infrared contamination for LOS overdensity below confirmation threshold.\tablefootmark{b} \\
            \hline
                \end{tabularx}
                \tablefoot{\tablefoottext{a}{The bright source's label in the \texttt{\fontfamily{pcr}\selectfont thumb} column comes from the center of the thumbnail and is therefore not guaranteed to equate to the IAU name of the source, in the event of the source's recorded location shifting from the precise center of the thumbnail.}\\ \tablefoottext{b}{For unconfirmed cluster candidates ($f_\textrm{cont}>0.2$), we provide properties of the most significant galaxy overdensity detected along the LOS, but do not report masses for these structures.}}
                \label{tab:columnlabels}
        \end{table*}
\end{appendix}

\end{document}